\documentclass[11pt,dvips,twoside,letterpaper]{article} 
\usepackage{graphicx} 
\usepackage{pslatex} 
\usepackage{fancyhdr} 
\usepackage{geometry}

\usepackage{epsfig} 
\usepackage{amsmath}
\usepackage{lscape}

\begin{document}
\title{
~\\[-1.2in]
{\normalsize\noindent
\begin{picture}(0,0)(208,-3.15)
\begin{tabular}{l p{0.10in} r}
In: New Developments in Ferromagnetism Research  & & ISBN 1-59454-461-1 \\
Editor: V.N. Murray, pp. 131-188  & & \copyright 2005 Nova Science Publishers, Inc.\\
\end{tabular}
\end{picture}}
{\begin{flushleft} ~\\[0.63in]{\normalsize\bfseries\textit{Chapter~6}} \end{flushleft} ~\\[0.13in] 
\bfseries\scshape The Classical Spectral Density Method\\ at Work: The Heisenberg Ferromagnet}}
\author{
\bfseries\itshape A. Cavallo, F. Cosenza, L. De Cesare\\
Dipartimento di Fisica "E.R. Caianiello", Universit\`{a} degli Studi di
Salerno,\\
Via S. Allende, 84081 Baronissi (SA), Italy \\
Istituto Nazionale di Fisica della Materia (INFM), Unit\`{a} di Salerno,
Italy}
\date{}
\maketitle
\thispagestyle{empty}
\setcounter{page}{131}

\pagestyle{fancy}
\fancyhead{}
\fancyhead[EC]{A. Cavallo, F. Cosenza, L. De Cesare}
\fancyhead[EL,OR]{\thepage}
\fancyhead[OC]{The Classical Spectral Density Method at Work: The Heisenberg Ferromagnet}
\fancyfoot{}
\renewcommand\headrulewidth{0.5pt}
\addtolength{\headheight}{2pt} 
\headsep=9pt

\begin{abstract}
In this article we review a less known unperturbative and powerful many-body
method in the framework of classical statistical mechanics and then we show
how it works by means of explicit calculations for a nontrivial classical
model. The formalism of two-time Green functions in classical statistical
mechanics is presented in a form parallel to the well known quantum
counterpart, focusing on the spectral properties which involve the important
concept of spectral density. Furthermore, the general ingredients of the
classical spectral density method (CSDM) are presented with insights for
systematic nonperturbative approximations to study conveniently the
macroscopic properties of a wide variety of classical many-body systems also
involving phase transitions. The method is implemented by means of key ideas
for exploring the spectrum of elementary excitations and the damping effects
within a unified formalism. Then, the effectiveness of the CSDM is tested
with explicit calculations for the classical $d$-dimensional spin-$S$
Heisenberg ferromagnetic model with long-range exchange interactions
decaying as $r^{-p}$ ($p>d$) with distance $r$ between spins and in the
presence of an external magnetic field. The analysis of the thermodynamic
and critical properties, performed by means of the CSDM to the lowest order
of approximation, shows clearly that nontrivial results can be obtained in a
relatively simple manner already to this lower stage. The basic spectral
density equations for the next higher order level are also presented and the
damping of elementary spin excitations in the low temperature regime is
studied. The results appear in reasonable agreement with available exact
ones and Monte Carlo simulations and this supports the CSDM as a promising
method of investigation in classical many-body theory.

\end{abstract}

\section{Introduction}

At the present a wide variety of methods exists to calculate the macroscopic
equilibrium and nonequilibrium quantities of many-body systems. However,
their potentialities and efficiencies differ sensibly especially when small
parameters are absent for intrinsic reasons and the ordinary perturbative
expansions appear inadequate. There exist, for instance, many systems of
experimental and theoretical interest which involve strongly interacting
microscopic degrees of freedom and exhibit anomalous behavior of the
specific heat, susceptibility and other macroscopic quantities which, in
turn, indicate various phase transitions to occur in these systems.
Therefore, there is always a great interest to search for reliable methods
going beyond the conventional perturbation theory to describe correctly
nonperturbative phenomena and, especially, the various anomalies in the
thermodynamic behavior of systems near a phase transition. In this
connection, the two-time Green function (GF) technique constitutes one of
the most powerful tools in quantum statistical mechanics and in condensed
matter physics to explore the thermodynamic and transport properties of a
wide variety of many-body systems. Within this framework, the equations of
motion method (EMM) and the spectral density method (SDM) allow to obtain
reliable approximations to treat typically nonperturbative problems \cite{Zubarev60,Tyablikov67,Majlis,Kalashnikov69,Kalashnikov73}.

When one mentions the two time GF technique, one refers usually to their
quantum many-body version which is well known from long time and widely and
successfully applied in quantum statistical physics. Nevertheless, the
pioneering introduction of the two-time GF's and the EMM in classical
statistical mechanics by Bogoljubov and Sadovnikov \cite{Bogoljubov62} has
opened the concrete possibility to describe classical and quantum many-body
systems on the same footing. In this context, it is worth mentioning that,
in many physical situations (when the quantum effects are negligible), the
use of the classical two-time GF formalism may offer substantial advantages
especially from the computational point of view because in the calculations
one handles only functions and not operators. In any case, the strong impact
of the two-time GF technique in many-body physics and the continuous efforts
for obtaining better approximations are well known and, hence, it is
unnecessary to stress again its effectiveness and potentialities. Rather,
the less known SDM merits further remarks.

The SDM was formulated some decades ago by Kalashnikov and Fradkin for
quantum many-body systems \cite{Kalashnikov69,Kalashnikov73}. It is
sufficiently simple and has been applied to a wide variety of quantum
systems, also exhibiting phase transitions \cite{Kalashnikov73}, as
superconductors \cite{Kalashnikov69,Kalashnikov73}, magnetic \cite{Kalashnikov69,Kalashnikov73,Campana79} and bosonic \cite{Caramico80}
systems, strongly correlated electron compounds \cite{Kalashnikov73,Nolting89,Nolting91,Hermann97}, and so on. It appears more
effective than the EMM for a direct calculation of two-time GF's, not only
because it allows to obtain reliable and systematic nonperturbative
approximations, but also because it assures the validity of a number of sum
rules irrespective either to the interaction strength or any other
characteristics inherent to the system under study.

Next, a classical version of the SDM (CSDM) has been also formulated in
extensive \cite{Caramico81,Campana83,Campana84} and nonextensive \cite{Cavallo01} classical statistical mechanics and applied to classical
magnetic systems with short and long-range exchange interactions. The last
method, strictly related to the classical two-time GF technique \cite{Bogoljubov62}, offers a robust technical instrument for systematic and well
tested nonperturbative approximations to explore the equilibrium and
transport properties of classical many-body systems as well, parallel to the
quantum counterpart.
Unfortunately, both the mentioned methods in the classical context have not
received the due consideration and further developments and applications are
desirable.

In this article, also with a pedagogical aim in mind, we first review the
basic formalism of two-time GF's in classical statistical mechanics in a
form strictly parallel to the quantum better known counterpart emphasizing
the concept of spectral density (SD), quite important for next purposes, and
its relation to the two-time GF's. Besides, as a convenient alternative to
the EMM for calculation of the GF's, the basic ingredients of the CSDM are
presented in some details, together with useful insights for systematic
nonperturbative approximations. Then, we clarify how this method can be
conveniently used to explore macroscopic properties of classical many-body
systems. We introduce also the key ideas for studying, in a conceptually
transparent way, the elementary excitation spectrum, or dispersion relation,
and the damping effects in a unified way within the formalism of CSDM. The
general presentation of the method in the classical many-body framework will
be performed in a form strictly related to the well known quantum
counterpart \cite{Zubarev60,Tyablikov67,Kalashnikov73}, giving the
possibility to use powerful calculation techniques, approximation methods
and the basic terminology which are typical of the quantum many-body theory.

Next, we will show, with explicit calculations, how the CSDM works emphasizing
its effectiveness already to the lowest order of approximation. As a
nontrivial theoretical laboratory, we choose the classical $d$-dimensional
spin-$S$ Heisenberg ferromagnet with long-range interactions (LRI's),
decaying as $r^{-p}$ ($p>d$) with distance $r$ between spins, in the
presence of an external magnetic field. For this model, besides some partial
exact results \cite{Kunz76,Frolich78,Roger81,Pfister81,Bruno01}, a lot of
new information about thermodynamic and critical properties have been
obtained by means of Monte Carlo (MC) simulations \cite{Romano89,Romano90}.
As we shall see, the CSDM allows us to obtain, in a relatively simple way,
untrivial results which appear to be in a reasonable agreement with the
available exact ones and MC simulations supporting the SDM as a promising
method of investigation also in classical statistical mechanics.

The paper is organized as follows. In Sec. 2, we introduce the two-time GF's
and the related SD's in classical statistical mechanics for two arbitrary
dynamic variables. Section 3 is devoted to the methods of calculation
focusing on the general formulation of the CSDM. Here, the method is
properly implemented for exploring also the damping of the classical
excitations within the same basic formalism. The method at work is presented
in Sec. 4 with a study of the thermodynamic properties and the critical
phenomena of the classical $d$-dimensional spin-$S$ Heisenberg ferromagnetic
model with LRI's. Here, after a summary of known theoretical and
simulational information about the model for a next checking of the
effectiveness of the CSDM, we present explicit calculations performed by
using the general many-body formalism developed before. As a successive
useful comparison, we first formulate the problem in the context of the EMM
for the two-time GF's appropriate to the spin model under study and the
basic equations in the Tyablikov-like approximation (classical version of
the well known quantum Tyablikov one \cite{Tyablikov67}) are presented.
Then, the basic SD and related moment equations (ME's) are derived, the
thermodynamic and critical properties of the model are presented working in
the so called one-pole approximation for the SD and a comparison with the
Tyablikov-like results is performed. To underline the effectiveness of this
lowest order approximation and the nature of difficulties arising to higher
orders, we deduce the basic equations for the two-poles approximation. As a
further insight, we use the implemented CSDM to obtain, in a unified and
relatively simple way, the dispersion relation and the damping of classical
spin excitations. Explicit results are given (for the short-range FM chain)
in the low-temperature limit. Finally, in Sec. 5 some conclusions are drawn.

\section{Two-Time Green Functions and Spectral Densities\\ in Classical
Statistical Mechanics}

\subsection{Basic Definitions and Spectral Representations}

In strict analogy with the quantum case \cite
{Zubarev60,Tyablikov67,Majlis,ZubarevB}, we define \cite{Bogoljubov62} the
two-time retarded $(\nu =r)$ and advanced $(\nu =a)$ GF's in classical
statistical mechanics for two arbitrary dynamical variables $A$ and $B$ as%
\footnote{%
The classical retarded GF, as in the quantum one, has a direct physical
meaning which arises from the theory of linear response for transport
phenomena in classical statistical mechanics. The advanced one is introduced
only for a full development of the two-time GF formalism.} 
\begin{equation}
G_{AB}^{(\nu )}(t,t^{\prime })=\theta _{\nu }(t-t^{\prime })\left\langle
\left\{ A(t),B\left( t^{\prime }\right) \right\} \right\rangle \equiv
\left\langle \left\langle A(t);B\left( t^{\prime }\right) \right\rangle
\right\rangle _{\nu } 
\begin{array}{cc}
, & (\nu =r,a),
\end{array}
\label{Eq2.1}
\end{equation}
where $\theta _{r}(t-t^{\prime })=\theta (t-t^{\prime })$, $\theta_{a}(t-t^{\prime })=-\theta (t^{\prime }-t)$ and $\theta (x)$ is the usual
step function. In Eq. (\ref{Eq2.1}), $\left\langle ...\right\rangle $
denotes an equilibrium or a nonequilibrium ensemble average \cite{ZubarevB}
and $\left\{ A,B\right\} $ is the Poisson bracket of $A$ and $B$. In the
following, we refer only to the equilibrium (generalized) ensemble average.
In this case one can easily prove \cite{ZubarevB} that the two-time GF's (%
\ref{Eq2.1}) and the corresponding correlations functions depend on times $t$%
, $t^{\prime }$ only through the difference $t-t^{\prime }$. So, one can
write 
\begin{equation}
G_{AB}^{(\nu )}(t-t^{\prime })=\left\langle \left\langle A(t-t^{\prime
});B\right\rangle \right\rangle _{\nu }=\left\langle \left\langle
A;B(t^{\prime }-t)\right\rangle \right\rangle _{\nu }.  \label{Eq2.2}
\end{equation}
In the previous equations, the dynamical variables $A$ and $B$ depend on
time via the conjugate canonical coordinates $(q(t),p(t))\equiv
(q_{1}(t),..,q_{\mathcal{N}}(t);p_{1}(t),..,p_{\mathcal{N}}(t))$, ($\mathcal{%
N}$ is the number of degrees of freedom of the classical system under
study), $X(t)=e^{iLt}X(0)$ with $X=A,B,q,p$, $L=i\{H,...\}$ is the Liouville
operator\footnote{%
In literature [23] one finds also the alternative definition $L=-i\{H,...\}$%
. However, the final conclusions remain unchanged.} and $H$ is the
Hamiltonian. Here, $e^{iLt}$ acts as a classical time-evolution operator
which transforms the dynamical variable $X(0)\equiv X(q(0),p(0))$ at the
initial time $t=0$ into the dynamical variable $X(t)\equiv X(q(t),p(t))$ at
the arbitrary time $t$. Of course, the time evolution of the generic
dynamical variable $X(t)$ is governed by the well known Liouville equation
of motion (EM) 
\begin{equation}
{\frac{{dX(t)}}{{dt}}}=\left\{ {X(t),H}\right\} .  \label{Eq2.3}
\end{equation}

With the previous definitions, one can show \cite{Bogoljubov62} that the
two-time correlation function $F_{AB}(t,t^{\prime })=F_{AB}(t-t^{\prime })={%
\left\langle A(t)B(t^{\prime })\right\rangle }={\left\langle A(\tau
)B\right\rangle }={\left\langle AB(-\tau )\right\rangle }$, with $\tau
=t-t^{\prime }$, is related to the classical GF's (\ref{Eq2.2}) by the
relation 
\begin{equation}
G_{AB}^{(\nu)}(\tau)=\beta\theta_{\nu}(\tau) \frac{d}{d\tau}\left\langle{A(\tau)B}\right\rangle=\beta\theta_{\nu}(\tau)\left\langle\left\{{A(\tau),H}\right\}{B}\right\rangle,  \label{Eq2.4}
\end{equation}
where $\beta =\left( {K_{B}T}\right) ^{-1}$, $T$ is the temperature and $K_{B}$ is the Boltzmann constant. In particular, we have also 
\begin{equation}
\left\langle {\left\{ {A(\tau ),B}\right\} }\right\rangle =\beta {\frac{d}{{%
d\tau }}}\left\langle {A(\tau )B}\right\rangle =\beta \left\langle {\left\{ {%
A(\tau ),H}\right\} B}\right\rangle ,  \label{Eq2.5}
\end{equation}
which connects the Poisson bracket of two dynamical variables and the
corresponding dynamical correlation function.

Working, as assumed before, within equilibrium Gibbs ensembles, one can
introduce for $G_{AB}^{(\nu )}(\tau )$ and $F_{AB}(\tau )$ the Fourier
transforms 
\begin{equation}
G_{AB}^{(\nu )}(\tau )=\int_{-\infty }^{+\infty }{{\frac{{d\omega }}{{2\pi }}%
}}G_{AB}^{(\nu )}(\omega )e^{-i\omega \tau },  \label{Eq2.6}
\end{equation}

\begin{equation}
F_{AB}(\tau )=\int_{-\infty }^{+\infty }{{\frac{{d\omega }}{{2\pi }}}}%
F_{AB}(\omega )e^{-i\omega \tau },  \label{Eq2.7}
\end{equation}
where $G_{AB}^{(\nu )}(\omega )=\left\langle {\left\langle {A(\tau );B}%
\right\rangle }\right\rangle _{\nu ,\omega }$ will be named the $\nu $-GF's
of $A$ and $B$ in the $\omega $-representation and $F_{AB}(\omega
)=\left\langle {A(\tau )B}\right\rangle _{\omega }$ is called the classical
spectral intensity of the time-dependent correlation function $F_{AB}(\tau )$%
, with $\Im (\omega )=\int_{-\infty }^{+\infty }d\tau e^{i\omega \tau }\Im
(\tau )$. Then, using Eq. (\ref{Eq2.4}) and the integral representations%
\begin{equation}
\theta (\tau )={i}\int_{-\infty }^{+\infty }\frac{{dx}}{{2\pi }}{{\frac{{%
e^{-ix\tau }}}{{x+i\varepsilon }}}} 
\begin{array}{ccc}
, & \varepsilon \rightarrow 0^{+}; & 
\end{array}
\delta (x)=\int_{-\infty }^{+\infty }\frac{{d\tau }}{{2\pi }}{e^{ix\tau },}
\label{Eq2.8}
\end{equation}
for the step function and the Dirac $\delta $-function, the Fourier
transforms of the two-time GF (\ref{Eq2.1}) for two dynamical variables $A$
and $B$ can be expressed in terms of the corresponding spectral intensity as:
\begin{equation}
G_{AB}^{(\nu )}(\omega )=\int_{-\infty }^{+\infty }{{\frac{{d\omega }%
^{\prime }}{{2\pi }}}}{\frac{{\beta \omega }^{\prime }{F_{AB}(\omega }%
^{\prime }{)}}{{\omega -\omega }^{\prime }{+(-1)^{\nu }i\varepsilon }}}\quad
,\quad \varepsilon \rightarrow 0^{+},  \label{Eq2.9}
\end{equation}
where the symbol $(-1)^{\nu }$ means $+1$ if $\nu =r$ and $-1$ if $\nu =a$.
It is interesting to compare Eq. (\ref{Eq2.9}) with the quantum
corresponding one for two operators $A$ and $B$ \cite
{Zubarev60,Tyablikov67,Majlis,ZubarevB} 
\begin{equation}
G_{AB}^{(\nu )}(\omega )=\int_{-\infty }^{+\infty }{{\frac{{d\omega }%
^{\prime }}{{2\pi }}}}{\frac{{(1+\eta e^{-\beta \hbar \omega ^{\prime
}})F_{AB}(\omega }^{\prime }{)}}{{\omega -\omega }^{\prime }{+(-1)^{\nu
}i\varepsilon }}}\quad ,\quad \varepsilon \rightarrow 0^{+},  \label{Eq2.10}
\end{equation}
where $\eta =-1$ and $\eta =+1$ for definition of quantum two-time GF's with
commutator or anticommutator, respectively, and $\hbar$ is the reduced
Planck constant. Notice that formally, as expected for internal consistency,
the function $C(\omega )=\beta \omega $ or $Q_{\eta }(\omega )=(1+\eta
e^{-\beta \hbar \omega })/\hbar $ characterizes the classical or quantum
nature of the problem under study, respectively.

In analogy to the quantum case \cite{Kalashnikov69,Kalashnikov73}, we now
introduce the time-dependent classical spectral density (CSD) for $A$ and $B$
\cite{Caramico81,Campana83,Campana84,Cavallo02} 
\begin{equation}
\Lambda _{AB}(\tau )=i\left\langle {\left\{ {A(\tau ),B}\right\} }%
\right\rangle .  \label{Eq2.11}
\end{equation}
Taking its Fourier transform, we get 

\begin{equation}
\Lambda _{AB}(\omega )=i\left\langle {\left\{ {A(\tau ),B}\right\} }%
\right\rangle _{\omega }=\int_{-\infty }^{+\infty }{d\tau }e^{i\omega \tau
}\Lambda _{AB}(\tau )=\beta \omega F_{AB}(\omega ).  \label{Eq2.12}
\end{equation}
Hence, from Eq. (\ref{Eq2.9}), one immediately obtains the spectral
representation 
\begin{equation}
G_{AB}^{(\nu )}(\omega )=\int_{-\infty }^{+\infty }{{\frac{{d\omega }%
^{\prime }}{{2\pi }}}}{\frac{{\Lambda _{AB}(\omega }^{\prime }{)}}{{\omega
-\omega }^{\prime }{+(-1)^{\nu }i\varepsilon }}}\quad ,\quad \varepsilon
\rightarrow 0^{+},  \label{Eq2.13}
\end{equation}
for the two-time GF's (\ref{Eq2.1}) in terms of the corresponding CSD $%
\Lambda _{AB}(\omega )$ in the $\omega $-representation. Also the dynamical
correlation function $\left\langle A(\tau )B\right\rangle $ can be easily
expressed in terms of $\Lambda _{AB}(\omega )$. From Eqs. (\ref{Eq2.5}), (%
\ref{Eq2.11}) and (\ref{Eq2.12}), we obtain indeed %
\begin{equation}
\left\langle A(\tau )B\right\rangle =\int_{-\infty }^{+\infty }{\frac{%
d\omega }{2\pi }}{\frac{\Lambda _{AB}(\omega )}{\beta \omega }}e^{-i\omega
\tau }.  \label{Eq2.14}
\end{equation}
Eqs. (\ref{Eq2.13}) and (\ref{Eq2.14}) assume a particular importance for
our purposes. If one determines independently the $\Lambda _{AB}(\omega )$,
these equations allow us to obtain the time-dependent correlation and Green
functions and therefore the macroscopic properties of classical many-body
systems. We will show below that, also in classical statistical mechanics,
it is possible to construct a formalism which allows for a systematic
calculation of the CSD.

From Eqs. (\ref{Eq2.11})-(\ref{Eq2.14}) some immediate exact results can be
easily obtained. First, Eqs. (\ref{Eq2.11}) and (\ref{Eq2.12}) yield 
\begin{equation}
\int_{-\infty }^{+\infty }{{\frac{{d\omega }}{{2\pi }}}}\Lambda _{AB}(\omega
)=i\left\langle {\left\{ {A,B}\right\} }\right\rangle .  \label{Eq2.15}
\end{equation}
Besides, from (\ref{Eq2.14}), it follows 
\begin{equation}
\int_{-\infty }^{+\infty }{{\frac{{d\omega }}{{2\pi }}}}{\frac{\Lambda
_{AB}(\omega )}{\beta \omega }}=\left\langle AB\right\rangle .
\label{Eq2.16}
\end{equation}
The relations (\ref{Eq2.15}) and (\ref{Eq2.16}) are formally exact and
constitute useful examples of the so called \textit{sum rules} of the CSD, $%
\Lambda _{AB}(\omega )$, which have great relevance for physical consistency
of practical calculations and approximations. Combining now Eqs. (\ref
{Eq2.13}) and (\ref{Eq2.15}), one can easily prove another general result
which plays an important role for calculation of the GF's. As $\omega
\rightarrow \infty $ we have indeed %
\begin{eqnarray}
G_{AB}^{(\nu )}(\omega ) &=&\omega ^{-1}\int_{-\infty }^{+\infty }{{\frac{{%
d\omega }^{\prime }}{{2\pi }}}}{\frac{{\Lambda _{AB}(\omega }^{\prime }{)}}{{%
1-{\frac{{\omega }^{\prime }{-(-1)^{\nu }i\varepsilon }}{\omega }}}}}  \notag
\\
&\approx &{\frac{1}{\omega }\int_{-\infty }^{+\infty }{{\frac{{d\omega
^{^{\prime }}}}{{2\pi }}}}\Lambda _{AB}(\omega ^{^{\prime }})\left[ {1+{%
\frac{{\omega }^{\prime }{-(-1)^{\nu }i\varepsilon }}{\omega }}+{\frac{1}{2}}%
{\frac{{\left( {\omega }^{\prime }{-(-1)^{\nu }i\varepsilon }\right) ^{2}}}{{%
\omega ^{2}}}}+...}\right] }  \notag \\
&=&{\frac{{i\left\langle {\left\{ {A,B}\right\} }\right\rangle }}{\omega }}+{%
\frac{1}{{\omega ^{2}}}}\int_{-\infty }^{+\infty }{{\frac{{d\omega }^{\prime
}}{{2\pi }}}}\Lambda _{AB}(\omega ^{\prime })\left( {\omega }^{\prime }{%
-(-1)^{\nu }i\varepsilon }\right) +O\left( {{\frac{1}{{\omega ^{3}}}}}%
\right) .  \notag \\
&&  \label{Eq2.17}
\end{eqnarray}
So, we can write 
\begin{equation}
G_{AB}^{(\nu )}(\omega )=\left\{ 
\begin{array}{ccc}
{\frac{{i\left\langle {\left\{ {A,B}\right\} }\right\rangle }}{\omega }}\sim
\omega ^{-1} & , & if\quad \left\langle {\left\{ {A,B}\right\} }%
\right\rangle \neq 0 \\ 
\sim \omega ^{-\alpha }\quad (\alpha \geq 2) & , & if\quad \left\langle {%
\left\{ {A,B}\right\} }\right\rangle =0,
\end{array}
\right.  \label{Eq2.18}
\end{equation}
which provide a relevant boundary conditions for the $\nu $-GF's.

Let's come back now to the relations (\ref{Eq2.13}) for classical retarded and
advanced GF's in the $\omega $-representation. As in the quantum counterpart 
\cite{Zubarev60,Tyablikov67}, using the definitions (\ref{Eq2.1}) and (\ref
{Eq2.2}), one can easily prove that $G_{AB}^{(r)}(\omega )$ and $%
G_{AB}^{(a)}(\omega )$, analytically continued in the $\omega $-complex
plane, are analytical functions in the upper and lower half-plane,
respectively. Then, combining these two analytical functions, one can
construct a single function\linebreak $G_{AB}(\omega )=\int_{-\infty }^{+\infty
}dte^{i\omega t}G_{AB}(t)$ of complex $\omega $ such that 
\begin{equation}
G_{AB}(\omega )=\left\{ 
\begin{array}{ccc}
G_{AB}^{(r)}(\omega ) & , & \text{Im }\omega >0 \\ 
G_{AB}^{(a)}(\omega ) & , & \text{Im }\omega <0.
\end{array}
\right.  \label{Eq2.19}
\end{equation}
Hence, Eq. (\ref{Eq2.13}) yields for $G_{AB}(\omega )$ the spectral
representation %
\begin{equation}
G_{AB}(\omega )=\int_{-\infty }^{+\infty }{{\frac{{d\omega }^{\prime }}{{%
2\pi }}}}{\frac{\Lambda _{AB}(\omega ^{\prime })}{\omega -\omega ^{\prime }}}%
.  \label{Eq2.20}
\end{equation}
This function is analytical in the whole complex $\omega $-plane with a cut
along the real axis where singularities for $G_{AB}(\omega )$ may occur. It
is worth noting that, in terms of $\Lambda _{AB}(\omega )$, no formal
differences exist for the spectral representations of $G_{AB}^{(\nu
)}(\omega )$ and $G_{AB}(\omega )$ in the classical and quantum context.
Hence, all the developments already known in quantum many-body theory remain
formally valid for the classical case. So, one has immediately the important
exact relation 
\begin{equation}
\Lambda _{AB}(\omega )=i\left[ G_{AB}(\omega +i\varepsilon )-G_{AB}(\omega
-i\varepsilon )\right] ,  \label{Eq2.21}
\end{equation}
which express the CSD in terms of the related two-time GF's in the $\omega$%
-representation. This allows us to state also that the cut for $%
G_{AB}(\omega )$ along the real axis in $\omega $-complex plane is
determined by Eq. (\ref{Eq2.21}) and its singularities are the points of the
real axis where the condition $\Lambda _{AB}(\omega )\neq 0$ is satisfied.
For the spectral intensity of classical systems, Eqs. (\ref{Eq2.12}) and (%
\ref{Eq2.21}) yield
\begin{equation}
F_{AB}(\omega )=\left\langle {A(\tau )B}\right\rangle _{\omega }=i{\frac{{%
G_{AB}(\omega +i\varepsilon )-G_{AB}(\omega -i\varepsilon )}}{{\beta \omega }%
}.}  \label{Eq2.22}
\end{equation}
Of course, other known quantum relations are formally valid for classical
many-body theory. For instance, when $\Lambda _{AB}(\omega )$ is real, the
classical Kramer-Kronig relations (classical dispersion relations) between
the real and imaginary parts of $G_{AB}^{(r)}(\omega )$ and $%
G_{AB}^{(a)}(\omega )$ for real values of $\omega $ are true
\begin{equation}
\text{Re }G_{AB}^{(\nu )}(\omega )={\frac{(-1)^{(\nu )}}{\pi }}\wp
\int_{-\infty }^{+\infty }d\omega ^{\prime }{\frac{\text{Im }G_{AB}^{(\nu
)}(\omega ^{\prime })}{\omega ^{\prime }-\omega },}  \label{Eq2.23}
\end{equation}
where the symbol $\wp $ denotes the main part of the integral defined as %
\begin{equation}
\wp \int_{-\infty }^{+\infty }{\frac{f(x)}{x-x_{0}}}=\lim_{\varepsilon
\rightarrow 0^{+}}\left[ {\int_{-\infty }^{x_{0}-\varepsilon }{dx{\frac{{f(x)%
}}{{x-x_{0}}}}+\int_{x_{0}+\varepsilon }^{+\infty }{{\frac{{f(x)}}{{x-x_{0}}}%
}}}}\right] .  \label{Eq2.24}
\end{equation}
We have also 
\begin{equation}
\Lambda _{AB}(\omega )=-2(-i)^{\nu }{\text{Im }G_{AB}^{(\nu )}(\omega ),}
\label{Eq2.25}
\end{equation}
and, in particular,
\begin{equation}
\Lambda _{AB}(\omega )=-2{\text{Im }G_{AB}^{(r)}(\omega )}.  \label{Eq2.26}
\end{equation}

Many physical information about the macroscopic properties of classical many
body systems can be obtained from the analytical properties of the GF's
defined above. In the Subsec. 2.2 we will derive a spectral decomposition
for $\Lambda _{AB}(\omega )$, and hence for $G_{AB}^{(\nu )}(\omega )$ and $%
F_{AB}(\omega )$, from which it becomes possible to analyze the nature of
the classical GF singularities and their connection with the physical
quantities.

\subsection{Spectral Decompositions, Classical Oscillations and Damping}

In this section we will show that, in strict analogy with quantum case, also in classical
statistical mechanics an exact spectral decomposition for the CSD is true.
With this aim, following Refs. \cite{Prigogine62,Forster75}, we introduce
the Hilbert space of the classical dynamical variables with a scalar product
defined as

\begin{equation}
\left\langle A\left| B\right. \right\rangle \overset{def}{=}Z\left\langle
A^{\ast }B\right\rangle =\int {d\Gamma e^{-\beta H}A^{\ast }(q,p)B(q,p),}
\label{Eq2.27}
\end{equation}
where $d\Gamma =dqdp$ and $Z=\int {d\Gamma e^{-\beta H}}$ is the
(generalized) partition function for the problem under study. In this space
one can consider the eigenvalue equation for the Hermitian operator $L$%
\begin{equation}
L\left| \psi _{k}\right\rangle =\omega _{k}\left| \psi _{k}\right\rangle ,
\label{Eq2.28}
\end{equation}
or, equivalently, in terms of the wave functions $\psi _{k}(q,p)$
corresponding to the vectors $\left| \psi _{k}\right\rangle $%
\begin{equation}
L\psi _{k}(q,p)=\omega _{k}\psi _{k}(q,p).  \label{Eq2.29}
\end{equation}
In Eqs. (\ref{Eq2.28}) and (\ref{Eq2.29}), $\left| \psi _{k}\right\rangle $ $%
\left( \psi _{k}\right) $ and $\omega _{k}$ denote the eigenvectors
(eigenfunctions) and the eigenvalues of the Liouville operator $L$.

An important property of the Liouville operator is that all its
eigenfunctions are (in general) complex \cite{Prigogine62} and one can
promptly prove that, if $\psi _{k}$ is an eigenfunction of $L$ with
eigenvalue $\omega _{k}$, then $\psi _{k}^{\ast }$ is also an eigenfunction
of $L$ with eigenvalue $-\omega _{k}$.

At this stage, if we assume that $\{\psi _{k}\}$ is a complete set of
orthonormal eigenfunctions ($\left\langle \psi _{k}\left| \psi _{k^{\prime
}}\right. \right\rangle =Z\left\langle \psi _{k}^{\ast }(q,p)\psi
_{k}(q,p)\right\rangle =\delta _{k,k^{\prime }}$), for two arbitrary
dynamical variables $A(q,p)$ and $B(q,p)$ we can consider the expansions:

\begin{equation}
\begin{array}{c}
A(p,q)=\sum_{k}\left\langle \psi_{k}^{\ast}\left| A\right. \right\rangle\psi_{k}^{\ast }(q,p), \\ 
B(p,q)=\sum_{k}\left\langle \psi_{k}\left| B\right. \right\rangle \psi
_{k}(q,p).
\end{array}
\label{Eq2.30}
\end{equation}

On the other hand, from the relations (\ref{Eq2.5}), (\ref{Eq2.11}) and (\ref
{Eq2.12}), it follows that 
\begin{eqnarray}
\Lambda _{AB}(\omega ) &=&\beta \omega \int_{-\infty }^{+\infty }{d\tau
e^{i\omega \tau }\left\langle {A(\tau )B}\right\rangle }  \notag
\label{Eq2.31} \\
&=&\beta \omega Z^{-1}\int d\Gamma e^{-\beta H(q,p)}{B(q,p)}\int_{-\infty
}^{+\infty }d\tau e^{i(\omega +L)\tau }A(q,p).  \notag \\
&&
\end{eqnarray}
Then, taking into account the definition (\ref{Eq2.27}), the expansions (\ref
{Eq2.30}) and the orthonormality and completeness conditions in the Hilbert
space, Eq. (\ref{Eq2.31}) yields
\begin{equation}
\Lambda _{AB}\left( \omega \right) =2\pi \beta \omega
Z^{-1}\sum_{k}\left\langle \psi _{k}^{\ast }\left| A\right. \right\rangle
\left\langle \psi _{k}\left| B\right. \right\rangle \delta (\omega -\omega
_{k}),  \label{Eq2.32}
\end{equation}
where we have used the relations

\begin{equation}
e^{i(\omega +L)\tau}\psi_{k}^{*}=e^{i(\omega -\omega _{k})\tau }\psi
_{k}^{\ast},  \label{Eq2.33}
\end{equation}
and
\begin{equation}
\int_{-\infty }^{+\infty }d\tau e^{i(\omega -\omega _{k})\tau }{=2\pi \delta
(\omega -\omega _{k}).}  \label{Eq2.34}
\end{equation}
Equation (\ref{Eq2.32}) is the desired spectral decomposition, or
classical Lehmann representation, for the $\Lambda _{AB}(\omega )$ to be
compared with the known quantum analog \cite{Kalashnikov69,Kalashnikov73}.
Of course, if $B=A^{\ast }$, we have
\begin{equation}
\Lambda _{AA^{\ast }}(\omega )=2\pi \beta \omega Z^{-1}\sum_{k}\left|
\left\langle \psi _{k}^{\ast }\left| A\right. \right\rangle \right|
^{2}\delta \left( \omega -\omega _{k}\right) ,  \label{Eq2.35}
\end{equation}
which is a real quantity. Equations (\ref{Eq2.32}) and (\ref{Eq2.35})
express formally $\Lambda _{AB}(\omega )$ as an infinite sum of
appropriately weighted $\delta $-functions. It may happen that, for special
physical cases, only a finite number of terms in the sum of Eqs. (\ref
{Eq2.32}) and (\ref{Eq2.35}) are different from zero. In such cases, the
many-body problem may be solvable. Usually, however, $\Lambda _{AB}(\omega )$
will be a continuous function of real frequency $\omega $.

It is easy to see that $\Lambda _{AB}(\omega )$ can be also written in the
compact form
\begin{equation}
\Lambda _{AB}(\omega )=2\pi \beta \omega \left\langle {B\delta (\omega +L)A}%
\right\rangle =2\pi i\left\langle {\left\{ {\delta (\omega +L)A,B}\right\} }%
\right\rangle ,  \label{Eq2.36}
\end{equation}
which, in view of definition (\ref{Eq2.27}), is just the operatorial version
of the spectral decomposition (\ref{Eq2.32}).

An important consequence of the general expression (\ref{Eq2.32}) for $%
\Lambda _{AB}(\omega )$ is that the $G_{AB}(\omega )$, defined by Eqs. (\ref
{Eq2.19}) and (\ref{Eq2.20}), can be written as 
\begin{equation}
G_{AB}(\omega )=\left\langle {\left\langle {A(\tau );B}\right\rangle }%
\right\rangle _{\omega }=\beta Z^{-1}\sum\limits_{k}{\left\langle {\psi
_{k}\left| B\right. }\right\rangle \left\langle {\psi _{k}^{\ast }\left|
A\right. }\right\rangle {\frac{{\omega _{k}}}{{\omega -\omega _{k}}}.}}
\label{Eq2.37}
\end{equation}
Hence, the eigenvalues of the Liouville operator $L=i\{H,...\}$, which is
related to the energy of the system, are just the real poles of the Green
function $G_{AB}(\omega )$. On the other hand, from Eq. (\ref{Eq2.14}), one
has for the $\tau $-dependent correlation function $F_{AB}(\tau )$ %
\begin{equation}
F_{AB}(\tau )=\left\langle {A(\tau )B}\right\rangle =Z^{-1}\sum\limits_{k}{%
\left\langle {\psi _{k}\left| B\right. }\right\rangle \left\langle {\psi
_{k}^{\ast }\left| A\right. }\right\rangle }e^{-i\omega _{k}\tau }.
\label{Eq2.38}
\end{equation}
Then, in strict analogy to the quantum case, Eqs. (\ref{Eq2.37}) and (\ref
{Eq2.38}) suggest that the real poles of $G_{AB}(\omega )$, \textit{i.e.}
the eigenvalues $\omega _{k}$ of the Liouville operator, represent the
classical frequency spectrum of undamped oscillations (elementary or
collective depending on the physical nature of the dynamical variables $A$
and $B$) in the system. A characteristic frequency of this type will be
called the \textit{oscillation dispersion relation}. Thus, in case of polar
singularities, the real poles of $G_{AB}(\omega )$ represent undamped
oscillations of the system associated to the time-dependent correlation
function for the dynamical variables $A$ and $B$. It is worth noting that,
in general, the $\delta $-poles in the exact classical Lehmann
representation (\ref{Eq2.31}) of the spectral density are expected lying
infinitesimally close, therewith defining a continuous function $\Lambda
_{AB}(\omega )$. It is then obvious that the damped oscillation concept
works only under the basic presumption that the CSD exhibits some pronounced
peaks, which have to be identified with oscillation poles for $G_{AB}(\omega
) $. The widths of these peaks are a direct measure of damping and,
therefore, of the finite lifetime of various quasioscillations in the
system. This can still better be seen by inspecting the connection of $%
\Lambda _{AB}(\omega )$ with the $G_{AB}^{(r)}(\omega )$. We start from the
observation that further complicated singularities may occur in the $\omega $%
-complex plane on the Riemann sheet of the analytical function $%
G_{AB}(\omega )$ below the real axis where $G_{AB}^{(r)}(\omega )$ is not an
analytical function\footnote{%
The advanced GF $G_{AB}^{(a)}(\omega )$ has not a direct physical meaning
and hence its possible singularities above the real axis will be not
considered.}. Then, in order to see what may happen, let us assume that, in
this region, at least approximately, a complex pole of the type $\tilde{%
\omega}_{AB}=\omega _{AB}-i\Gamma _{AB}$ ($\Gamma _{AB}>0$) may take place
for $G_{AB}^{(r)}(\omega )$. In this simple situation, one can write%
\begin{equation}
G_{AB}^{(r)}(\omega )\approx {\frac{{C_{AB}}}{{\omega -\omega _{AB}+i\Gamma
_{AB}}}}  \label{Eq2.39}
\end{equation}
for real $\omega $ very close to the singularity value $\omega _{AB}$. In
Eq. (\ref{Eq2.39}), $C_{AB}$ is an inessential constant depending on the
nature of the dynamical variables $A$ and $B$. Besides, we suppose also that 
$\Lambda _{AB}(\omega )$ is a continuous real function of $\omega $ so that
we can use the exact relation (\ref{Eq2.26}). Then, from Eq. (\ref{Eq2.39}),
we obtain the (approximate) expression
\begin{equation}
\Lambda _{AB}(\omega )\simeq 2\frac{{\Gamma _{AB}}\text{Re }{C_{AB}-(\omega
-\omega _{AB})\text{Im }C_{AB}}}{{(\omega -\omega _{AB})^{2}+\Gamma _{AB}^{2}%
}}.  \label{Eq2.40}
\end{equation}
If $\Lambda _{AB}(\omega )>0$, we should have necessarily $\text{Im }%
C_{AB}=0 $ and hence $\Lambda _{AB}(\omega )$ should have a Lorentzian
shape. Unfortunately, according to the exact spectral decompositions (\ref
{Eq2.32}) and (\ref{Eq2.36}), $\Lambda _{AB}(\omega )$ is not a definite
positive function in the whole range of real $\omega $. Hence, in principle,
for each polar singularity for $G_{AB}^{(r)}(\omega )$ one must assume the
quasi-Lorentzian or modified Lorentzian expression (\ref{Eq2.40}) which
reduces to a $\delta $-function singularity for $\omega $ close to $\omega
_{AB}$ and $\Gamma _{AB}\rightarrow 0$. In any case, using the
representation (\ref{Eq2.8}) for $\theta (\tau )$, from Eq. (\ref{Eq2.39})
and the inverse Fourier transform for $G_{AB}^{(r)}(\omega )$, the time
behavior of the retarded GF is given by%
\begin{equation}
G_{AB}^{(r)}(\tau )=\theta (\tau )\left\langle {\left\{ {A(\tau ),B}\right\} 
}\right\rangle \approx \theta (\tau )\mathcal{A}_{AB}(\tau )e^{-i\omega
_{AB}\cdot \tau },  \label{Eq2.41}
\end{equation}
where
\begin{equation}
\mathcal{A}_{AB}(\tau )=\mathcal{A}_{AB}e^{-\Gamma _{AB}\cdot \tau },
\label{Eq2.42}
\end{equation}
with $\mathcal{A}_{AB}$ an inessential constant. Similarly, using the exact
relation (\ref{Eq2.5}) and the approximate expression for $\left\langle
\left\{ {A(\tau ),B}\right\} \right\rangle $ which can be extracted from Eq.
(\ref{Eq2.41}), one easily finds:
\begin{equation}
\left\langle {A(\tau )B}\right\rangle -\left\langle A\right\rangle
\left\langle B\right\rangle \simeq \mathcal{C}_{AB}(\tau )e^{-i\omega
_{AB}\tau }=(\mathcal{C}_{AB}e^{-\Gamma _{AB}\tau })e^{-i\omega _{AB}\tau }.
\label{Eq2.43}
\end{equation}
Eq. (\ref{Eq2.43}) is based on the boundary condition $\mathop {\lim }%
\limits_{\tau \rightarrow \infty }\left\langle {A(\tau )B}\right\rangle
=\left\langle A\right\rangle \left\langle B\right\rangle $ related to the
physical feature that, for all real systems, there is always attenuation of
the correlations on time. This is also supported by the rigorous
Riemann-Lebesgue lemma in theory of analytical functions \cite{ZubarevB} as
applied to the spectral representation (\ref{Eq2.7}) where the spectral
intensity $F_{AB}(\omega )$, and hence $\Lambda _{AB}(\omega )$, is a
continuous function.

In conclusion, when $G_{AB}^{(r)}(\omega )$ has a pole below the real axis,
with $\omega $ close to $\omega _{AB}$ and $\Gamma _{AB}/\left| {\omega _{AB}%
}\right| \ll 1$, $\Lambda _{AB}(\omega )$ has a pronounced quasi-Lorentzian
peak at $\omega =\omega _{AB}$ and both $G_{AB}^{(r)}(\omega )$ and $%
\left\langle {A(\tau )B}\right\rangle -\left\langle A\right\rangle
\left\langle B\right\rangle $ are characterized by damped oscillations with
frequency $\omega _{AB}$. Here $\Gamma _{AB}$, which is related to the width
of the peak, measures the damping of oscillations. Of course, in this
picture, $\tau _{AB}=\Gamma _{AB}^{-1}$ defines the life-time of the
oscillations in the system. It is also worth noting that, if $\Gamma _{AB}=0$
(absence of damping), $\Lambda _{AB}(\omega )$ has a $\delta $-function peak
and undamped oscillations with frequency $\omega _{AB}$ occur in the system.

The considerations made above for a single pole for $G_{AB}^{(r)}(\omega )$
very close the real $\omega $-axis from below can be easily extended to
several singularities of this type. In such a case, $\Lambda _{AB}(\omega )$
will result in a superposition of modified Lorentzian peaks whose widths
measure the damping of the related oscillations. If the widths reduce to
zero, $\Lambda _{AB}(\omega )$ becomes a superposition of $\delta $%
-functions and this signals the occurrence of undamped oscillations in the
system.

In the next section, where we will consider the problem to search reliable
approximations within the classical GF formalism, we will see that the
quasi-Lorentzian shape for $\Lambda _{AB}(\omega )$ is not mathematically
adequate and another more appropriate representation, capturing the
essential physics, will be necessary.

\setcounter{equation}{0}

\section{Methods of Calculation}

All the previous general considerations suggest that, also in classical
statistical mechanics, the formalism of the two-time GF's may be a useful
tool to study the macroscopic properties of classical many-body systems. The
basic ingredients are $G_{AB}(\omega )$ and $\Lambda _{AB}(\omega )$ for two
properly chosen dynamical variables $A$ and $B$. Remarkably, if we are able
to calculate $G_{AB}(\omega )$ $(\Lambda _{AB}(\omega ))$, the exact
relations established in the previous sections allow us to obtain $\Lambda
_{AB}(\omega )$ $(G_{AB}(\omega ))$. In the present section we present the
classical formulation of two general methods in strict analogy to the
quantum counterpart, \textit{i.e.} the classical EMM, for calculating the
GF's, and the CSDM for a direct calculation of the SD. In principle, both
the methods should give exactly the GF's and the related SD's and hence, in
this sense, they are completely equivalent. Nevertheless, in practical
calculations, previous experiences in quantum many-body theory \cite
{Kalashnikov69,Kalashnikov73,Campana79,Caramico80,Nolting89,
Nolting91,Hermann97,Caramico81} and, although reduced, in classical
many-body theory \cite
{Caramico81,Campana83,Campana84,Cavallo01,Romano90,Cavallo04}, suggest that
the SDM has several advantages to make more reliable nonperturbative
approximations in a systematic and controllable manner.

\subsection{The Classical Equations of Motion Method}

Differentiating Eq. (\ref{Eq2.2}) with respect to $\tau =t-t^{^{\prime }}$
yields
\begin{equation}
\frac{d}{d\tau }\left\langle {\left\langle {A(\tau );B}\right\rangle }%
\right\rangle _{\nu }={\frac{{d\theta _{\nu }(\tau )}}{{d\tau }}}%
\left\langle {\left\{ {A(\tau ),B}\right\} }\right\rangle +\left\langle {%
\left\langle {{\frac{{dA(\tau )}}{{d\tau }}};B}\right\rangle }\right\rangle
_{\nu }.  \label{Eq3.1}
\end{equation}
Then, using the EM (\ref{Eq2.3}) for dynamical variables and the obvious
relation $d\theta _{\nu }(\tau )/d\tau =\delta (\tau )$, Eq. (\ref{Eq3.1})
becomes
\begin{equation}
{\frac{d}{{d\tau }}}\left\langle {\left\langle {A(\tau );B}\right\rangle }%
\right\rangle _{\nu }=\delta (\tau )\left\langle {\left\{ {A,B}\right\} }%
\right\rangle +\left\langle {\left\langle {\left\{ {A(\tau ),H}\right\} ;B}%
\right\rangle }\right\rangle _{\nu }.  \label{Eq3.2}
\end{equation}
This is the basic EM for the GF $\left\langle {\left\langle {A(\tau );B}%
\right\rangle }\right\rangle _{\nu }$ which, however, is not a closed
differential equation. Indeed, in the right-hand side of Eq. (\ref{Eq3.2}) a
new higher order $\nu $-GF occurs involving Poisson brackets of a greater
number of dynamical variables. Then, one needs to consider a new EM for the
two-time $\nu $-GF $\left\langle {\left\langle {\left\{ {A(\tau ),H}\right\}
;B}\right\rangle }\right\rangle _{\nu }$. The $\tau $-derivative of this
function yields an additional equation, formally identical to Eq. (\ref
{Eq3.2}) with $A(\tau )$ replaced by ${\left\{ {A(\tau ),H}\right\} }$, the
right-hand side of which contains the new $\nu $-GF $\left\langle {%
\left\langle \left\{ {\left\{ {A(\tau ),H}\right\} ,H}\right\} {;B}%
\right\rangle }\right\rangle _{\nu }$. By iteration of this procedure, we
obtain an infinite chain of coupled EM's for GF's of increasing order which
can be written in a compact form as%

\begin{equation}
\frac{d}{d\tau }\left\langle {\left\langle {\mathcal{L}_{H}^{m}A(\tau );B}%
\right\rangle }\right\rangle _{\nu }=\delta (\tau )\left\langle {\left\{ {%
\mathcal{L}_{H}^{m}A,B}\right\} }\right\rangle +\left\langle {\left\langle {%
\mathcal{L}_{H}^{m+1}A(\tau );B}\right\rangle }\right\rangle _{\nu } 
\begin{array}{cc}
, & (m=0,1,2,...).
\end{array}
\label{Eq3.3}
\end{equation}
Here $\mathcal{L}_{H}=iL=\{...,H\}$ and $\mathcal{L}_{H}^{m}A$ means $%
\mathcal{L}_{H}^{0}A=A$, $\mathcal{L}_{H}^{1}=\{A,H\}$, $\mathcal{L}%
_{H}^{2}A=\{\{A,H\},H\}$, and so on. Notice that the chain of EM's (\ref
{Eq3.3}) is formally the same for different types of GF's and hence one can
eliminate the index $\nu $ when the physical context is clear.

In the practical calculations it is generally more convenient to work in the 
$\omega $-Fourier space. With $i\int_{-\infty }^{+\infty }d\tau e^{i\omega
\tau }\left( d{f(\tau )/d\tau }\right) =\omega f(\omega)$, the chain of
equations, in the $\omega $-representation, assumes the form:
\begin{equation}
\omega \left\langle {\left\langle {\mathcal{L}_{H}^{m}A(\tau );B}%
\right\rangle }\right\rangle _{\nu ,\omega }=i\left\langle {\left\{ {%
\mathcal{L}_{H}^{m}A,B}\right\} }\right\rangle +i\left\langle {\left\langle {%
\mathcal{L}_{H}^{m+1}A(\tau );B}\right\rangle }\right\rangle _{\nu ,\omega } 
\begin{array}{cc}
, & (m=0,1,2,...).
\end{array}
\label{Eq3.4}
\end{equation}
To solve the chain of EM's in the form (\ref{Eq3.3}) or (\ref{Eq3.4}), we
must add appropriate boundary conditions. For this, it is again more
convenient to use the $\omega $-representation (\ref{Eq3.4}) for which they
can be obtained in the form of spectral representations or dispersion
relations for the GF's (see Subsec. 2.1 and, in particular, Eq. (\ref{Eq2.18}%
)), in strict analogy with the quantum case. Anyway, although Eqs. (\ref
{Eq3.3}) and (\ref{Eq3.4}) are exact, a free solution for interacting
systems is, of course, impossible. In practical calculations one is forced
to introduce decoupling procedures, and hence approximate methods, to reduce
the infinite chain of coupled equations in a finite closed one which,
although approximate, may be solved. However, in general, systematic and
controllable decouplings are not easy to find and one must check the
reliability of a given approximation by comparing the results with
experiments, simulations or other types of approaches. The CSDM, which will
be the subject of the next subsection, seems more flexible in such a
direction.

\subsection{The Classical Spectral Density Method}

Here, we present a general formulation of the CSDM \cite
{Caramico81,Campana84} for a systematic calculation of the CSD. For this
aim, it is convenient to start from the expression (\ref{Eq2.11}) of the SD
in the $\tau $-representation. By successive derivatives of ${\Lambda
_{AB}(\tau )}$ with respect to $\tau $ and using the EM (\ref{Eq2.3}), one
has
\begin{equation}
{\frac{{d^{m}\Lambda _{AB}(\tau )}}{{d\tau ^{m}}}=}i\left\langle {\left\{ {%
\mathcal{L}_{H}^{m}A(\tau ),B}\right\} }\right\rangle 
\begin{array}{cc}
, & (m=0,1,2,...).
\end{array}
\label{Eq3.5}
\end{equation}
Then, taking the Fourier transform of Eq. (\ref{Eq3.5}), we easily get%
\begin{equation}
(-i\omega )^{m}\Lambda_{AB}(\omega)=i\int_{-\infty }^{+\infty } d\tau
 e^{i\omega \tau }\left\langle \left\{ {\mathcal{L}_{H}^{m}A(\tau ),B}
\right\} \right\rangle .  \label{Eq3.6}
\end{equation}
Finally, integration over $\omega $ yields:
\begin{equation}
\int_{-\infty }^{+\infty }{{\frac{{d\omega }}{{2\pi }}}\omega ^{m}\Lambda
_{AB}(\omega )=-i^{m-1}\left\langle {\left\{ \mathcal{L}{_{H}^{m}A,B}%
\right\} }\right\rangle } 
\begin{array}{cc}
, & (m=0,1,2,...).
\end{array}
\label{Eq3.7}
\end{equation}

The quantity on the left-hand side of Eq. (\ref{Eq3.7}) is called the 
\textit{m-moment} of $\Lambda _{AB}(\omega )$ and the relations (\ref{Eq3.7}%
) constitute an infinite set of exact ME's or sum rules for the CSD. Notice 
that, for $m=0$, the sum rule (\ref{Eq2.15}) is reproduced.

Equations (\ref{Eq3.7}) can be seen in a different way. Indeed, since the 
Poisson brackets and hence the ensemble averages involved in the right-hand 
side of relations (\ref{Eq3.7}) can be evaluated, at least in principle, it 
is quite remarkable that the $m$-moments of the SD can be explicitly 
obtained without the \textit{a priori} knowledge of the function $\Lambda 
_{AB}(\omega )$. This important result implies that the sequence of Eqs. (%
\ref{Eq3.7}) represents a typical \textit{moment problem.} Its solution 
would yield the unknown SD and hence all the related quantities (GF's, 
correlation functions and other observables). Unfortunately, also this 
problem, in general, cannot be solved exactly and one must look for an 
approximate solution. Suggested by the exact classical spectral 
decomposition (\ref{Eq2.32}) or (\ref{Eq2.35}), we seek for an approximation for 
$\Lambda _{AB}(\omega )$ as a finite sum of properly weighted $\delta $%
-functions of the form (polar approximation)%
\begin{equation} 
\Lambda _{AB}(\omega )=2\pi \sum\limits_{k=1}^{n}{\lambda _{k}(A,B)\delta 
\left( {(\omega -\omega _{k}(A,B)}\right) ,}  \label{Eq3.8} 
\end{equation} 
where $n$ is a finite integer number. The unknown parameters $\lambda 
_{k}(A,B)$ and $\omega _{k}(A,B)$, depending on the nature of the dynamical 
variables A and B, have to be determined as a solution of the finite set of $%
2n$ (generally integral) equations obtained by inserting the expression (\ref 
{Eq3.8}) in the first $2n$ ME's (\ref{Eq3.7}). This is the basic idea of the 
CSDM. Physically, the parameters $\omega _{k}(A,B)$ in Eq. (\ref{Eq3.8}) 
play the role of effective eigenvalues of the Liouville operator and each of 
them, as a real (approximate) pole of the GF $G_{AB}(\omega )$ (see the 
spectral representation (\ref{Eq2.20}) or (\ref{Eq2.37})), corresponds to a 
possible mode of undamped oscillations for the correlation function $%
\left\langle A(\tau )B\right\rangle $ (see Eq. (\ref{Eq2.38})).  

As discussed at the end of Sec. 2, there are physical situations where the 
damping of classical oscillations in the system under study may be important 
and hence the polar approximation (\ref{Eq3.8}) is inadequate. In these 
cases, the basic idea of the SDM, related to the moment problem (\ref{Eq3.7}%
), remains still valid but it is necessary to chose a more appropriate 
functional structure for the SD which allows us to determine the modes of 
oscillations in the system and their damping or life-time. An extension of 
the SDM in this sense was first proposed by Nolting and Oles \cite{Nolting80} 
for Fermi systems and by Campana et al. \cite{Campana83} for Bose and 
classical systems whose SD's are not positive definite in the whole range of 
$\omega $. Here we will focus on the classical case. In Sec. 2, we have 
shown that information about the spectrum and the damping of the 
oscillations in the system can be obtained if the SD is assumed as a 
superposition of the quasi-Lorentzian peaks. The peaks would correspond to 
the frequencies of oscillations and the widths would measure their damping 
or life-time. However, we are now in a position to easily check that 
modified Lorentzian shapes for $\Lambda _{AB}(\omega )$ cannot be valid over 
the whole $\omega $ range since all the SD moments of order $m\geq 2$ would 
diverge. This feature clearly contradicts Eq. (\ref{Eq3.7}). To assure the 
convergency of the SD moments at any order and to preserve the intrinsic 
physical character of the CSD's connected with the not everywhere-positive 
factor $\omega $ in the exact spectral decomposition (\ref{Eq2.32}) or (\ref 
{Eq2.35}), one can assume the \textit{modified Gaussian ansatz} for the $%
\Lambda _{AB}(\omega )$ \cite{Campana83}
\begin{equation} 
\Lambda _{AB}(\omega )=2\pi \omega \sum\limits_{k=1}^{n}{\lambda _{k}(A,B){%
\frac{{\exp \left[ {{\frac{{-\left( {\omega -\omega _{k}(AB)}\right) ^{2}}}{{%
\Gamma _{k}(A,B)}}}}\right] }}{\sqrt{\pi \Gamma _{k}(A,B)}}}.}  \label{Eq3.9} 
\end{equation} 
Notice that, with the functional representation (\ref{Eq3.9}) for $\Lambda 
_{AB}(\omega )$, the parameter ${\Gamma _{k}(A,B)}$ has not to be identified 
directly with the parameter ${\Gamma (A,B)}$ in Eqs. (\ref{Eq2.39}) for $%
G_{AB}(\omega )$ and (\ref{Eq2.40}) for $\Lambda _{AB}(\omega )$ but, 
rather, with ${\Gamma }^{2}{(A,B)}$. Then, the width of the peak in $\omega 
=\omega _{k}(A,B)$ is represented here by ${\Gamma _{k}^{1/2}(A,B)}$, the 
condition ${\Gamma (A,B)/\left| {\omega (A,B)}\right| }\ll 1$ must be 
replaced by ${\Gamma _{k}(A,B)/\left[ {\omega _{k}(A,B)}\right] ^{2}}\ll 1$ 
and the life-time of the classical excitations with frequency $\omega 
_{k}(A,B)$ is identified by $\tau _{k}(A,B)={\Gamma _{k}^{1/2}(A,B)}$. The 
choose (\ref{Eq3.9}) is only motivated by the fact that it takes direct 
contact with the notation used in literature \cite{Campana83,Nolting80} and 
simplifies the algebra in explicit calculations. Here the parameters $\omega 
_{k}(A,B)$ represent the oscillation frequencies and the new parameters $%
\Gamma _{k}(A,B)$, which describe the broadening of the $\delta $-poles due 
to the finite life-times of the respective oscillations, give a measure of 
the damping effects. In any case, the representation (\ref{Eq3.9}) assures 
that $\Lambda _{AB}(\omega )$ is a superposition of sharp quasi-$\delta $%
-function peaks and reduces to the polar one (\ref{Eq3.8}) in the limit $%
\Gamma _{k}(A,B)\rightarrow 0$, as expected. This functional structure 
allows us to investigate classical oscillations and damping effects 
systematically just as the $\delta $-function ansatz (\ref{Eq3.8}) does for 
case of undamped oscillations. Of course, with the ansatz (\ref{Eq3.9}), the 
parameters $\lambda _{k}(A,B)$, $\omega _{k}(A,B)$ and $\Gamma _{k}(A,B)$ 
have to be calculated solving the first $3n$ ME's (\ref{Eq3.7}) consistently 
with the basic condition ${\Gamma _{k}(A,B)/\left[ {\omega _{k}(A,B)}\right] 
^{2}}\ll 1$.  

As in the EMM, also in the SDM the problem remains to close the truncated 
finite set of ME's arising from the polar ansatz (\ref{Eq3.8}) or the 
modified Gaussian ansatz (\ref{Eq3.9}). In any case, the evaluation of the 
right-hand sides of Eqs. (\ref{Eq3.7}) generally involves higher order SD's.  
Therefore, higher order moment problems should be considered, but the 
difficulty of calculations will increase considerably. So, in order to solve 
self-consistently the finite set of ME's, which arises from Eq. (\ref{Eq3.7}%
) using the ansatz (\ref{Eq3.8}) or (\ref{Eq3.9}), it is usually necessary 
to use some decoupling procedures and thus to introduce, in a systematic 
way, additional approximations in the CSDM which, however, will be 
automatically consistent with an increasing number of sum rules for the SD.  

\setcounter{equation}{0} 

\section{The Classical $d$-dimensional Spin-$S$ Heisenberg Ferromagnet with 
Long-Range Interactions: A Many-body Approach} 

\subsection{An Introduction to the Model} 

The classical $d$-dimensional spin-$S$ Heisenberg ferromagnet with LRI's 
decaying as $r^{-p}$ ($p>d$) with the distance $r$ between spins, in the 
presence of an external magnetic field, is described by the Hamiltonian%
\begin{equation} 
H=-{\frac{1}{2}}\sum\limits_{i,j=1}^{N}{J_{ij}\vec{S}_{i}\cdot \vec{S}_{j}}%
-h\sum\limits_{i=1}^{N}{S_{i}^{z}}.  \label{Eq4.1} 
\end{equation} 
Here $N$ is the number of sites of a hypercubic lattice with unitary 
spacing,\linebreak $\left\{ {\vec{S}_{j}\equiv (S_{j}^{x},S_{j}^{y},S_{j}^{z});}\text{ 
}{j=1,...,N}\right\} $ are the classical spins, $h$ is the external magnetic 
field and the exchange interaction, in view of the thermodynamic limits as $%
N\rightarrow \infty $, is assumed to be $J_{ij}=J/r_{ij}^{p}$, where $%
r_{ij}=\left| {\vec{r}_{i}-\vec{r}_{j}}\right| $ and $J$ measures the 
strength of the spin-spin coupling. The extreme case $p\rightarrow \infty $ 
corresponds to the standard nearest-neighbor interaction while the mean 
field approximation is obtained when $p=0$ (replacing $J$ by $J/N$). For 
this type of interaction, the thermodynamic limit $N\rightarrow \infty $ is 
well defined only for $p>d$, while for $p\leq d$ the ground state of the 
system has an infinite energy per spin as $N\rightarrow \infty $ and the 
conventional statistical mechanics cannot be directly applied.  

The classical spin model (\ref{Eq4.1}) can be properly described by the set 
of $2N$ canonical variables $\{\varphi _{j},S_{j}^{z}\}$ where $\varphi _{j}$ 
is the angle between the projection of the spin vector $\vec{S}_{j}$ in the $%
(x-y)$-plane and the $x$-axis. The Poisson bracket of two arbitrary 
classical dynamical variables $A(\varphi ,S^{z})$ and $B(\varphi ,S^{z})$ is 
then defined by
\begin{equation} 
\left\{ {A,B}\right\} =\sum\limits_{j=1}^{N}{\left( {{\frac{{\partial A}}{{%
\partial \varphi _{j}}}}{\frac{{\partial B}}{{\partial S_{j}^{z}}}}-{\frac{{%
\partial A}}{{\partial S_{j}^{z}}}}{\frac{{\partial B}}{{\partial \varphi 
_{j}}}}}\right) }.  \label{Eq4.2} 
\end{equation} 
For the following calculations, we find convenient to introduce the new spin 
variables $S_{j}^{\pm }=S_{j}^{x}\pm iS_{j}^{y}$, so that $%
S^{2}=(S_{j}^{z})^{2}+S_{j}^{+}S_{j}^{-}.$ Then, if one defines the Fourier 
transforms of the spin vectors and the exchange interaction as%
\begin{equation} 
\vec{S}_{\vec{k}}=\sum\limits_{j=1}^{N}{e^{-i\vec{k}\cdot \vec{r}_{j}}\vec{S}%
_{j}}\quad ,\quad J(\vec{k})=\sum\limits_{j=1}^{N}{e^{-i\vec{k}\cdot (\vec{r}%
_{i}-\vec{r}_{j})}J_{ij},}  \label{Eq4.3} 
\end{equation} 
where $\vec{k}$ denotes a wave vector in the $d$-dimensional Fourier space, 
the Hamiltonian (4.1) can be rewritten as (with ${J(\vec{k})=J(-\vec{k})}$)%
%
%
\begin{equation} 
H=-{\frac{1}{{2N}}}\sum\limits_{\vec{k}}{J(\vec{k})\left( {S_{\vec{k}%
}^{+}S_{-\vec{k}}^{-}+S_{\vec{k}}^{z}S_{-\vec{k}}^{z}}\right) }-hS_{0}^{z}.  
\label{Eq4.4} 
\end{equation} 
The sum in Eq. (\ref{Eq4.4}) is restricted to the first Brillouin zone ($1BZ$%
) of the lattice. Within this representation, the Poisson brackets for the 
spin Fourier components relevant for us are
\begin{equation} 
\left\{ {S_{\vec{k}}^{\pm },S_{k^{\prime }}^{z}}\right\} =\pm iS_{\vec{k}+%
\vec{k}^{\prime }}^{\pm }\quad ,\quad \left\{ {S_{\vec{k}}^{+},S_{k^{\prime 
}}^{-}}\right\} =-2iS_{\vec{k}+\vec{k}^{\prime }}^{z}.  \label{Eq4.5} 
\end{equation} 

On the basis of the previous definitions, it is now easy to establish exact 
general formulas for the internal energy and the free energy per spin from 
which all the thermodynamic properties of the spin model (\ref{Eq4.1}) can 
be obtained when the appropriate GF's or SD's have been determined using the 
many-body formalism presented before. For this aim, the Fourier 
representation (\ref{Eq4.4}) of the Hamiltonian in terms of the dynamical 
variables $S_{\vec{k}}^{+}S_{-\vec{k}}^{-}$ and $S_{\vec{k}}^{z}S_{-\vec{k}%
}^{z}$ is particularly convenient since, in general, all the thermodynamic 
quantities can be easily expressed exactly in terms of the correlation 
functions $\left\langle S_{\vec{k}}^{+}S_{-\vec{k}}^{-}\right\rangle $ and $%
\left\langle S_{\vec{k}}^{z}S_{-\vec{k}}^{z}\right\rangle $. To see this, we 
rewrite Eq. (\ref{Eq4.2}) as :
\begin{equation} 
H=H_{0}+H_{I}(J),  \label{Eq4.6} 
\end{equation} 
with
\begin{equation} 
H_{0}=-hS_{0}^{z},  \label{Eq4.7} 
\end{equation} 
and
\begin{equation} 
H_{I}(J)=-{\frac{J}{{2N}}}\sum\limits_{\vec{k}}{\gamma (}\vec{k})\left[ {S_{%
\vec{k}}^{+}S_{-\vec{k}}^{-}+S_{\vec{k}}^{z}S_{-\vec{k}}^{z}}\right] , 
\label{Eq4.8} 
\end{equation} 
where $\gamma (\vec{k})=J({\vec{k}})/J=\sum\limits_{j=1}^{N}{e^{i\vec{k}%
\cdot (\vec{r}_{i}-\vec{r}_{j})}/r_{ij}^{p}}$. Then, the internal energy of 
the spin model is formally given by the ensemble average
\begin{equation} 
U=U_{0}(J)+U_{I}(J), 
\end{equation} 
where
\begin{equation} 
U_{0}(J)=\left\langle H_{0}\right\rangle =-h\left\langle 
S_{0}^{z}\right\rangle ,  \label{Eq4.10} 
\end{equation} 
and
\begin{equation} 
U_{I}(J)=\left\langle {H_{I}}\right\rangle \left( J\right) =-{\frac{J}{{2N}}}%
\sum\limits_{\vec{k}}{\gamma (}\vec{k})\left[ {\left\langle {S_{\vec{k}%
}^{+}S_{-\vec{k}}^{-}}\right\rangle +\left\langle {S_{\vec{k}}^{z}S_{-\vec{k}%
}^{z}}\right\rangle }\right] .  \label{Eq4.11} 
\end{equation} 
For the free energy $F=-\beta ^{-1}\text{ln}Z$, we have %
\begin{equation} 
{\frac{{\partial F}}{{\partial J}}}={\frac{1}{Z}}\int {d\Gamma }\left( {{%
\frac{{\partial H_{I}(J)}}{{\partial J}}}}\right) e^{-\beta H}=\left\langle {%
{\frac{{\partial H_{I}(J)}}{{\partial J}}}}\right\rangle =\left\langle {{%
\frac{{H_{I}(J)}}{J}}}\right\rangle .  \label{Eq4.12} 
\end{equation} 
Then, by integrating Eq. (\ref{Eq4.12}) between $0$ and $J$ with the initial 
condition $F(J=0)=F_{0}=-\beta ^{-1}\text{ln}\left[ {\int d\Gamma e^{-\beta {%
H_{0}}}}\right] $, we get
\begin{equation} 
F=F_{0}+\int_{0}^{J}{{\frac{{dJ^{\prime }}}{{J^{\prime }}}}\left\langle {%
H_{I}}\right\rangle (J^{\prime }).}  \label{Eq4.13} 
\end{equation} 
From these general expressions, it is easy to obtain, for the internal 
energy $u=U/N$ and the free energy $f=F/N$ per spin, the desired exact 
relations (with $N\rightarrow \infty $)
\begin{equation} 
u(T,h)={\frac{{\left\langle H\right\rangle }}{N}}=-{\frac{h}{N}}\left\langle 
{S_{0}^{z}}\right\rangle -{\frac{J}{{2N^{2}}}}\sum\limits_{\vec{k}}{\gamma (%
\vec{k})\left[ {\left\langle {S_{\vec{k}}^{+}S_{-\vec{k}}^{-}}\right\rangle 
+\left\langle {S_{\vec{k}}^{z}S_{-\vec{k}}^{z}}\right\rangle }\right] }, 
\label{Eq4.14} 
\end{equation} 
and
\begin{eqnarray} 
f(T,h) &=&f_{0}+{\frac{1}{N}}\int_{0}^{J}{{\frac{{dJ^{\prime }}}{{J^{\prime }%
}}}\left\langle {H_{I}}\right\rangle }(J^{\prime })  \notag  \label{Eq4.15} 
\\ 
&=&f_{0}+{\frac{1}{{2N^{2}}}}\sum\limits_{\vec{k}}{\gamma (\vec{k}%
)\int_{0}^{J}{dJ^{\prime }}\left[ {\left\langle {S_{\vec{k}}^{+}S_{-\vec{k}%
}^{-}}\right\rangle _{J^{\prime }}+\left\langle {S_{\vec{k}}^{z}S_{-\vec{k}%
}^{z}}\right\rangle _{J^{\prime }}}\right] }.  \notag \\ 
&& 
\end{eqnarray} 
Here, $f_{0}$ is the free energy per spin for a magnetic model without 
interactions and $\left\langle {...}\right\rangle _{J^{\prime }}$ denotes an 
ensemble average as function of the interaction strength $J^{\prime }$. So, 
when the correlation functions $\left\langle S_{\vec{k}}^{+}S_{-\vec{k}%
}^{-}\right\rangle $ and $\left\langle S_{\vec{k}}^{z}S_{-\vec{k}%
}^{z}\right\rangle $ are known, Eqs. (\ref{Eq4.14}) and (\ref{Eq4.15}) allow 
us to determine $u$ an $f$ and hence all the thermodynamic quantities using 
standard relations.  

Also the transverse $(\xi _{\perp })$ and longitudinal $(\xi _{z})$ 
correlation lengths can be obtained in terms of the previous exact 
relations. This can be performed by using the definitions
\begin{equation} 
\xi _{\perp }^{2}=-{\frac{1}{2}}\mathop {\lim }\limits_{k\rightarrow 
0}\left\{ \frac{1}{{\left\langle {S_{\vec{k}}^{+}S_{-\vec{k}}^{-}}%
\right\rangle }}{{\frac{d^{2}}{{dk^{2}}}}}\left\langle {S_{\vec{k}}^{+}S_{-%
\vec{k}}^{-}}\right\rangle \right\}  \label{Eq4.16} 
\end{equation} 
and
\begin{equation} 
\xi _{z}^{2}=-{\frac{1}{2}}\mathop {\lim }\limits_{k\rightarrow 0}\left\{ {{%
\frac{1}{{\left\langle {S_{\vec{k}}^{z}S_{-\vec{k}}^{z}}\right\rangle }}%
\frac{d^{2}}{{dk^{2}}}}\left\langle {S_{\vec{k}}^{z}S_{-\vec{k}}^{z}}%
\right\rangle }\right\} .  \label{Eq4.17} 
\end{equation} 
It is worth nothing that, since the SDM allows to calculate $\left\langle S_{%
\vec{k}}^{+}S_{-\vec{k}}^{-}\right\rangle $ and $\left\langle S_{\vec{k}%
}^{z}S_{-\vec{k}}^{z}\right\rangle $ solving integral ME's, all the 
thermodynamic quantities can be obtained without the explicit calculation of 
the partition function. This is a relevant aspect of the SDM as a moment 
problem both in quantum and classical statistical mechanics.  

As a conclusion of this subsection, to fully appreciate the effectiveness of 
the CSDM, it may be useful to summarize the main known features about the 
model (\ref{Eq4.1})-(\ref{Eq4.4}) obtained by means of rigorous methods, MC 
simulations and alternative microscopic techniques. For a future comparison, 
we include also some information for the corresponding quantum spin model 
which attracted recently much interest.  

The one and two dimensional long-range quantum spin-$1/2$ Heisenberg 
ferromagnets (for brevity reasons, models of the type (\ref{Eq4.1}) will be 
also named \textit{long-range spin models}) in absence of an external 
magnetic field were investigated by Nakano and Takahashi using the so called 
modified spin-wave theory \cite{Nakano94a} and the Schwinger-boson mean 
field approximation \cite{Nakano94b}. Further information were derived for 
the $d$-dimensional case by means of the EMM for the two-time GF's using the 
Tyablikov decoupling procedure \cite{Nakano95} (in the next section we will 
present shortly the corresponding classical version for model (\ref{Eq4.1}%
)). Monte Carlo simulations for the two-dimensional quantum spin-$1/2$ 
Heisenberg model have been performed for $2<p\leq 6$ \cite{Vassiliev01}.  
This scenario has been recently enriched by an extension \cite{Bruno01} of 
the Mermin-Wagner theorem \cite{MerminWagner66} for the existence of 
ferromagnetic (FM) long-range order (LRO) at finite temperature in quantum 
Heisenberg and $XY$ models in $d$($=1,2$) dimensions with $r^{-p}-$ and 
oscillating$-$ exchange interactions.  

Classical long-range spin-$S$ Heisenberg FM models have attracted great 
attention, too. It has been rigorously proved the LRO exists in d(=1,2) 
dimension when $d<p<2d$ \cite{Kunz76,Frolich78} and is destroyed at all 
finite temperatures for $p\geq 2d$ \cite{Roger81,Pfister81}. Similar results 
were obtained for the spherical model \cite{Joyce66} and the present 
scenario of the critical properties is largely based on renormalization 
group calculations for the classical $n$-vector model \cite 
{Fisher72,Kosterlitz76}. Classical long-range antiferromagnetic (AFM) models 
have been studied less extensively. The available rigorous results \cite 
{Kunz76,Frolich78,Roger81} suggest orientational disorder at all finite 
temperatures when $p\geq 2d$, but no theorem exists entailing existence or 
absence of LRO for $d<p<2d$. Monte Carlo simulations have been also 
performed for both $d$($=1,2$)-dimensional classical FM (for $p=2d$ \cite 
{Romano89}) and AFM (for $p={\frac{3}{2}}$ and $p=1$ with $d=1$ and $d=2$, 
respectively \cite{Romano90}) Heisenberg long-range models. The results 
confirm that FM-LRO survives at finite temperature provided $d<p<2d$ and 
allow to conjecture that no AFM-LRO exists at all finite temperatures for $%
p>d$. Spin-wave studies \cite{Romano90} agree with the last conjecture but 
no definitive statement can be drawn at the present stage.  

Recent studies for model (\ref{Eq4.1}) \cite{Cavallo02,Cavallo04} via the 
CSDM to lowest order approximation are in good agreement with previous 
available analytical and numerical investigations. Below we present the 
problem in a detailed and more systematic way with the inclusion of 
additional results showing the efficiency of the SDM also in classical 
statistical mechanics. In next subsection, as mentioned before, we present 
also the essential results for the model (\ref{Eq4.1}) using the classical 
EMM and the Tyablikov-like approximation, although our main interest is 
devoted to the CSDM which will be used extensively in the remaining part of 
the article. In any case, these results will be very useful as a comparison 
and to underline the potentiality of the classical many-body formalism 
presented in Secs. 2 and 3.  

\subsection{The Classical Equations of Motion Method and the Tyablikov-Like\\ 
Approximation} 

For the Hamiltonian model in the form (\ref{Eq4.4}), we introduce the GF 
(without distinction between the retarded and advanced one)
\begin{equation} 
G_{\vec{k}}(t-t^{\prime })=\left\langle {\left\langle {S_{\vec{k}%
}^{+}(t);S_{-\vec{k}}^{-}(t^{\prime })}\right\rangle }\right\rangle , 
\label{Eq4.18} 
\end{equation} 
with the Fourier transform
\begin{equation} 
G_{\vec{k}}(\omega )=\left\langle {\left\langle {S_{\vec{k}}^{+}(\tau );S_{-%
\vec{k}}^{-}}\right\rangle }\right\rangle _{\omega }=\int_{-\infty 
}^{+\infty }d\tau G_{\vec{k}}(\tau )e^{i\omega \tau }.  \label{Eq4.19} 
\end{equation} 
From Eq. (\ref{Eq3.2}), we have for $G_{\vec{k}}(\omega )$ the EM%
\begin{equation} 
\omega G_{\vec{k}}(\omega )=i\left\langle {\left\{ {S_{\vec{k}}^{+},S_{-\vec{%
k}}^{-}}\right\} }\right\rangle +i\left\langle {\left\langle {\left\{ {S_{%
\vec{k}}^{+}(\tau ),H}\right\} ;S_{-\vec{k}}^{-}}\right\rangle }%
\right\rangle _{\omega }.  \label{Eq4.20} 
\end{equation} 
Then, using the spin Poisson brackets (\ref{Eq4.5}), Eq. (\ref{Eq4.20}) 
reduces to
\begin{eqnarray} 
&&(\omega -h)G_{\vec{k}}(\omega )=2Nm+  \notag  \label{Eq4.21} \\ 
&&-{\frac{1}{N}}\sum\limits_{\vec{p}}{J(\vec{p})}\left\{ {\left\langle {%
\left\langle {S_{\vec{k}-\vec{p}}^{z}(\tau )S_{\vec{p}}^{+}(\tau );S_{-\vec{k%
}}^{-}}\right\rangle }\right\rangle _{\omega }-\left\langle {\left\langle {%
S_{\vec{p}}^{z}(\tau )S_{\vec{k}-\vec{p}}^{+}(\tau );S_{-\vec{k}}^{-}}%
\right\rangle }\right\rangle _{\omega }}\right\} ,  \notag \\ 
&& 
\end{eqnarray} 
where $m=\left\langle S_{j}^{z}\right\rangle $ is the magnetization per 
spin. This equation, as explained in Subsec. 3.1, is not closed and hence a 
decoupling procedure is necessary to determine $G_{\vec{k}}(\omega )$. In 
strict analogy with the quantum Tyablikov decoupling \cite{Tyablikov67}, we 
introduce the classical Tyablikov-like decoupling (or random phase like 
approximation) for which one can approximate
\begin{equation} 
\begin{array}{c} 
\left\langle {\left\langle {S_{\vec{k}-\vec{p}}^{z}(\tau )S_{\vec{p}%
}^{+}(\tau );S_{-\vec{k}}^{-}}\right\rangle }\right\rangle _{\omega }\simeq 
\left\langle {S_{\vec{k}-\vec{p}}^{z}}\right\rangle \left\langle {%
\left\langle {S_{\vec{p}}^{+}(\tau );S_{-\vec{k}}^{-}}\right\rangle }%
\right\rangle _{\omega }, \\ 
\left\langle {\left\langle {S_{\vec{p}}^{z}(\tau )S_{\vec{k}-\vec{p}%
}^{+}(\tau );S_{-\vec{k}}^{-}}\right\rangle }\right\rangle _{\omega }\simeq 
\left\langle {S_{\vec{p}}^{z}}\right\rangle \left\langle {\left\langle {S_{%
\vec{k}-\vec{p}}^{+}(\tau );S_{-\vec{k}}^{-}}\right\rangle }\right\rangle 
_{\omega }.  
\end{array} 
\end{equation} 
Then, Eq. (\ref{Eq4.21}) yields for $G_{\vec{k}}(\omega )$ the closed 
equation
\begin{equation} 
(\omega -h)G_{\vec{k}}(\omega )=2Nm+m\left( {J(0)-J(\vec{k})}\right) G_{\vec{%
k}}(\omega ).  \label{Eq4.23} 
\end{equation} 
This gives ($(Ty)$ stands for Tyablikov):
\begin{equation} 
G_{\vec{k}}^{(Ty)}(\omega )=\frac{2Nm}{\omega -\omega _{\vec{k}}^{(Ty)}}%
\text{ },  \label{Eq4.24} 
\end{equation} 
where
\begin{equation} 
\omega _{\vec{k}}^{(Ty)}=h+mJ\left( {\gamma (0)-\gamma (\vec{k})}\right) , 
\label{Eq4.25} 
\end{equation} 
is the Tyablikov-like dispersion relation. From Eq. (\ref{Eq4.24}) and the $%
\delta $-function representation
\begin{equation} 
\mathop {\lim }\limits_{\varepsilon \rightarrow 0^{+}}{\frac{1}{{2\pi i}}}%
\left[ {{\frac{1}{{x-i\varepsilon }}}-{\frac{1}{{x+i\varepsilon }}}}\right] 
=\delta (x),  \label{Eq4.25bis} 
\end{equation} 
one immediately has (see Eq. (\ref{Eq4.21}))
\begin{equation} 
\Lambda _{k}^{(Ty)}(\omega )=i\left[ {G_{\vec{k}}(\omega +i\varepsilon )-G_{%
\vec{k}}(\omega -i\varepsilon )}\right] =2\pi (2Nm)\delta (\omega -\omega _{%
\vec{k}}^{(Ty)}).  \label{Eq4.26} 
\end{equation} 
Notice that the expression (\ref{Eq4.26}) for the spectral density $\Lambda 
_{k}^{(Ty)}(\omega )$ satisfies the sum rule (see Eq. (\ref{Eq2.15}))%
\begin{equation} 
\int_{-\infty }^{+\infty }\frac{{d\omega }}{{2\pi }}\Lambda _{\vec{k}%
}^{(Ty)}(\omega ){=i\left\langle {\left\{ {S_{\vec{k}}^{+},S_{-\vec{k}}^{-}}%
\right\} }\right\rangle }=2Nm.  \label{Eq4.27} 
\end{equation} 
Finally, from the spectral representation (\ref{Eq2.14}), we find the 
time-dependent correlation function $\left\langle {S_{\vec{k}}^{+}(\tau )S_{-%
\vec{k}}^{-}}\right\rangle $ to be (hereafter we put $K_{B}=1$)%
%
\begin{equation} 
\left\langle {S_{\vec{k}}^{+}(\tau )S_{-\vec{k}}^{-}}\right\rangle 
^{(Ty)}=T\int_{-\infty }^{+\infty }{{\frac{{d\omega }}{{2\pi }}}}{\frac{{%
\Lambda _{\vec{k}}^{(Ty)}(\omega )}}{\omega }}e^{-i\omega \tau }={\frac{{2NmT%
}}{{\omega _{\vec{k}}^{(Ty)}}}}e^{-i\omega _{\vec{k}}^{(Ty)}\tau }.  
\label{Eq4.28} 
\end{equation} 
In the static case ($\tau =0$), one has $\left\langle {S_{\vec{k}}^{+}S_{-%
\vec{k}}^{-}}\right\rangle ^{(Ty)}=2NmT/\omega _{\vec{k}}^{(Ty)}$.  

The previous relations contain the unknown mean value $m=\left\langle 
S^{z}\right\rangle $ and hence, in order to determine the thermodynamic 
properties of our spin model, we must obtain a suitable expression for $m$.  
This constitutes a serious difficulty for classical spin systems because one 
cannot write, also for $S=1/2$, a classical counterpart of the quantum 
kinematic rule for the $z$-component of the spin. However, working in the 
low-temperature regime where the angular momentum is nearly saturated $%
(S_{j}^{z}\simeq S)$, from the identity $%
S^{2}=(S_{j}^{z})^{2}+S_{j}^{+}S_{j}^{-}$ we have $\left| 
S_{j}^{+}S_{j}^{-}\right| /S^{2}\ll 1$ and the magnetization per spin $m$ 
can be approximately expressed in the form
\begin{equation} 
m\simeq S-{\frac{1}{{2S}}}\left\langle {S_{j}^{+}S_{j}^{-}}\right\rangle =S-{%
\frac{1}{{N^{2}}}}\sum\limits_{\vec{k}}{{\frac{{\left\langle {S_{\vec{k}%
}^{+}S_{-\vec{k}}^{-}}\right\rangle }}{{2S}}.}}  \label{Eq4.29} 
\end{equation} 
Then, from Eq. (\ref{Eq4.28}), one should have:
\begin{equation} 
m\simeq S\left\{ {1-}\frac{Tm}{S^{2}N}{\sum\limits_{\vec{k}}}\frac{1}{\omega 
_{\vec{k}}^{(Ty)}}\right\} .  \label{Eq4.30} 
\end{equation} 
Equations (\ref{Eq4.25}) and (\ref{Eq4.30}) constitute a closed 
self-consistent system for ($\omega _{\vec{k}}^{(Ty)},m$) which can be 
solved to obtain the thermodynamic properties under near saturation 
condition.  

A general and reliable expression for $m$ in terms of the dispersion 
relation, valid for arbitrary temperature and including the near saturation 
expression (\ref{Eq4.30}), was suggested in Refs. \cite{Campana84,Cavallo02}%
. However, since our main purpose is to focus on the CSDM predictions, we 
postpone the derivation of such general expression to the next subsection.  
Then, working to the lowest order in the CSDM, we will obtain a 
self-consistent system of equations which differs from that in the 
Tyablikov-like approximation only for a new expression of $\omega _{\vec{k}}$%
. Hence, once these more general equations are studied, one can similarly discuss 
the classical Tyablikov ones simply replacing $\omega _{\vec{k}}$ 
with $\omega _{\vec{k}}^{(Ty)}$.  


\subsection{The Classical Spectral Density Method and the Moment Equations} 

Let us introduce the $\omega $-dependent CSD 
\begin{equation} 
\Lambda _{\vec{k}}(\omega )=i\left\langle {\left\{ {S_{\vec{k}}^{+}(\tau 
),S_{-\vec{k}}^{-}}\right\} }\right\rangle _{\omega }=i\int_{-\infty 
}^{+\infty }d\tau e^{i\omega \tau }\left\langle {\left\{ {S_{\vec{k}%
}^{+}(\tau ),S_{-\vec{k}}^{-}}\right\} }\right\rangle .  
\end{equation} 
The system of ME's for $\Lambda _{\vec{k}}(\omega )$, with $A=S_{\vec{k}%
}^{+} $ and $B=S_{-{\vec{k}}}^{-}$ in Eq. (\ref{Eq3.7}), is given by 
\begin{equation} 
\int_{-\infty }^{+\infty }{{\frac{{d\omega }}{{2\pi }}}\omega ^{m}\Lambda _{%
\vec{k}}(\omega )=-i^{m-1}}\left\langle {\left\{ {\mathcal{L}_{H}^{m}S_{\vec{%
k}}^{+},S_{-\vec{k}}^{-}}\right\} }\right\rangle 
\begin{array}{cc} 
, & (m=0,1,2,...).  
\end{array} 
\label{Eq4.32} 
\end{equation} 
Focusing on the first three ME's for next considerations, the use of the 
basic Poisson brackets (\ref{Eq4.5}) yields 
\begin{equation} 
\int_{-\infty }^{+\infty }{{\frac{{d\omega }}{{2\pi }}}\Lambda _{\vec{k}%
}(\omega )=i}\left\langle {\left\{ {S_{\vec{k}}^{+},S_{-\vec{k}}^{-}}%
\right\} }\right\rangle =2Nm,  \label{Eq4.33} 
\end{equation} 
\begin{eqnarray} 
&&\int_{-\infty }^{+\infty }{{\frac{{d\omega }}{{2\pi }}}\omega \Lambda _{%
\vec{k}}(\omega )=-}\left\langle {\left\{ {\left\{ {S_{\vec{k}}^{+},H}%
\right\} ,S_{-\vec{k}}^{-}}\right\} }\right\rangle  \notag  \label{Eq4.34} \\ 
&=&{\frac{1}{N}}\sum\limits_{\vec{k}}{\left[ {J(\vec{k}^{\prime })-J(\vec{k}-%
\vec{k}^{\prime })}\right] }\left( {\left\langle {S_{\vec{k}^{\prime 
}}^{+}S_{-\vec{k}^{\prime }}^{-}}\right\rangle +2\left\langle {S_{\vec{k}%
^{\prime }}^{z}S_{-\vec{k}^{\prime }}^{z}}\right\rangle }\right) +2Nmh, 
\notag \\ 
&& 
\end{eqnarray} 
\begin{eqnarray} 
&&\int_{-\infty }^{+\infty }{{\frac{{d\omega }}{{2\pi }}}\omega ^{2}\Lambda 
_{\vec{k}}(\omega )=-}i\left\langle {\left\{ {\left\{ {\left\{ {S_{\vec{k}%
}^{+},H}\right\} ,H}\right\} ,S_{-\vec{k}}^{-}}\right\} }\right\rangle = 
\notag  \label{Eq4.35} \\ 
&=&{\frac{1}{{N^{2}}}}\sum\limits_{\vec{k}_{1},\vec{k}_{2}}{\left[ {J(\vec{k}%
_{1})-J(\vec{k}_{1}+\vec{k})}\right] }\left[ {J(\vec{k}_{1}-\vec{k}_{2})-J(%
\vec{k}_{2})}\right] \times  \notag \\ 
&&\left\{ \left\langle {S_{\vec{k}_{2}-\vec{k}_{1}}^{+}S_{-\vec{k}%
_{2}}^{-}S_{\vec{k}_{1}}^{z}}\right\rangle -\left\langle {S_{\vec{k}_{2}-%
\vec{k}_{1}}^{+}S_{-\vec{k}-\vec{k}_{2}}^{-}S_{\vec{k}+\vec{k}_{1}}^{z}}%
\right\rangle -\left\langle {S_{\vec{k}_{2}-\vec{k}_{1}}^{+}S_{\vec{k}%
_{1}}^{-}S_{-\vec{k}_{2}}^{z}}\right\rangle +\right.  \notag \\ 
&&+\left. \left\langle {S_{\vec{k}+\vec{k}_{1}}^{+}S_{-\vec{k}_{2}}^{-}S_{%
\vec{k}_{2}-\vec{k}_{1}-\vec{k}}^{z}}\right\rangle +2\left\langle {S_{\vec{k}%
+\vec{k}_{1}}^{z}S_{-\vec{k}_{2}}^{z}S_{\vec{k}_{2}-\vec{k}_{1}-\vec{k}}^{z}}%
\right\rangle \right\} +  \notag \\ 
&&+{\frac{{2h}}{N}}\sum\limits_{\vec{k}_{1}}{\left[ {J(\vec{k}_{1})-J(\vec{k}%
_{1}-\vec{k})}\right] }\left\{ {\left\langle {S_{\vec{k}_{1}}^{+}S_{-\vec{k}%
_{1}}^{-}}\right\rangle +2\left\langle {S_{\vec{k}_{1}}^{z}S_{-\vec{k}%
_{1}}^{z}}\right\rangle }\right\} +2h^{2}Nm,  \notag \\ 
&& 
\end{eqnarray} 
where we have made frequent use of the standard properties of the Poisson 
brackets.  

Notice that, in view of the general Eq. (\ref{Eq2.16}), in the previous ME's 
only the correlation function $\left\langle S_{\vec{k}}^{+}S_{-\vec{k}%
}^{-}\right\rangle $ can be exactly expressed in terms of ${\Lambda _{\vec{k}%
}(\omega )}$ as
\begin{equation} 
\left\langle S_{\vec{k}}^{+}S_{-\vec{k}}^{-}\right\rangle =\int_{-\infty 
}^{+\infty }\frac{{d\omega }}{{2\pi }}\frac{\Lambda _{\vec{k}}(\omega )}{%
\beta \omega }.  \label{Eq4.35bis} 
\end{equation} 

As explained in Subsec. 3.2, in order to give an approximate solution to the 
``moment problem`` (\ref{Eq4.32}) (or (\ref{Eq4.33})-(\ref{Eq4.35})), 
according to the CSDM one must choose an appropriate functional structure 
for the SD. The aim is to truncate the system of moments (\ref{Eq4.32}) at 
different levels and to determine $\Lambda _{\vec{k}}(\omega )$ as a 
solution of a finite set of equations. In the next sections we will consider 
some of the numerous possibilities suggested in Subsec. 3.2. Although the 
chooses are very simple in order to avoid prohibitive calculations, they are 
able to capture the essential physics of the spin model under study. In any 
case, one must remember that finding $\Lambda _{\vec{k}}(\omega )$ as a 
solution of a finite set of ME's means to have a SD which satisfies two or 
more sum rules.  


\subsubsection{One $\protect\delta$-function Polar Ansatz} 

As suggested by the exact spectral decomposition (\ref{Eq2.35}), according 
to the spirit of the CSDM, we try to determine $\Lambda _{\vec{k}}(\omega )$%
, to the lowest order, in the form \cite{Campana83,Cavallo02,Cavallo04} 
\begin{equation} 
\Lambda _{\vec{k}}(\omega )=2\pi \lambda _{\vec{k}}\delta (\omega -\omega _{%
\vec{k}}),  \label{Eq4.36} 
\end{equation} 
involving two unknown parameters $\lambda _{\vec{k}}$ and $\omega _{\vec{k}}$%
. Then, to determine these parameters only the first two ME's (\ref{Eq4.33}) 
and (\ref{Eq4.34}) are necessary. Taking into account the polar ansatz (\ref 
{Eq4.36}), these reduce to the integral equations:
\begin{eqnarray}  \label{Eq4.37b} 
\lambda _{\vec{k}} &=&2Nm  \label{Eq4.37a} \\ 
\lambda _{\vec{k}}\omega _{\vec{k}} &=&2Nmh+\frac{1}{N}\sum\limits_{\vec{k}%
^{\prime }}\left[ J(\vec{k}^{\prime })-J(\vec{k}-\vec{k}^{\prime })\right] 
\left( {\left\langle {S_{\vec{k}^{\prime }}^{+}S_{-\vec{k}^{\prime }}^{-}}%
\right\rangle +2\left\langle {S_{\vec{k}^{\prime }}^{z}S_{-\vec{k}^{\prime 
}}^{z}}\right\rangle }\right) .  \notag \\ 
&&  \label{Eq4.37c} 
\end{eqnarray} 
To close this system, we must express the involved unknown quantities in 
terms of $\Lambda _{\vec{k}}(\omega )$. The transverse correlation function $%
{\left\langle {S_{\vec{k}}^{+}S_{-\vec{k}}^{-}}\right\rangle }$, which 
enters the right-hand side of Eq. (\ref{Eq4.37c}), can easily expressed in 
terms of the SD and hence in terms of the parameters $\lambda _{\vec{k}}$ 
and $\omega _{\vec{k}}$. Indeed, Eq. (\ref{Eq2.16}) and the ansatz (\ref 
{Eq4.36}) yield%
%
\begin{equation} 
{\left\langle {S_{\vec{k}}^{+}S_{-\vec{k}}^{-}}\right\rangle }={\frac{2Nm}{%
\beta \omega _{\vec{k}}}}.  \label{Eq4.38} 
\end{equation} 
Unfortunately, there is not a simple way to express the longitudinal 
correlation function ${\left\langle {S_{\vec{k}}^{z}S_{-\vec{k}}^{z}}%
\right\rangle }$ in terms of $\Lambda _{\vec{k}}(\omega )$. Hence, to close 
the truncated system of ME's (\ref{Eq4.33})-(\ref{Eq4.34}), one should 
introduce the longitudinal SD $\Lambda _{\vec{k}}^{(z)}({\vec{\omega}}%
)=i\left\langle {\left\{ {S_{\vec{k}}^{z}(\tau ),S_{-\vec{k}}^{z}}\right\} }%
\right\rangle $ and then formulate and solve another moment problem 
according to the basic idea of the CSDM. However, this problem would be 
coupled to that for the transverse SD $\Lambda _{\vec{k}}(\omega )$ and the 
difficulties would be sensibly amplified also if a one $\delta $-function 
structure for both $\Lambda _{\vec{k}}(\omega )$ and $\Lambda _{\vec{k}%
}^{(z)}(\omega )$ is assumed. The simplest way to overcome this difficulty, 
frequently used in literature \cite 
{Kalashnikov69,Kalashnikov73,Campana79,Caramico80,Nolting89,Nolting91,Hermann97, Caramico81,Campana83,Campana84,Cavallo01}%
, is to resort, as a first level, to the decoupling procedure $\left\langle 
\ S_{\vec{k}}^{z}S_{-\vec{k}}^{z}\right\rangle \simeq \left\langle S_{\vec{k}%
}^{z}\right\rangle \left\langle S_{-\vec{k}}^{z}\right\rangle 
=N^{2}m^{2}\delta _{\vec{k},0}$ which corresponds to neglect the 
correlations between the Fourier transforms of the longitudinal spin 
components. With this approximation, Eqs. (\ref{Eq4.37a})-(\ref{Eq4.38}) 
immediately yield the self-consistent equation for the frequency dispersion 
relation 
\begin{equation} 
\omega _{\vec{k}}=h+m\left( J(0)-J(\vec{k})\right) +\frac{T}{N}\sum\limits_{%
\vec{k}^{\prime }}\frac{J(\vec{k}^{\prime })-J(\vec{k}-\vec{k}^{\prime })}{%
\omega _{\vec{k}^{\prime }}}{.}  \label{Eq4.39} 
\end{equation} 
It is worth noting that the first two terms in the right-hand side of Eq. (%
\ref{Eq4.39}) constitutes just the expression for the dispersion relation 
obtained in Subsec. 4.2 using the Tyablikov-like decoupling within the EMM.  
Of course, the previous decoupling and the Eq. (\ref{Eq4.39}) are 
appropriate to describe only thermodynamic regimes with a finite 
magnetization as under near saturation conditions. In particular, one cannot 
use the basic Eq. (\ref{Eq4.39}) to explore near-zero magnetization domains 
in the phase diagram as the critical region and the paramagnetic phase in 
zero external magnetic field. To overcome this difficulty, one is forced to 
find a more appropriate decoupling procedure for the longitudinal 
correlation function which allows us to obtain self-consistent ME's 
appropriate for describing regimes when $m\rightarrow 0$ and preserves also 
the simplicity of the one $\delta $-function ansatz for the transverse SD. A 
possible and successful solution to this crucial question was suggested many 
years ago for Heisenberg spin models with short-range interactions (SRI's) 
\cite{Campana84}. It was shown that a suitable decoupling procedure when the 
magnetization approaches to zero (see also Refs. \cite 
{Kalashnikov69,Kamieniarz77} for the quantum counterpart) consists in 
writing in the ME (\ref{Eq4.34})
\begin{eqnarray} 
&&\frac{1}{N}\sum\limits_{\vec{k}^{\prime }}{\left[ {J(\vec{k}^{\prime })-J(%
\vec{k}-\vec{k}^{\prime })}\right] \left\langle {S_{\vec{k}^{\prime 
}}^{z}S_{-\vec{k}^{\prime }}^{z}}\right\rangle }\simeq   \notag 
\label{Eq4.40} \\ 
&\simeq &\frac{1}{N}\sum\limits_{\vec{k}^{\prime }}{\left[ {J(\vec{k}%
^{\prime })-J(\vec{k}-\vec{k}^{\prime })}\right] }\left\{ {\left\langle {S_{%
\vec{k}^{\prime }}^{z}}\right\rangle \left\langle {S_{-\vec{k}^{\prime }}^{z}%
}\right\rangle -{\frac{1}{2}}\left( {1-{\frac{{\left\langle {S_{0}^{z}}%
\right\rangle }}{{N^{2}S^{2}}}}}\right) \left\langle {S_{\vec{k}^{\prime 
}}^{+}S_{-\vec{k}^{\prime }}^{-}}\right\rangle }\right\} .  \notag \\ 
&& 
\end{eqnarray} 
This equation only involves $\Lambda _{\vec{k}}(\omega )$ (see Eqs. (\ref 
{Eq2.16}) and (\ref{Eq4.38})) and hence, inserting in the ME (\ref{Eq4.34}) 
one finds, for the dispersion relation, the new expression
\begin{equation} 
\omega _{\vec{k}}=h+m\left( {J(0)-J(\vec{k})}\right) +\frac{m^{2}}{S^{2}}%
\frac{T}{N}\sum\limits_{\vec{k}^{\prime }}\frac{J(\vec{k}^{\prime })-J(\vec{k%
}-\vec{k}^{\prime })}{{\omega _{\vec{k}^{\prime }}}}.  \label{Eq4.41} 
\end{equation} 
Notice that the effect of the decoupling (\ref{Eq4.40}) corresponds 
essentially to perform in Eq. (\ref{Eq4.39}) the transformation $%
\sum\limits_{\vec{k}^{\prime }}(...)\rightarrow m^{2}/S^{2}\sum\limits_{\vec{%
k}^{\prime }}(...)$. Of course, alternative decoupling procedures can be 
conjectured for taking into account the effect of the longitudinal spin 
correlations working in terms of the single $\Lambda _{\vec{k}}(\omega )$.  
An example, appropriate to near saturation regimes, will be presented in 
Subsec. 4.5.2.  

\subsubsection{Magnetization per Spin in Terms of the Transverse Spectral 
Density} 

To close the system of ME's (\ref{Eq4.33})-(\ref{Eq4.35}) for $\Lambda _{%
\vec{k}}(\omega )$ (and also the EM (\ref{Eq4.21}) for $G_{\vec{k}}(\omega )$%
), we must find an explicit expression for $m=\left\langle 
S_{j}^{z}\right\rangle $ in terms of the SD. A reliable and successful 
expression of $m$, valid for any $T$ and $h$, can be obtained by means of 
the following procedure within the spirit of the SDM \cite 
{Campana84,Cavallo02}. Let us introduce the higher-order SD:
\begin{equation} 
\Omega (\omega )={\frac{1}{{N^{2}}}}\sum\limits_{\vec{k},\vec{k}^{\prime }}{%
\left[ {i\left\langle {\left\{ {S_{\vec{k}+\vec{k}^{\prime }}^{+}(\tau ),S_{-%
\vec{k}}^{-}S_{-\vec{k}^{\prime }}^{z}}\right\} }\right\rangle _{\omega }}%
\right] ,}  \label{Eq4.42} 
\end{equation} 
where the summand is constructed by $\Lambda _{\vec{k}}(\omega 
)=i\left\langle {\left\{ {S_{\vec{k}}^{+}(\tau ),S_{-\vec{k}}^{-}}\right\} }%
\right\rangle _{\omega }$ with the change $S_{\vec{k}}^{+}(\tau )\rightarrow 
S_{\vec{k}+\vec{k}^{\prime }}^{+}(\tau )$ and $S_{-\vec{k}}^{-}\rightarrow 
S_{-\vec{k}}^{-}S_{-\vec{k}^{\prime }}^{z}$. The SD (\ref{Eq4.42}) is 
associated to the higher-order GF
\begin{equation} 
G^{(r)}(\omega )=\sum\limits_{\vec{k},\vec{k}^{\prime }}{\left[ {\theta 
(\tau )\left\langle {\left\{ {S_{\vec{k}+\vec{k}^{\prime }}^{+}(\tau ),S_{-%
\vec{k}}^{-}S_{-\vec{k}^{\prime }}^{z}}\right\} }\right\rangle _{\omega }}%
\right] .}  \label{Eq4.43} 
\end{equation} 
We now perform in Eq. (\ref{Eq4.42}) the decoupling procedure%
\begin{equation} 
\left\langle {\left\{ {S_{\vec{k}+\vec{k}^{\prime }}^{+}(\tau ),S_{-\vec{k}%
}^{-}S_{-\vec{k}^{\prime }}^{z}}\right\} }\right\rangle _{\omega }\approx 
\left\langle {S_{-\vec{k}^{\prime }}^{z}}\right\rangle (1+a)\left\langle {%
\left\{ {S_{\vec{k}+\vec{k}^{\prime }}^{+}(\tau ),S_{-\vec{k}}^{-}}\right\} }%
\right\rangle ,  \label{Eq4.44} 
\end{equation} 
where $a$ is an unknown parameter to be properly determined (notice that, 
with $a=0$, Eq. (\ref{Eq4.44}) reduces to a Tyblikov-like decoupling). It is 
easy to see that this allows us to reduce $\Omega (\omega )$ to the simplest 
form
\begin{equation} 
\Omega (\omega )=m(1+a){\frac{1}{N}}\sum\limits_{\vec{k}}{\Lambda _{\vec{k}%
}(\omega ),}  \label{Eq4.45} 
\end{equation} 
in terms of the original $\Lambda _{\vec{k}}(\omega )$. The parameter $a$ 
can be now determined by imposing a sum rule for $\Omega (\omega )$, and 
precisely, by requiring that the zeroth-ME for the exact $\Omega (\omega )$ 
is preserved in the decoupling procedure. Taking into account the general 
Eq. (\ref{Eq2.15}) and the identity $(S_{j}^{z})^{2}=S^{2}-S_{j}^{+}S_{j}^{-} 
$, a straightforward use of the Poisson bracket properties yields, for $%
\Omega (\omega )$ in Eq. (\ref{Eq4.42}), 
\begin{equation} 
\int_{-\infty }^{+\infty }{{\frac{{d\omega }}{{2\pi }}}\Omega (\omega )={%
\frac{1}{N^{2}}}\sum\limits_{\vec{k},\vec{k}^{\prime }}{\left[ {%
i\left\langle {\left\{ {S_{\vec{k}+\vec{k}^{\prime }}^{+}(\tau ),S_{-\vec{k}%
}^{-}S_{-\vec{k}^{\prime }}^{z}}\right\} }\right\rangle }\right] 
=2N^{2}S^{2}-3N\left\langle {S_{j}^{+}S_{j}^{-}}\right\rangle }}.  
\label{Eq4.46} 
\end{equation} 
On the other hand, the zeroth-ME for the reduced form (\ref{Eq4.45}) for $%
\Omega (\omega )$ is given by (see Eq. (\ref{Eq4.33}))
\begin{equation} 
\int_{-\infty }^{+\infty }{{\frac{{d\omega }}{{2\pi }}}\Omega (\omega 
)\approx m(1+a){\frac{1}{N}}\sum\limits_{\vec{k}}{\int_{-\infty }^{+\infty }{%
{\frac{{d\omega }}{{2\pi }}}}}}\Lambda _{\vec{k}}(\omega )=2Nm^{2}(1+a).  
\label{Eq4.47} 
\end{equation} 
Then, by imposing that Eqs. (\ref{Eq4.46}) and (\ref{Eq4.47}) coincide, we 
obtain
\begin{equation} 
m^{2}(1+a)=S^{2}-{\frac{3}{2}}\left\langle {S_{j}^{+}S_{j}^{-}}\right\rangle 
.  \label{Eq4.48} 
\end{equation} 
From this equation, it follows that in the high-temperature regime, when $%
m\rightarrow 0$, the isotropy condition $\left\langle {S_{j}^{+}S_{j}^{-}}%
\right\rangle ={3}S^{2}/2$ is consistently reproduced regardless of the 
value of $a$. On the other hand, in the regimes where the spins $S_{j}^{z}$ 
are nearly saturated ($S_{j}^{z}\simeq S$ or $\left| 
S_{j}^{+}S_{j}^{-}\right| /S^{2}\ll 1$), the magnetization per spin can be 
approximately expressed in the form (\ref{Eq4.29}). Then, according to Eq. (%
\ref{Eq4.48}), the parameter $a$ must go to zero in an appropriate way so 
that Eq. (\ref{Eq4.29}) is recovered. Bearing this in mind, a direct 
comparison between Eqs. (\ref{Eq4.29}) and (\ref{Eq4.48}) provides for $a$ 
the expression
\begin{equation} 
a={\frac{{\left\langle {S_{j}^{+}S_{j}^{-}}\right\rangle }}{{2S^{2}}}.} 
\label{Eq4.49} 
\end{equation} 
Inserting it in Eq. (\ref{Eq4.48}) leads to the required formula for the 
magnetization per spin $m$ suitable for our aims (with $h>0$)%
\begin{equation} 
m=S\left[ {{\frac{{\displaystyle{1-{\frac{3}{2}}{\frac{{\left\langle {S_{j}^{+}S_{j}^{-}}%
\right\rangle }}{{S^{2}}}}}}}{\displaystyle{{1-{\frac{1}{2}}{\frac{{\left\langle {%
S_{j}^{+}S_{j}^{-}}\right\rangle }}{{S^{2}}}}}}}}}\right] ^{{\frac{1}{2}}}, 
\label{Eq4.50} 
\end{equation} 
where
\begin{equation} 
\left\langle {S_{j}^{+}S_{j}^{-}}\right\rangle ={\frac{1}{{N^{2}}}}%
\sum\limits_{\vec{k}}{\left\langle {S_{\vec{k}}^{+}S_{-\vec{k}}^{-}}%
\right\rangle }.  \label{Eq4.51} 
\end{equation} 
One can immediately check that, under near saturation condition, Eq. (\ref 
{Eq4.50}) reproduces the relation (\ref{Eq4.29}). In view of Eq. (\ref 
{Eq4.51}) and the exact expression of ${\left\langle {S_{\vec{k}}^{+}S_{-%
\vec{k}}^{-}}\right\rangle }$ in terms of $\Lambda _{\vec{k}}(\omega )$ (see 
Eq. (\ref{Eq4.35bis})), Eq. (\ref{Eq4.50}) for $m$ allows us to take 
immediate contact with the SDM and, in particular, with the one $\delta $%
-function for $\Lambda _{\vec{k}}(\omega )$ introduced in the previous 
subsection.  

Postponing this problem to the next subsection, we wish to outline here how 
the expression (\ref{Eq4.50}), which is valid for arbitrary temperature and 
magnetic field, can be used in the EMM for $G_{\vec{k}}(\omega )$ within the 
Tyablikov-like approximation (see Subsec. 4.2). In this case the GF has a 
real pole at $\omega _{\vec{k}}^{(Ty)}=h+mJ(\gamma (0)-\gamma ({\vec{k}})$ 
which corresponds to a one $\delta $-function for the SD. This feature and 
the general expression (\ref{Eq4.50}) for $m$ allows us to express $m$ as a 
function of $\omega _{\vec{k}}^{(Ty)}$ yielding to a system of two 
self-consistent equations. When this system is solved, all the thermodynamic 
properties can be obtained.  

It is worth emphasizing that the one $\delta $-pole ansatz for $\Lambda _{%
\vec{k}}(\omega )$ within the CSDM, involving two ME's, is not equivalent to 
the classical Tyablikov decoupling. This becomes clear by comparing the two 
different expressions (\ref{Eq4.25}) and (\ref{Eq4.39}) for the dispersion 
relation, although one assumes the same expression (\ref{Eq4.50}) (or (\ref 
{Eq4.27})) for $m$. The main reason is that, while $\Lambda _{\vec{k}%
}^{(Ty)}(\omega )$ (and hence $\omega _{\vec{k}}^{(Ty)}$) satisfies only the 
first of the sum rules (\ref{Eq4.33})-(\ref{Eq4.34}), $\Lambda _{\vec{k}%
}(\omega )$ (and hence $\omega _{\vec{k}}$) is obtained to satisfy both the 
sum rules (\ref{Eq4.33}) and (\ref{Eq4.34}). In this sense, one can claim 
that the one $\delta $-function ME's solution for $\Lambda _{\vec{k}}(\omega 
)$ in the CSDM is better than the corresponding one derived by means of the 
Tyablikov-like decoupling in the classical EMM for $G_{\vec{k}}(\omega )$ 
and constitutes a one step beyond such an approximation.  

\subsubsection{Moment Equations and Dispersion Relation for one $\protect%
\delta $-pole Ansatz} 

We now come back to the Subsec. 4.3.1 and add, to the Eq. (\ref{Eq4.39}) or (%
\ref{Eq4.41}) for the dispersion relation, the corresponding equation for $m$ 
in terms of the parameter $\lambda _{\vec{k}}$ and $\omega _{\vec{k}}$ in $%
\Lambda _{\vec{k}}(\omega )=\lambda _{\vec{k}}\delta (\omega -\omega _{\vec{k%
}})$. From Eq. (\ref{Eq4.50}), we easily have
\begin{equation} 
{\frac{{m^{2}}}{{S^{2}}}}={\frac{{1-3\displaystyle{{\frac{{Tm}}{{S^{2}N}}}\sum\limits_{%
\vec{k}}{{\frac{1}{{\omega _{\vec{k}}}}}}}}}{{1-\displaystyle{{\frac{{Tm}}{{S^{2}N}}}%
\sum\limits_{\vec{k}}{{\frac{1}{{\omega _{\vec{k}}}}}}}}},}  \label{Eq4.53} 
\end{equation} 
which, in the near saturation regime, reduces to
\begin{equation} 
m\simeq S\left\{ {1-{\frac{{Tm}}{{S^{2}N}}}\sum\limits_{\vec{k}}{{\frac{1}{{%
\omega _{\vec{k}}}}}}}\right\} .  
\end{equation} 
Eq. (\ref{Eq4.39}) (for $m\neq 0$) or (\ref{Eq4.41}) (for $m\rightarrow 0$) 
and Eq. (\ref{Eq4.53}) represent the closed system of ME's to be solved 
self-consistently. For this purpose it is convenient to introduce the 
dimensionless variables
\begin{equation} 
\sigma ={\frac{m}{S}}\quad ,\quad \bar{T}={\frac{{T}}{{JS^{2}}}}\quad ,\quad 
\bar{h}={\frac{h}{{JS}}}\quad ,\quad \bar{\omega}_{\vec{k}}={\frac{{\omega _{%
\vec{k}}}}{{JS}}}.  \label{Eq4.55} 
\end{equation} 
Then, taking the thermodynamic limit $N\rightarrow \infty $ with ${\frac{1}{N%
}}\sum\limits_{\vec{k}}(...)\rightarrow \int_{1BZ}d^{d}k(...)/(2\pi )^{d}$, 
our self-constistent equations can be written as
\begin{eqnarray} 
\bar{\omega}_{\vec{k}} &=&\bar{h}+\sigma \Omega ^{(P)}(\vec{k})R(\vec{k}) 
\label{Eq4.56} \\ 
\sigma ^{2} &=&{\frac{{1-3\displaystyle{\bar{T}\sigma \int_{1BZ}{{\frac{{d^{d}k}}{{(2\pi 
)^{d}}}}{\frac{1}{{\bar{\omega}_{\vec{k}}}}}}}}}{{1-3\displaystyle{\bar{T}\sigma \int_{1BZ}{%
{\frac{{d^{d}k}}{{(2\pi )^{d}}}}{\frac{1}{{\bar{\omega}_{\vec{k}}}}}}}}}}%
\text{ },  \label{Eq4.57} 
\end{eqnarray} 
with
\begin{equation} 
R({\vec{k}})=\left\{ 
\begin{array}{lcl} 
1+{\frac{{\bar{T}}}{\sigma }}\displaystyle{\int_{1BZ}{{\frac{{d^{d}k^{\prime }}}{{(2\pi 
)^{d}}}}{\frac{{\Omega ^{(p)}(\vec{k}-\vec{k}^{\prime })-\Omega ^{(p)}(\vec{k%
}^{\prime })}}{{\Omega ^{(p)}(\vec{k})\bar{\omega}_{\vec{k}^{\prime }}}}}}} & 
, & \sigma \neq 0 \\ 
1+\bar{T}\sigma \displaystyle{\int_{1BZ}{{\frac{{d^{d}k^{\prime }}}{{(2\pi )^{d}}}}{\frac{{%
\Omega ^{(p)}(\vec{k}-\vec{k}^{\prime })-\Omega ^{(p)}(\vec{k}^{\prime })}}{{%
\Omega ^{(p)}(\vec{k})\bar{\omega}_{\vec{k}^{\prime }}}}}}} & , & \sigma 
\rightarrow 0, 
\end{array} 
\right.  \label{Eq4.58} 
\end{equation} 
and
\begin{equation} 
\Omega ^{(p)}(\vec{k})=\gamma (0)-\gamma ({\vec{k}})=\sum\limits_{\vec{r}}{{%
\frac{{1-\cos (\vec{k}\cdot \vec{r})}}{{\left| {\vec{r}}\right| ^{p}}}}}.  
\label{Eq4.59} 
\end{equation} 

The system of self-consistent integral Eqs. (\ref{Eq4.56})-(\ref{Eq4.58}) 
for the unknown parameters $\bar{\omega}_{\vec{k}}$ (the dimensionless or 
reduced excitation dispersion relation) and $\sigma $ \c{(}the dimensionless 
or reduced magnetization per spin) is very hard to solve. For obtaining 
explicit results one is forced to consider asymptotic thermodynamic regimes 
or resort to numerical calculations.  


\subsection{Thermodynamic and Critical Properties within the one $\protect%
\delta $-pole Ansatz\\ for the Transverse Spectral Density} 

\subsubsection{Low-Temperature Properties and Long-Range Order} 

Our primary purpose is to solve the self-consistent system of Eqs. (\ref 
{Eq4.56})-(\ref{Eq4.58}) and then, using the general relations in terms of $%
\Lambda _{\vec{k}}(\omega )$, to determine the relevant thermodynamic 
properties of the classical Heisenberg FM model (\ref{Eq4.1}). Since any 
attempt to obtain explicit analytical solutions for all the allowed values 
of $T$ and $h$ is hopeless, we first examine the possibility to have 
analytical results in the low-temperature regime. We can expand Eqs. (\ref 
{Eq4.56}) and (\ref{Eq4.57}) in power series of ${\bar{T}}$ for allowed 
values of $\bar{h}$ which prevent the occurrence of divergences. Focusing on 
the first order expressions in the reduced temperature, which are sufficient 
to capture the relevant features of the low-temperature physics of the model 
(\ref{Eq4.1}), we have:
\begin{equation} 
\bar{\omega}_{\vec{k}}\cong \bar{h}+\Omega ^{(p)}(\vec{k})-\bar{T}\left\{ 
\mathcal{F}{_{1}^{(p)}(\bar{h})\Omega ^{(p)}(\vec{k})+}\mathcal{F}{%
_{2}^{(p)}(\bar{h},\vec{k})}\right\} +O(\bar{T}^{2}),  \label{Eq4.60} 
\end{equation} 
and
\begin{equation} 
\sigma \approx 1-\bar{T}\mathcal{F}_{1}^{(p)}(\bar{h})+O(\bar{T}^{2}), 
\label{Eq4.61} 
\end{equation} 
where
\begin{equation} 
\mathcal{F}_{1}^{(p)}(\bar{h})=\int_{1BZ}{{\frac{{d^{d}k}}{{(2\pi )^{d}}}}{%
\frac{1}{{\bar{h}+\Omega ^{(p)}(\vec{k})}}}}\quad ,  \label{Eq4.62} 
\end{equation} 
and
\begin{equation} 
\mathcal{F}_{2}^{(p)}(\bar{h},\vec{k})=\int_{1BZ}{{\frac{{d^{d}k^{\prime }}}{%
{(2\pi )^{d}}}}{\frac{{\Omega ^{(p)}(\vec{k}^{\prime })-\Omega ^{(p)}(\vec{k}%
-\vec{k}^{\prime })}}{{\bar{h}+\Omega ^{(p)}(\vec{k}^{\prime })}}}}\quad .  
\label{Eq4.63} 
\end{equation} 
At first, we assume ${\bar{h}}\neq 0$ so that no convergency problems occur 
for the $\vec{k}$-integrals in Eqs. (\ref{Eq4.62}) and (\ref{Eq4.63}).  
Unfortunately, since one cannot obtain an explicit expression of the 
function $\Omega ^{(p)}({\vec{k}})$ in terms of elementary functions for $%
\vec{k}$ in the whole $1BZ$ for arbitrary values of the interaction 
parameter $p>d$, the integrations in Eqs. (\ref{Eq4.62}) and (\ref{Eq4.63}) 
cannot be performed explicitly. Nevertheless, for a sufficiently low 
external magnetic field, the dominant contribution to the integrals arises 
from the low wave-vector excitations. Hence, one can obtain an explicit 
estimate of $\bar{\omega}_{\vec{k}}$ and $\sigma $ assuming the dominant 
behavior of $\Omega ^{(p)}({\vec{k}})$ in the $1BZ$ as $k=\left| {\vec{k}}%
\right| \rightarrow 0$, provided that the $\bar{h}$-dependent coefficient of 
$\bar{T}$ in Eqs. (\ref{Eq4.60}) and (\ref{Eq4.61}) remain finite. Bearing 
this in mind, one can show \cite{Bruno01,Romano89,Romano90,Nakano94b} that, 
for $p>d$, we have for $\Omega ^{(p)}({\vec{k}})$ the low-$k$ expansions%
\begin{equation} 
\Omega ^{(p)}({\vec{k}})\simeq \left\{ 
\begin{array}{lcl} 
A_{d}k^{p-d}+B_{d}k^{2}+O(k^{4}) & , & p\neq d+2 \\ 
\\
C_{d}k^{2}\ln ({\frac{1}{\Lambda }})+O(k^{4}) & , & p=d+2\quad .  
\end{array} 
\right.  \label{Eq4.64} 
\end{equation} 
Here, the coefficients $A_{d}$, $B_{d}$ and $C_{d}$ depend in a cumbersome 
way on the dimensionality $d$, the interaction exponent $p$ and the 
wave-vector cutoff $\Lambda $ related to the geometrical definition of $1BZ$%
. Their explicit general expressions are inessential for our purposes and 
will be omitted. However, for case $d=2$, they will be explicitly presented 
in Subsec. 4.4.2 where our analytical predictions near criticality are 
compared with those obtained by MC simulations.  

The most relevant analytical and numerical results for case $d=1$ will be 
considered explicitly after a discussion of the low-temperature properties 
for general $d$ with $p>d$.  

Taking into account the asymptotic behavior (\ref{Eq4.64}), the integral $%
\mathcal{F}_{1}^{(p)}({\bar{h}})$ can be easily estimated and we get, to 
leading order in $\bar{T}$, the following near saturation representations 
for the reduced magnetization per spin
\begin{equation} 
\sigma ({\bar{T}},{\bar{h}})\simeq 1-{\bar{T}}\cdot \left\{ 
\begin{array}{lcl} 
\overline{{h}}^{^{-1}}{}_{2}\displaystyle{F_{1}\left( {1,{\frac{d}{{p-d}}},{\frac{p}{{p-d}}%
};-{\frac{{A_{d}\Lambda ^{p-d}}}{{\bar{h}}}}}\right)} & , & d<p<d+2 \\ 
\\
K_{d}\displaystyle{\int_{0}^{\Lambda }dkk^{d-1}\left[ {\bar{h}+C_{d}k^{2}\ln \left( {{%
\frac{1}{k}}}\right) }\right] ^{-1}} & , & p=d+2 \\
\\ 
\overline{{h}}^{^{-1}}{}_{2}\displaystyle{F_{1}\left( {1,{\frac{d}{2}},{\frac{{d+2}}{2}};-{%
\frac{{B_{d}\Lambda ^{2}}}{{\bar{h}}}}}\right)} & , & p>d+2 
\end{array} 
\right.  \label{Eq4.65} 
\end{equation} 
where ${}_{2}F_{1}(a,b,c;z)$ is the hypergeometric function, $%
K_{d}=2^{1-d}\pi ^{-{\frac{d}{2}}}/\Gamma \left( d/2\right) $ and $\Gamma 
(z) $ is the gamma function.  

An explicit estimate of $\bar{\omega}_{\vec{k}}$ for small $k$ can be 
obtained by means of an analogous but rather complicated and tedious 
calculation of $\mathcal{F}_{2}^{(p)}({\bar{h},{\vec{k}}})$. However, since 
an explicit estimate of $\bar{\omega}_{\vec{k}}$ for arbitrary $p$ and $d$ 
(with $p>d$) is inessential for next developments, we avoid to present the 
related cumbersome result. Information about the reduced excitation 
dispersion relation $\bar{\omega}_{\vec{k}}$ in the whole $1BZ$ will be 
given below for $d=1$ varying the interaction parameter $p>1$.  

The low-temperature susceptibility $\chi =\left( {\partial m/\partial h}%
\right) _{T}$ can be now easily obtained from Eq. (${}$\ref{Eq4.65}).  
Indeed, for the reduced susceptibility $\bar{\chi}=\chi /J$ we have%
\begin{eqnarray} 
\bar{\chi}(\bar{T},\bar{h}) &=&\left( {{\frac{{\partial \sigma }}{{\partial 
\bar{h}}}}}\right) _{\bar{T}}\simeq   \notag  \label{Eq4.66} \\ 
&\simeq &{\bar{T}}\left\{ 
\begin{array}{lcl} 
\displaystyle{\overline{{h}}^{^{-2}}{}_{2}F_{1}\left( {2,{\frac{d}{{p-d}}},{\frac{p}{{p-d}}%
};-{\frac{{A_{d}\Lambda ^{p-d}}}{{\bar{h}}}}}\right)}  & , & d<p<d+2 \\
\\ 
\displaystyle{K_{d}\int_{0}^{\Lambda }{dkk^{d-1}\left[ {\bar{h}+C_{d}k^{2}\ln \left( {{%
\frac{1}{k}}}\right) }\right] }^{-2}} & , & p=d+2 \\
\\ 
\displaystyle{\overline{{h}}^{^{-2}}{}_{2}F_{1}\left( {1,{\frac{d}{2}},{\frac{{d+2}}{2}};-{%
\frac{{B_{d}\Lambda ^{2}}}{{\bar{h}}}}}\right)}  & , & p>d+2\quad .  
\end{array} 
\right.   \notag \\ 
&& 
\end{eqnarray} 
Of course, the above expressions have a physical meaning only for values of $%
\bar{T}$ and $\bar{h}$ such that the near saturation condition ($\sigma 
\simeq 1$) is assured and hence also for $\bar{h}\rightarrow 0$ if the 
long-range order (LRO) occurs. In particular, if we write in general $\sigma 
\simeq 1-a(\bar{h})\bar{T}$, this condition implies necessary $a(\bar{h})%
\bar{T}\ll 1$ and hence $\bar{T}a(\bar{h})$ becomes the natural expansion 
parameter in the problem. As already mentioned before, the integrals (${}$%
\ref{Eq4.62}) and (${}$\ref{Eq4.63}), and hence the functions in Eqs. (${}$%
\ref{Eq4.65})-(${}$\ref{Eq4.66}), could diverge as $\bar{h}\rightarrow 0$ 
for particular values of $p$ and $d$. When this does not happen, Eq. (${}$%
\ref{Eq4.65}) should yield a spontaneous magnetization at a finite 
temperature signaling that our spin model exhibits LRO. From the low-$k$ 
behaviors (${}$\ref{Eq4.64}) it is easy to see that the integral $\mathcal{F%
}_{1}^{(p)}(\bar{h})$ converges in the limit $\bar{h}\rightarrow 0$ only for 
$d<p<2d$ with $d\leq 2$ and for $p>d$ with $d>2$. This means that for these 
values of $p$ and $d$ a spontaneous magnetization per spin $m_{0}(T)=\sigma 
_{0}(T)S$, and hence LRO, exists at small but finite $T$ with%
\begin{equation} 
\sigma _{0}(T)\simeq 1-\bar{T}\mathcal{F}_{1}^{(p)}(0),  \label{Eq4.67} 
\end{equation} 
where $\mathcal{F}_{1}^{(p)}(0)$ is a finite quantity given by Eq. (${}$\ref 
{Eq4.62}) for $\bar{h}=0$. Of course, an estimate of this constant 
(depending on $p$ and $d$) can be immediately derived from Eq. (${}$\ref 
{Eq4.65}). For $p\geq 2d$ with $d<2$, the integral $\mathcal{F}_{1}^{(p)}(%
\bar{h})$ diverges as $\bar{h}\rightarrow 0$ so that no finite solution for $%
\sigma $ exists at $\bar{T}\neq 0$. Then, for these values of $p$ and $d$, 
Eq. (${}$\ref{Eq4.61}) is satisfied only if $\sigma =0$ and hence no LRO 
occurs at finite temperature. In Fig. 1 we present the ($p-d$)-plane version 
of the phase diagram of our FM model displaying the scenario discussed 
above. Here we also show the regions where, as we shall see in the next 
Subsec. 4.4.2: i) the system exhibits a critical behavior like for the 
Heisenberg model with SRI's and ii) the LRI's become effective.  

\begin{figure}[tbh] 
\centerline{ 
\includegraphics*[width=9cm]{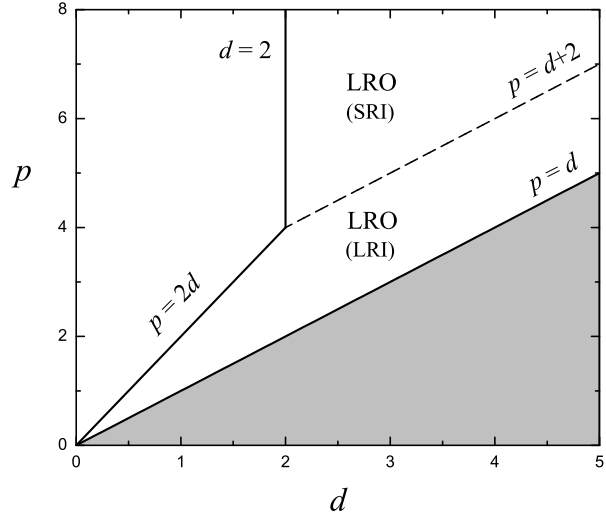}} 
\caption{Regions of the ($p-d$)-plane where long-range order (LRO) takes 
place for a $d$-dimensional spin-$S$ Heisenberg FM model with interactions 
decaying as $r^{-p}$. The dashed line separates the domains where long-range 
interaction (LRI)- and short-range interaction (SRI)- regimes occur. The 
dark region corresponds to nonextensive thermodynamics not considered 
through this paper.} 
\end{figure} 

We can conclude that our low-temperature results suggest that a transition 
to a FM phase at a finite temperature occurs in the regions of the ($p-d$%
)-plane where LRO takes place. In the remaining domains a different scenario 
happens with absence of a phase transition. These predictions agree with the 
recent extensions \cite{Bruno01} of the well known Mermin theorem \cite 
{MerminWagner66} to spin models with LRI's of the type here considered.  

Other low-temperature thermodynamic properties follow from the general exact 
expressions (${}$\ref{Eq4.14}) and (${}$\ref{Eq4.15}). Within our 
approximations, the correlation functions in these equations can be 
expressed in terms of the transverse SD and hence the reduced internal 
energy $\bar{u}=u/JS^{2}$ and free energy $\overline{f}=f/JS^{2}${\ }per 
spin assume the forms
\begin{equation} 
\bar{u}=-\bar{h}\sigma -\bar{T}\sigma \int_{1BZ}{{\frac{{d^{d}k}}{{(2\pi 
)^{d}}}}}{\frac{{\gamma (\vec{k})}}{{\bar{\omega}_{\vec{k}}}}}-{\frac{{1}}{{2%
}}\gamma (0)}\sigma ^{2},  \label{Eq4.68} 
\end{equation} 
and
\begin{equation} 
\bar{f}=\bar{f}_{0}-\bar{T}\int_{1BZ}{{\frac{{d^{d}k}}{{(2\pi )^{d}}}}}%
\gamma (\vec{k})\left( {\int_{0}^{J}{dJ^{\prime }{\frac{{\sigma (J^{\prime })%
}}{{\bar{\omega}_{\vec{k}}(J^{\prime })}}}}}\right) -{\frac{1}{2}}\gamma 
(0)\int_{0}^{J}{dJ^{\prime }\sigma ^{2}(J^{\prime })}.  \label{Eq4.69} 
\end{equation} 
In particular, in the low-temperature limit, we find
\begin{equation} 
\bar{u}\simeq \bar{T}-\bar{h}-{\frac{1}{2}}\gamma (0)+O(\bar{T}^{2}g(\bar{h}%
)),  \label{Eq4.70} 
\end{equation} 
where the explicit expression of the term $O(\bar{T}^{2}g(\bar{h}))$ can be 
obtained from Eqs. (${}$\ref{Eq4.60}), (${}$\ref{Eq4.61}) and (${}$\ref 
{Eq4.68}) in a straightforward way. A similar low-$T$ expression can be 
easily obtained for $\bar{f}$. From the result (${}$\ref{Eq4.70}), the 
reduced specific heat at constant magnetic field $\bar{C}_{\bar{h}%
}=(\partial \bar{u}/\partial \bar{T})_{\bar{h}}=C_{h}/S$ is given by%
\begin{equation} 
\bar{C}_{\bar{h}}\simeq 1+O(\bar{T}^{2}g(\bar{h})),  \label{Eq4.71} 
\end{equation} 
as expected for a classical spin model \cite{Fisher64}. We now focus on the 
case $d=1$ with $p>1$ for which explicit analytical results can be obtained 
\cite{Cavallo02} allowing a comparison with recent MC simulations and 
transfer-matrix predictions. In this case, in the thermodynamic limit $%
N\rightarrow \infty $, we use the transformation $\sum\limits_{k}{(...)/N}%
\rightarrow \int_{0}^{2\pi }{dk}(...)/{2\pi }=\int_{0}^{\pi }{dk}(...)/{\pi } 
$ (due to the symmetry of the problem) and, for small $k$ \cite 
{Nakano94a,Nakano94b,Nakano95}
\begin{equation} 
\Omega ^{(p)}(k)=\sum\limits_{n=1}^{\infty }{{\frac{{1-\cos (kn)}}{{n^{p}}}}}%
\simeq \left\{ 
\begin{array}{lcl} 
\displaystyle{{\frac{1}{2}}\eta (p)k^{p-1}} & , & 1<p<3 \\
\\ 
\displaystyle{{\frac{1}{2}}k^{2}\ln \left( {{\frac{1}{k}}}\right)}  & , & p=3\\
\\ 
\displaystyle{{\frac{1}{2}}\zeta (p-2)k^{2}} & , & p>3, 
\end{array} 
\right.   \label{Eq4.72} 
\end{equation} 
where $\eta (p)=\pi \Gamma ^{-1}\left( p\right) {/\sin [\pi (p-2)]}$ and $%
\zeta (z)$ is the Riemann's zeta function. In particular, for $p=2$, we get 
the exact results \cite{Gradshteyn80}
\begin{equation} 
\Omega ^{(2)}(k)={\frac{\pi }{2}}k-{\frac{k^{2}}{4}},\quad \quad 0\leq k\leq 
2\pi \quad .  \label{Eq4.73} 
\end{equation} 
With the low-$k$ expansions (\ref{Eq4.72}), for the reduced magnetization 
per spin $\sigma $, we have the low-temperature representation (see Eq. (\ref 
{Eq4.65}))
\begin{eqnarray} 
\sigma (\bar{T},\bar{h}) &\approx &1-\bar{T}\int_{0}^{\pi }{{\frac{{dk}}{\pi 
}}{\frac{1}{{\bar{h}+\Omega ^{(p)}(k)}}}}  \notag  \label{Eq4.74} \\ 
&\approx &1-\bar{T}\left\{ 
\begin{array}{lcl} 
\displaystyle{\overline{{h}}^{^{-2}}{}_{2}F_{1}\left( {1,{\frac{1}{{p-1}}},{\frac{p}{{p-1}}%
};-{\frac{{\pi ^{p-1}}}{2}}{\frac{{\eta (p)}}{{\bar{h}}}}}\right)}  & , & 
1<p<3 \\ 
\\
\displaystyle{\int_{0}^{\pi }{{\frac{{dk}}{\pi }}\left[ {\bar{h}+{\frac{1}{2}}k^{2}\ln 
\left( {{\frac{1}{k}}}\right) }\right] ^{-1}}} & , & p=3 \\ 
\\
\displaystyle{{\frac{2}{{\pi \zeta (p-2)}}}\sqrt{{\frac{\zeta {(p-2)}}{{2\bar{h}}}}}%
\arctan \left( {\pi \sqrt{{\frac{\zeta {(p-2)}}{{2\bar{h}}}}}}\right)}  & , & 
p>3, 
\end{array} 
\right.   \notag \\ 
&& 
\end{eqnarray} 
where the hypergeometric function, when $\bar{h}\rightarrow 0$ behaves as%
\begin{eqnarray} 
&&\overline{{h}}^{^{-2}}{}_{2}F_{1}\left( {1,{\frac{1}{{p-1}}},{\frac{p}{{p-1%
}}};-{\frac{{\pi ^{p-1}\eta (p)}}{{2\bar{h}}}}}\right) \approx   \notag \\ 
&\approx &\left\{ 
\begin{array}{lcl} 
\displaystyle{{\frac{{\eta (p)\pi ^{p-2}}}{{2(2-p)}}}+O(\bar{h})} & , & 1<p<2\\
\\ 
\displaystyle{{\frac{2}{{\pi ^{2}}}}\ln \left[ {1+{\frac{{\pi ^{2}}}{{2h}}}}\right]}  & , & 
p=2\\
\\ 
\displaystyle{{\frac{2}{{\eta (p)(p-1)}}}\Gamma \left( {{\frac{1}{{p-1}}}}\right) \Gamma 
\left( {{\frac{{p-2}}{{p-1}}}}\right) \left( {{\frac{{2\bar{h}}}{{\eta (p)}}}%
}\right) ^{{\frac{{2-p}}{{p-1}}}}} & , & 2<p<3.  
\end{array} 
\right.   \label{Eq4.75} 
\end{eqnarray} 
According to Eqs. (\ref{Eq4.74})-(\ref{Eq4.75}), if we write, as before, $%
\sigma =1-\bar{T}a(\bar{h})$, the nearly saturation condition imposes that $%
\bar{T}a(\bar{h})\ll 1$. For instance, with $p>3$, this condition is 
expressed by $\bar{T}/\sqrt{\bar{h}}\ll 1$ and hence the natural expansion 
parameter is $\bar{T}/\sqrt{\bar{h}}$. Of course, more accurate estimates of 
the integral in Eq. (\ref{Eq4.74}) can be performed including higher-order 
terms in the low-$k$ expansions (\ref{Eq4.72}). A check of the reliability 
of the estimates (\ref{Eq4.74}) can be obtained by calculating the reduced 
magnetization per spin to the first order in $\bar{T}$ with $p=2$ when $%
\Omega ^{(2)}(k)$ has the exact expression (\ref{Eq4.73}). In this case, for 
$\bar{h}>0$, we have
\begin{equation} 
\sigma (\bar{T},\bar{h})\simeq 1-{\frac{{\bar{T}}}{{\pi \sqrt{{\frac{{\pi 
^{2}}}{4}}+\bar{h}}}}}\ln \left[ {{\frac{{\sqrt{{\frac{{\pi ^{2}}}{4}}+\bar{h%
}}+{\frac{\pi }{2}}}}{{\sqrt{{\frac{{\pi ^{2}}}{4}}+\bar{h}}-{\frac{\pi }{2}}%
}}}}\right] .  \label{Eq4.76} 
\end{equation} 
For small $\bar{h}$, this reduces to
\begin{equation} 
\sigma (\bar{T},\bar{h})\simeq 1-{\frac{{\ 2}}{\pi ^{2}}}\bar{T}\ln \left[ 1+%
{\frac{\pi ^{2}}{2\bar{h}}}\right] ,  \label{Eq4.77} 
\end{equation} 
which is just the result obtained from the hypergeometric function 
representation in Eq. (\ref{Eq4.74}) with $p=2$ and $\eta (2)=\pi $ assuming 
the low-$k$ expression $\Omega ^{(2)}(k)\approx {\frac{\pi }{2}}k$ (see Eqs.  
\ref{Eq4.72}) and (\ref{Eq4.75})). Eq. (\ref{Eq4.76}) (or (\ref{Eq4.77})) 
shows that the near saturation condition for $\bar{h}\rightarrow 0$ is 
satisfied only if $\bar{T}\ln (1/\bar{h})\ll 1$. It is worth noting that, 
consistently with the previous results for general $d$ (see Fig. 1), from 
Eq. (\ref{Eq4.74}), provided that, for small values of $\bar{h}$ (for finite 
$\bar{h}$ no problem occurs), the $\bar{T}$-expansion preserves its physical 
meaning, the following low-$\bar{T}$ features arise: 

(i) for $1<p<2$, LRO occurs with a spontaneous magnetization 

%
\begin{equation} 
\sigma _{0}(\bar{T})=1-{\frac{{2\pi ^{1-p}}}{{(2-p)\eta (p)}}}\bar{T}+O(\bar{%
T}^{2});  \label{Eq4.78} 
\end{equation} 

(ii) for $p\geq 2$ and $\bar{h}\rightarrow 0$, the coefficients in the $\bar{%
T}$ expansions in Eqs. (\ref{Eq4.74})-(\ref{Eq4.77}) diverge and no LRO 
takes place at finite temperature.  

From the previous equations one can immediately obtain also the reduced 
susceptibility. In particular, for the simplest case $p=2$ we get, exactly%
\begin{equation} 
\bar{\chi}={\frac{{\bar{T}}}{{2\pi \left( {{\frac{{\pi ^{2}}}{4}}+\bar{h}}%
\right) }}}\left\{ {{\frac{\pi }{{\bar{h}}}}+{\frac{1}{\sqrt{{\frac{{\pi ^{2}%
}}{4}}+\bar{h}}}}\ln \left[ {{\frac{{\sqrt{{\frac{{\pi ^{2}}}{4}}+\bar{h}}+{%
\frac{\pi }{2}}}}{{\sqrt{{\frac{{\pi ^{2}}}{4}}+\bar{h}}-{\frac{\pi }{2}}}}}}%
\right] }\right\} +O(\bar{T}^{2}).  \label{Eq4.79} 
\end{equation} 
Besides, the internal and free energies per spin, and hence other 
thermodynamic quantities, can be obtained from Eqs. (\ref{Eq4.68}) and (\ref 
{Eq4.69}). In particular, for reduced internal energy per spin, we find 
(with $\gamma (0)=\zeta (p)$)
\begin{equation} 
\bar{u}=\bar{T}-\bar{h}-{\frac{1}{2}}\zeta (p)+O(\bar{T}^{2}g(\bar{h})) 
\label{Eq4.80} 
\end{equation} 
and hence $\bar{C}_{\bar{h}}=1+O(\bar{T}^{2}g(\bar{h}))$ as for a generic $d$%
.  

Consistently with the previous results and exact theorems for the $d$%
-dimensional spin model, our analytical near saturation calculations for the 
classical long-range Heisenberg FM chain with $1<p<2$ predict, in a 
transparent way, a transition to a FM phase at finite temperature (see Fig.  
1). For $p\geq 2$, a thermodynamic scenario at finite temperature, similar 
to that for SRI's \cite{Campana84}, takes place with absence of a phase 
transition.  

Additional information about the excitation dispersion relation and 
thermodynamic properties for the FM chain, within nearly saturated regimes 
beyond the linear expansion in $\bar{T}a(\bar{h})\ll 1$, can be obtained 
solving numerically the full set of self-consistent Eqs. (\ref{Eq4.56})-(\ref 
{Eq4.58}) with respect to $\bar{\omega}_{k}$ and $\sigma $ for $d=1$ \cite 
{Cavallo02}. We present here the most relevant results including comparisons 
with predictions obtained by means of different (and in a some sense, exact) 
methods. As we will see, all the numerical results confirm the previous 
scenario based on the estimates involving low-temperature expansions (for 
appropriate values of $\bar{h}$) and the dominant contributions to the 
dispersion relation as $k\rightarrow 0$.  
In Fig. 2, the reduced dispersion relation $\bar{\omega}_{k}$ is plotted as 
a function of $k$ in the interval $0\leq k\leq \pi $ for $\bar{T}=0$ (Fig.  
2(a)) and $\bar{T}=0.5$ (Fig. 2(b)) at $\bar{h}=0.5$ with selected values of 
the interaction exponent $p$. The reduced magnetization $\sigma $ and 
susceptibility $\bar{\chi}$ in terms of $\bar{T}$ at a fixed value of $\bar{h%
}$ with some values of $p$ are shown in Fig. 3(a,b) where, as a comparison, 
transfer-matrix results for the Heisenberg FM chain with SRI's ($p=\infty $) 
\cite{Campana84,Balucani82} are also presented.  
The plots of the reduced magnetization per spin $\sigma $ as a function of $%
\bar{h}$ for some values of $\bar{T}$ with $p=1.5$ in Fig. 4(a) and for 
different values of $p$ with $\bar{T}=0.1$ in Fig. 4(b) are particularly 
meaningful.  

\begin{figure}[h] 
\centerline{ 
\includegraphics*[width=7.5cm]{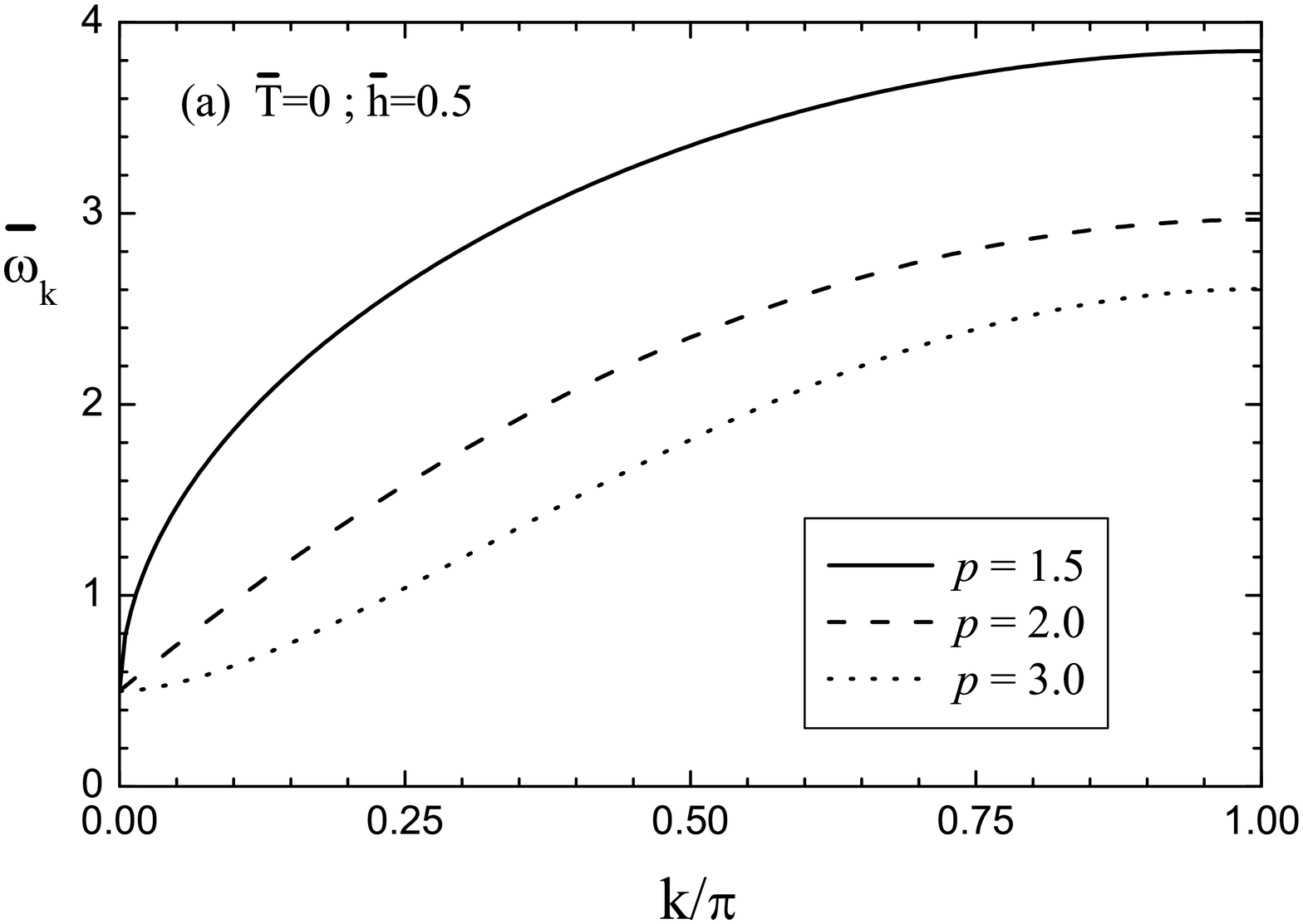} 
\includegraphics*[width=7.5cm]{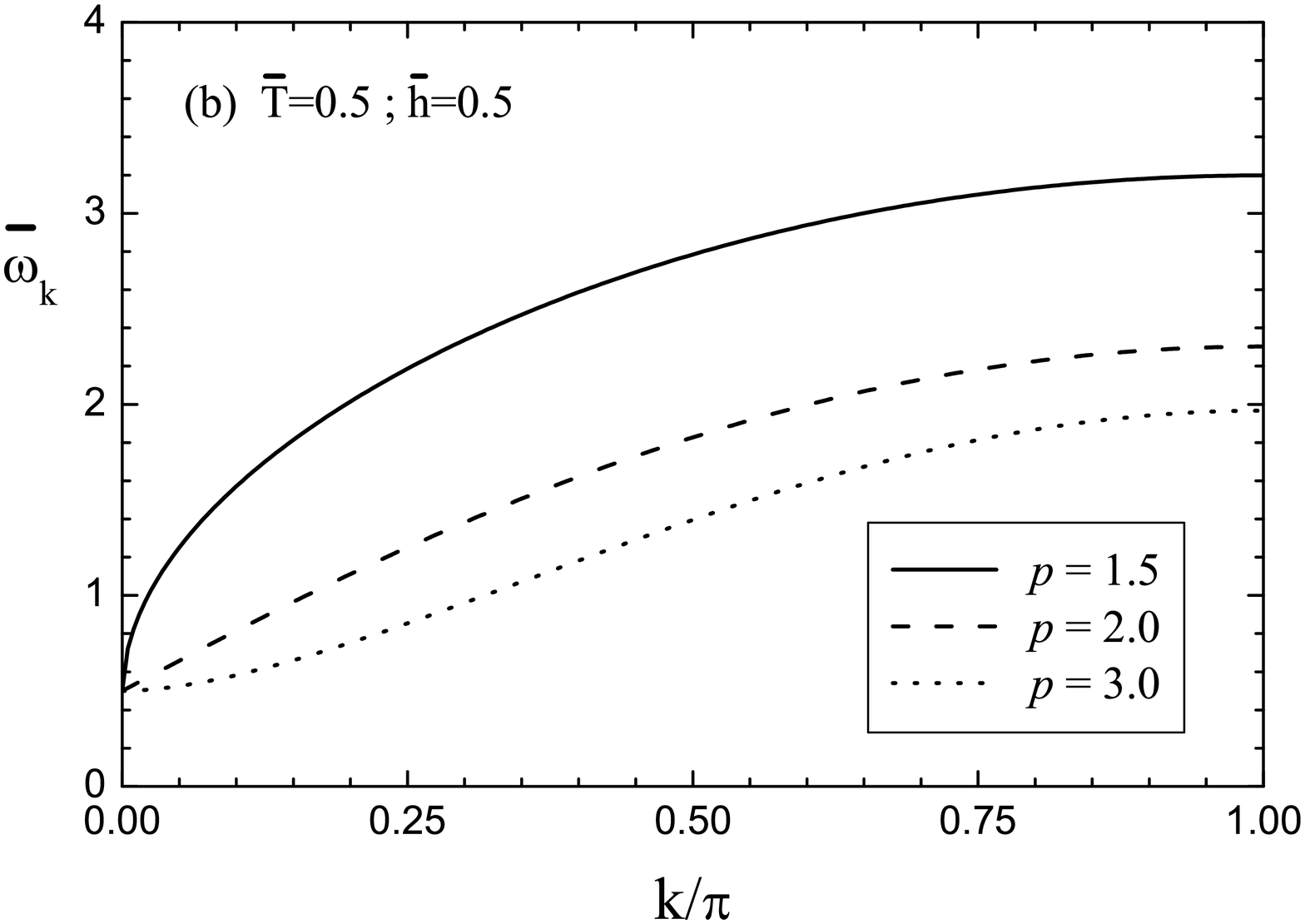}} 
\caption{Dispersion relation $\overline{\protect\omega }_{k}$ of an 
Heisenberg FM chain as a function of $k$ in the interval $\left[ 0,\protect%
\pi \right] $ for reduced magnetic field $\overline{h}=0.5$ and temperature $%
\overline{T}=0$ (a) and $\overline{T}=0.5$ (b) with different values of the 
decaying parameter $p$.} 
\end{figure} 

\begin{figure}[h] 
\centerline{ 
\includegraphics*[width=7cm]{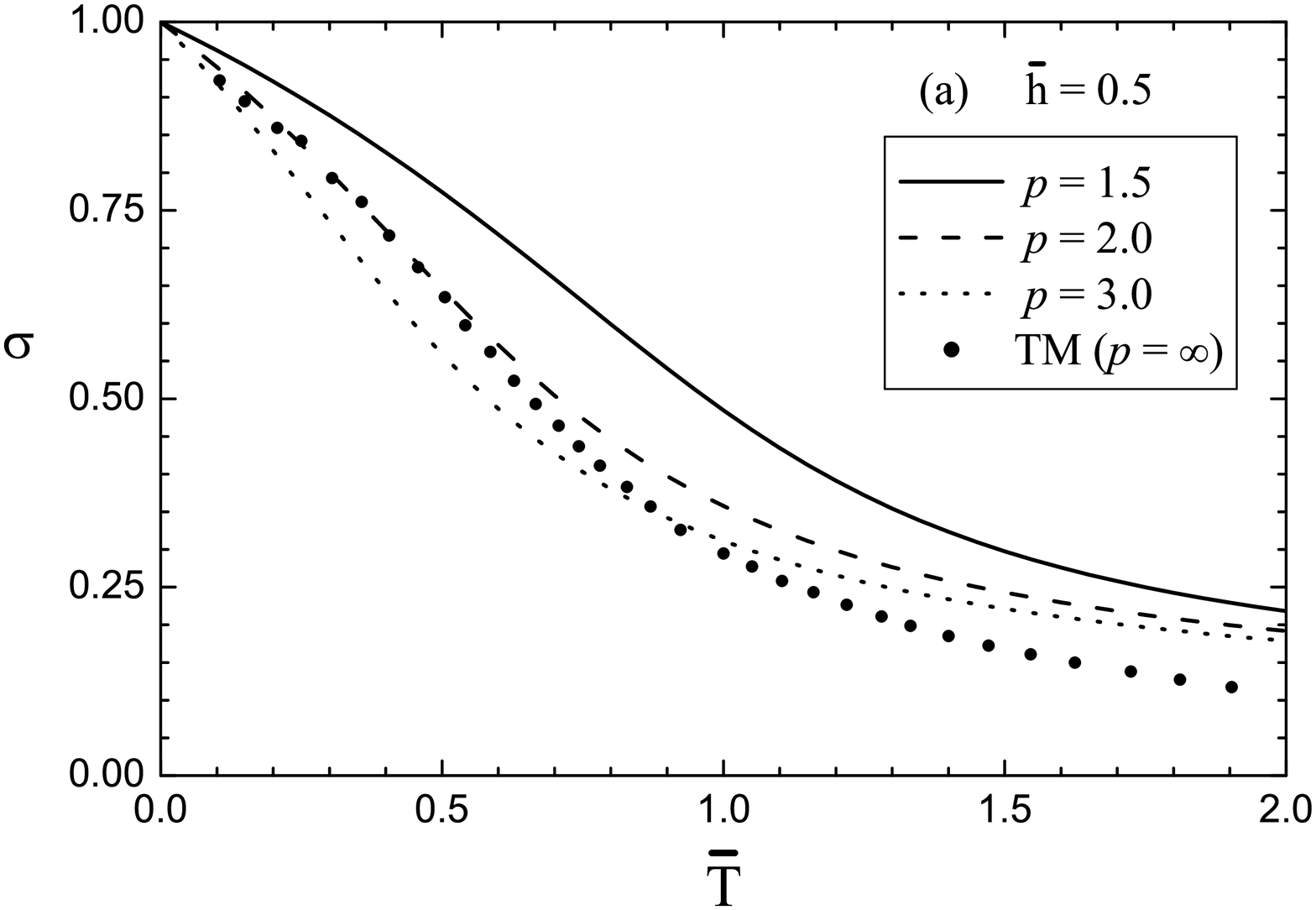} 
\includegraphics*[width=7cm]{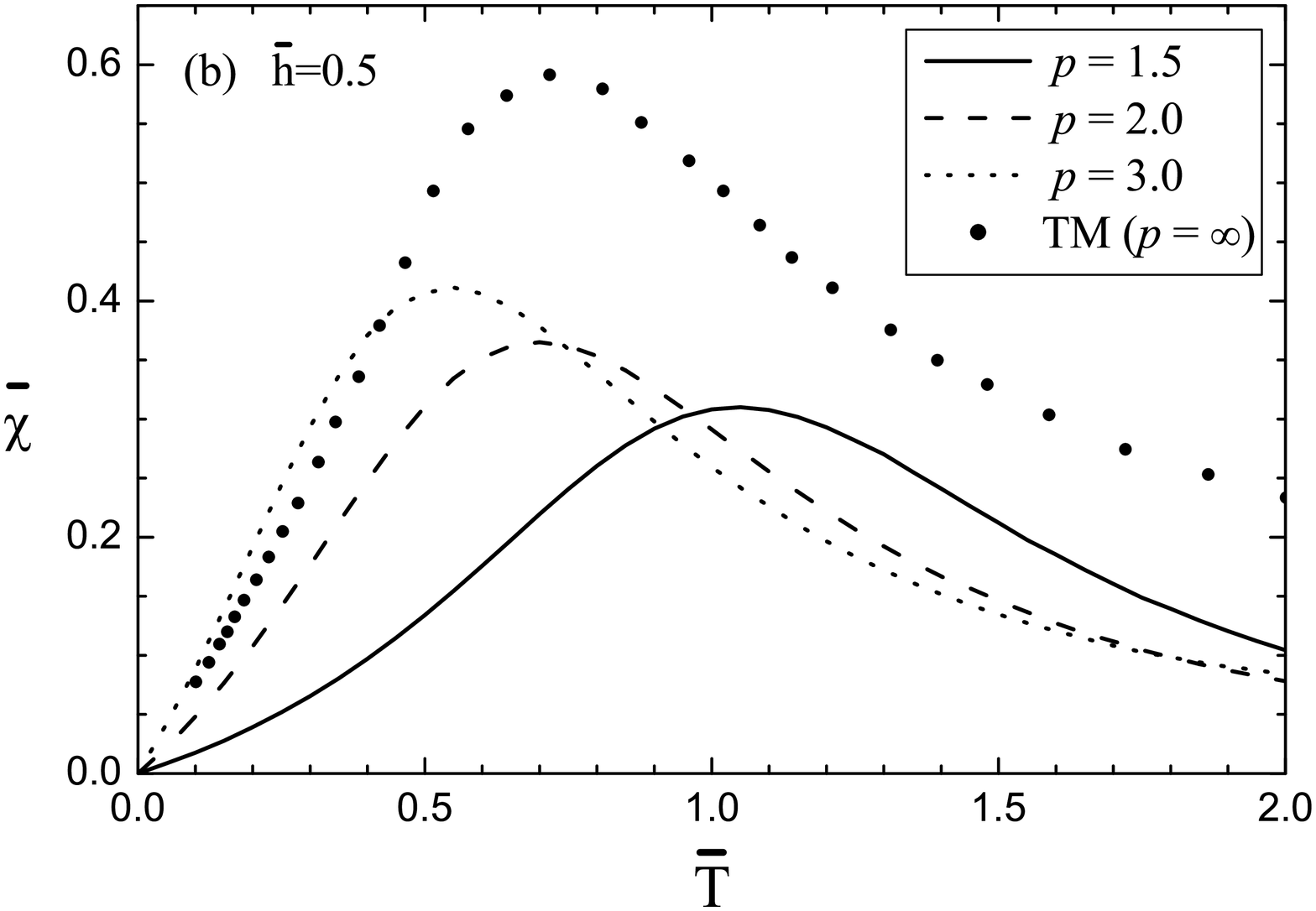}} 
\caption{Temperature dependence of the reduced magnetization $\protect\sigma 
$\ (a) and susceptibility $\overline{\protect\chi}$ (b) of an Heisenberg FM 
chain for $\overline{h}=0.5$ and different values of $p$. The dots represent 
the transfer-matrix (TM) results for SRI's ($p=\infty $).} 
\end{figure} 

\begin{figure}[h] 
\centerline{ 
\includegraphics*[width=7cm]{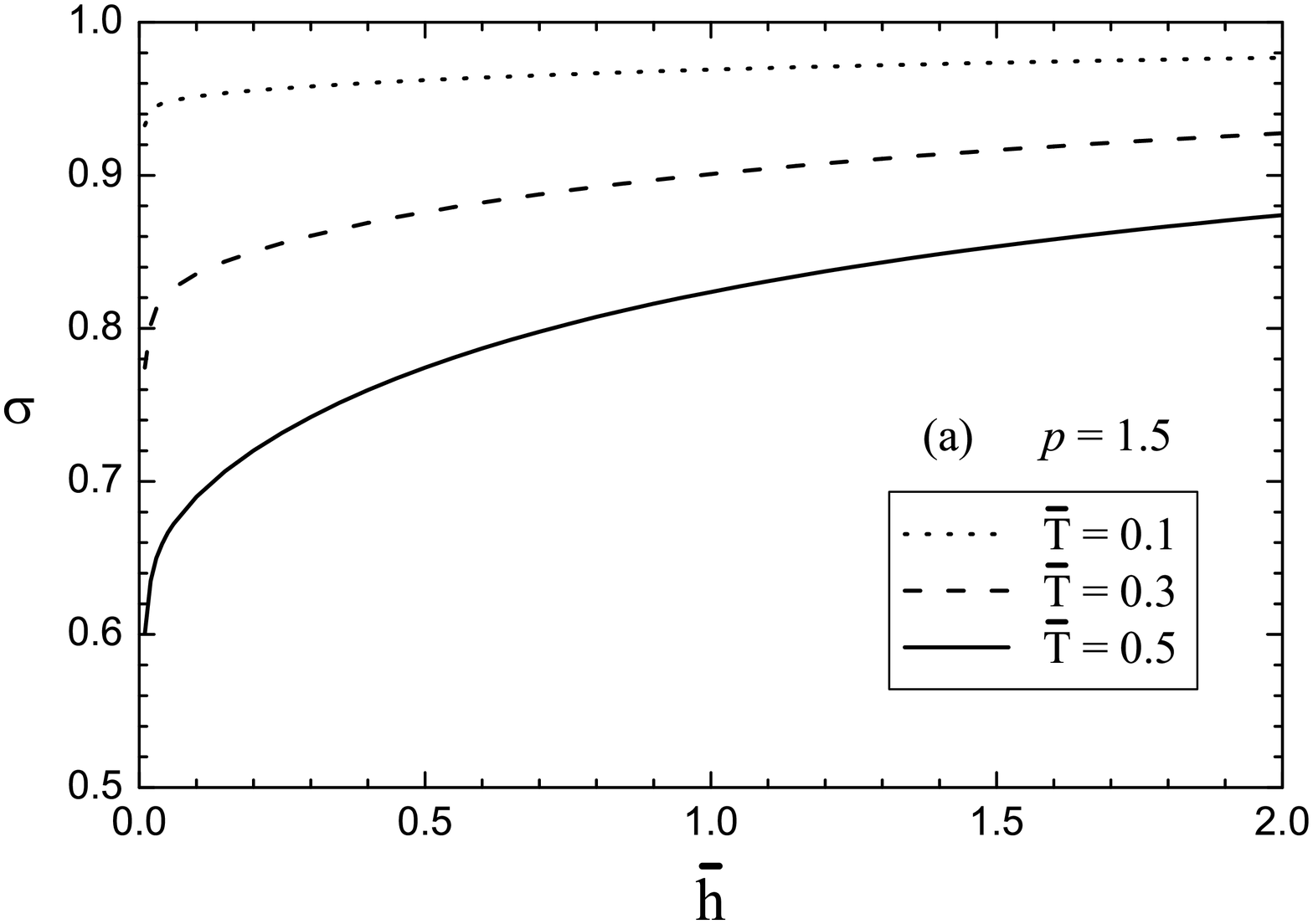} 
\includegraphics*[width=7cm]{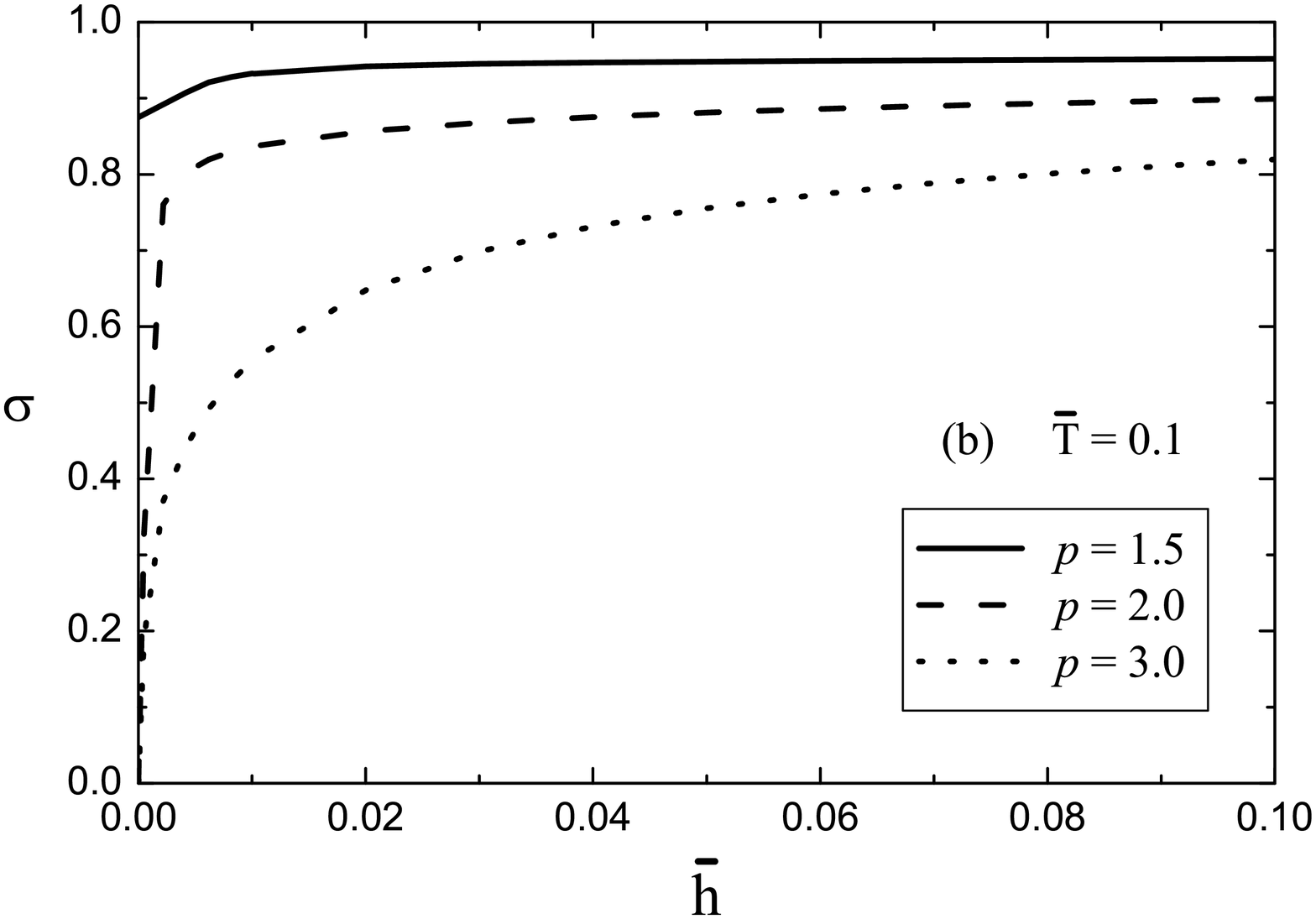}} 
\caption{Reduced magnetization $\protect\sigma $ of an Heisenberg FM chain 
as a function of $\overline{h}$ for (a) $p=1.5$ and different values of $%
\overline{T}$ and (b) $\overline{T}=0.1$\ and different values of $p$.} 
\end{figure} 

\begin{figure}[h] 
\centerline{ 
\includegraphics*[width=7.5cm]{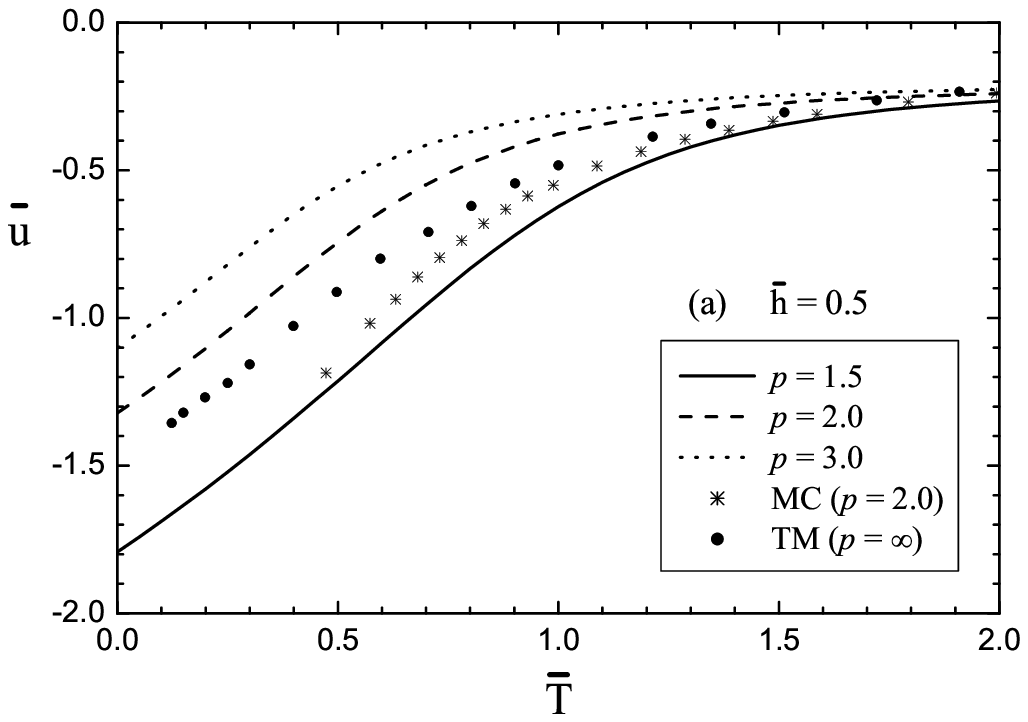} 
\includegraphics*[width=7.5cm]{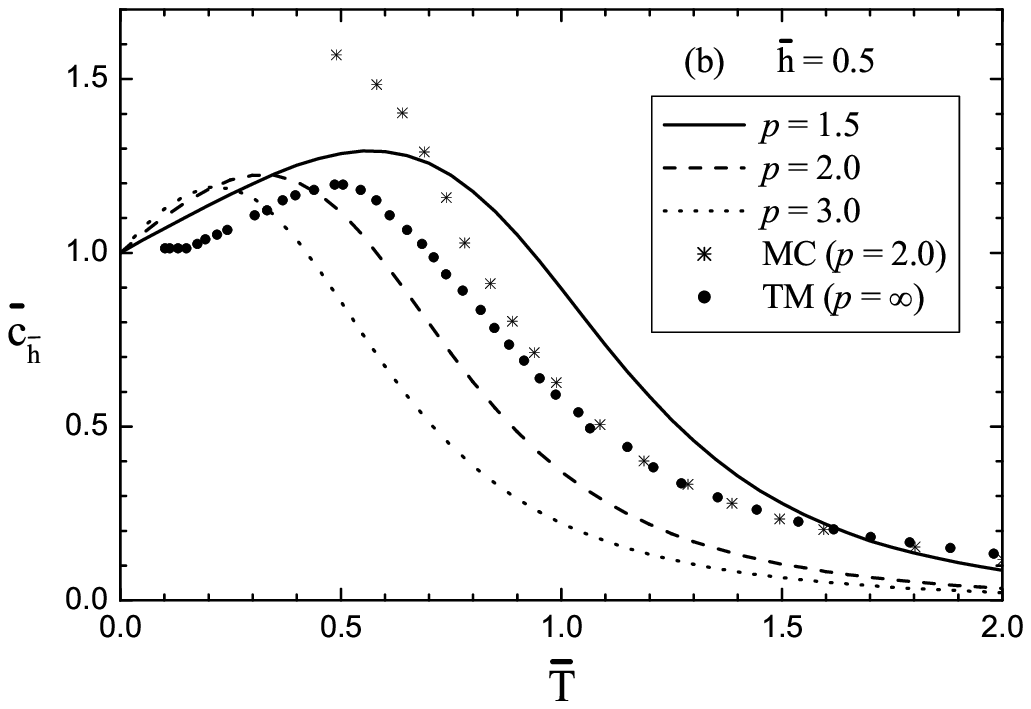}} 
\caption{Temperature dependence of the reduced\ (a) internal energy $%
\overline{u}$ and (b) specific heat $\overline{C}_{\overline{h}}$ of an 
Heisenberg FM chain for $\overline{h}=0.5$ and different values of $p$. The 
stars represent Monte Carlo (MC) results for $p=2$ and the dots TM 
predictions for SRI's ($p=\infty $).} 
\end{figure}

They show clearly our previous analytical finding that, for $1<p<2$, a FM 
phase occurs at finite temperature. Of course, at this stage, our ME's do 
not allow to study regimes with $\sigma \rightarrow 0 $ as $\bar{h}%
\rightarrow 0$ and hence to find the transition point to the FM phase where $%
1<p<2$. Information about this transition and the related critical 
properties will be given in the next subsection. The reduced internal energy 
$\bar{u}$ and the specific heat $\bar{C}_{\bar{h}}$, evaluated from the 
solutions of the $(d=1)$-ME's (\ref{Eq4.56})-(\ref{Eq4.58}), are also 
plotted in Figs. 5(a) and 5(b) as functions of $\bar{T}$ for different 
values of $p$.  
In both these figures, as a further comparison, we have also reported the 
transfer-matrix results for the short-range chain \cite{Campana84,Balucani82} 
and MC simulations recently obtained \cite{Romano89} for a classical 
long-range FM Heisenberg chain with $p=2$. Finally, in Fig.6, we present the 
behavior of the transverse correlation length $\xi _{\perp }$ as a function 
of $\bar{T}$ for $\bar{h}=0.5$ and different values of $p$. Notice that a 
substantial difference clearly occurs for $1<p<2$ and $p\geq 2$, 
respectively. Again, in the same figure, we have plotted also 
transfer-matrix results for a FM chain with SRI's \cite{Campana84,Balucani82}.%

\begin{figure}[h] 
\centerline{ 
\includegraphics*[width=8cm]{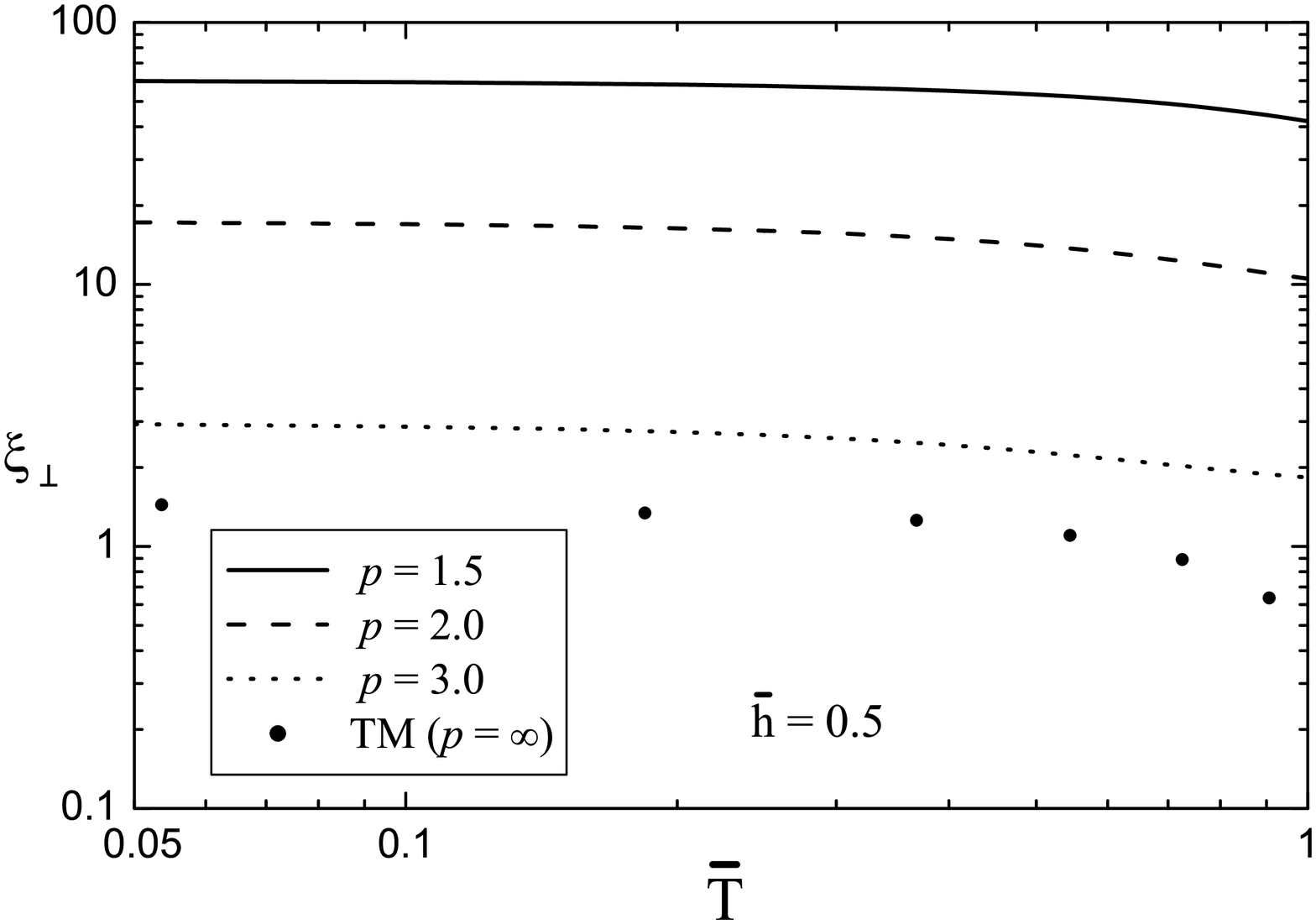}} 
\caption{Temperature dependence of the transverse correlation length $%
\protect\xi _{\perp }$ of an Heisenberg FM chain for $\overline{h}=0.5$ and 
different values of $p$ (with logarithmic scales). The dots represent TM 
results for SRI's ($p=\infty $).} 
\end{figure}


\subsubsection{Thermodynamic Regimes with Near-zero Magnetization} 

As discussed in Subsec. 4.3.1, Eq. (\ref{Eq4.39}) for the dispersion 
relation, or the first expression for $R(\vec{k})$ in Eq (\ref{Eq4.58}), is 
inadequate to obtain suitable results for thermodynamics regimes with 
near-zero magnetization ($\sigma \rightarrow 0$). However, this case can be 
successfully explored using the self-consistent ME's (\ref{Eq4.56})-(\ref 
{Eq4.57}) with the simple modification for $R(\vec{k})$ given in Eq. (\ref 
{Eq4.58}) when $\sigma \rightarrow 0$. Below, we show indeed that the 
modified ME's can be properly used for an estimate of the main critical 
properties of our classical spin model, when it starts to exhibit LRO, and 
of the low-temperature paramagnetic susceptibility in the remaining regions 
of the ($p-d$)-plane when no LRO exists.  


\paragraph{\textit{A. Transition temperature and critical properties}} 

In the limit $\overline{h}\rightarrow 0$ with $\sigma \geq 0$, the system of 
Eqs. (\ref{Eq4.56})-(\ref{Eq4.57}), with $R(\vec{k})$ appropriate to include 
the case $\sigma \rightarrow 0$, assumes the form \cite{Cavallo04} 
\begin{eqnarray} 
\bar{\omega}_{\vec{k}} &=&\sigma \Omega ^{(p)}({\vec{k}})R({\vec{k}})\ , 
\label{Eq4.81} \\ 
R({\vec{k}}) &=&=1+{\frac{\overline{T}}{\Omega ^{(p)}({\vec{k}})}}\int_{1BZ}{%
\frac{d^{d}k^{\prime }}{(2\pi )^{d}}}{\frac{\Omega ^{(p)}({\vec{k}}-{\vec{k}}%
^{\prime })-\Omega ^{(p)}({\vec{k}}^{\prime })}{\Omega ^{(p)}({\vec{k}}%
^{\prime })R({\vec{k}}^{\prime })}}\ ,  \label{Eq4.82} \\ 
\sigma ^{2} &=&{\frac{1-3{\overline{T}}Q({\overline{T}})}{{1-{\overline{T}}Q(%
{\overline{T}})}}}\ ,  \label{Eq4.83} 
\end{eqnarray} 
where 
\begin{equation} 
Q({\overline{T}})=\int_{1BZ}{\frac{d^{d}k}{(2\pi )^{d}}}{\frac{1}{\Omega 
^{(p)}({\vec{k}})R({\vec{k}})}}\ .  \label{Eq4.84} 
\end{equation} 
From Eq. (\ref{Eq4.83}) we can immediately obtain an equation which 
determines the critical temperature $T_{c}$ (when it exists). Indeed, the 
reduced critical temperature ${\overline{T}}_{c}$ (with $T_{c}=JS^{2}{%
\overline{T}}_{c}$) can be obtained setting $\sigma ({\overline{T}}_{c})=0$.  
This yields the equation
\begin{equation} 
1-3{\overline{T}}_{c}Q({\overline{T}}_{c})=0.  \label{Eq4.85} 
\end{equation} 
It is not simple to solve exactly this equation and one must resort again to 
numerical calculations. However, it is instructive to get an explicit 
estimate of ${\overline{T}}_{c}$ by calculating the integrals in Eqs. (\ref 
{Eq4.81})-(\ref{Eq4.84}) taking for $\Omega ^{(p)}({\vec{k}})$ the dominant 
contribution as ${\vec{k}}\rightarrow 0$. Then, solving by iteration our 
self-consistent ME's, to first level of iteration we have
\begin{equation} 
Q({\overline{T}})\simeq \frac{\mathcal{F}_{1}^{(p)}(0)}{1+\overline{T}%
\mathcal{F}_{1}^{(p)}(0)},  \label{Eq4.86} 
\end{equation} 
and
\begin{equation} 
\bar{\omega}_{\vec{k}}\simeq \Omega ^{(p)}({\vec{k}})\sigma ({\overline{T}})%
\left[ 1+\overline{T}\mathcal{F}_{1}^{(p)}(0)\right] , 
\end{equation} 
where $\mathcal{F}_{1}^{(p)}(0)$ and $\sigma ({\overline{T}})$ are given by 
Eqs. (\ref{Eq4.62}) (for $\overline{h}=0$) and (\ref{Eq4.83}), respectively.  
With $Q({\overline{T}})$ explicitly known, Eq. (\ref{Eq4.85}) yields for ${%
\overline{T}}_{c}={\overline{T}}_{c}(p,d)$
\begin{equation} 
{\overline{T}}_{c}={\frac{1}{2\mathcal{F}_{1}^{(p)}(0)}}\ .  \label{Eq4.88} 
\end{equation} 
From this simple expression, one immediately sees that a critical 
temperature exists only when $\mathcal{F}_{1}^{(p)}(0)$ is finite and hence 
in the regions of the $(p-d)$- plane where LRO occurs. Here we explicitly consider 
the one-dimensional case for $1<p<2$ and the two-dimensional one 
for $2<p<4$ for which a comparison with recent alternative analytical and MC 
results can be performed.  

For $d=1$ and $1<p<2$ one finds \cite{Cavallo02}
\begin{equation} 
{\overline{T}}_{c}(d=1)={\frac{(2-p)\eta (p)}{4\pi ^{1-p}}}\ .  
\label{Eq4.89} 
\end{equation} 
In particular, for $p\rightarrow 2^{-}$, Eq. (\ref{Eq4.89}) yields $%
T_{c}=JS^{2}{\overline{T}}_{c}\approx \pi ^{2}JS^{2}(2-p)/4$ to be compared 
with $T_{c}\approx \pi ^{2}JS(S+1)(2-p)/3$, obtained in Ref. \cite{Nakano95} 
for the quantum case by means of the EMM, and with $T_{c}\propto JS^{2}(2-p)$ 
derived by Kosterlitz \cite{Kosterlitz76} for the classical chain using a 
perturbative renormalization group approach.  

For the case $d=2$ with $2<p<4$, the coefficients of interest in the low-$k$ 
expansions (\ref{Eq4.64}) for $\Omega ^{(p)}({\vec{k}})$ are given by \cite 
{Nakano95}
\begin{equation} 
\left\{ 
\begin{array}{l} 
\displaystyle{A_{2}=2^{2-p}\pi ^{2}\Gamma ^{-2}(p)/\sin [\pi (p-2)/2]} \\
\\ 
\displaystyle{B_{2}=2^{p-2}\zeta \left( {\frac{p-2}{2}}\right) \left[ \zeta \left( {\frac{%
2-p}{2}},{\frac{1}{4}}\right) -\zeta \left( {\frac{2-p}{2}},{\frac{3}{4}}%
\right) \right] }
\end{array} 
\right. ,~2<p<4,  \label{Eq4.90} 
\end{equation} 
where $\zeta (z,a)=\sum_{n=0}^{\infty }(n+a)^{-z}$ is the generalized 
Riemann zeta function and $\zeta (z)\equiv \zeta (z,0)$ is the ordinary 
Riemann zeta function. Then, in Eq. (\ref{Eq4.88}), we have
\begin{equation} 
\left. \mathcal{F}_{1}^{(p)}(0)\right| _{d=2}\simeq \frac{\Lambda ^{2-p}}{{%
4-p}}\frac{\Gamma ^{2}(p)}{2^{1-p}\pi ^{2}}\sin \left[ \pi (p-2)/2\right] 
\mathcal{\,,}  \label{Eq4.91} 
\end{equation} 
and hence
\begin{equation} 
{\overline{T}}_{c}(d=2)={\frac{(4-p)\pi ^{2}\Lambda ^{p-2}}{2^{p}\Gamma 
^{2}(p)\sin \left[ \pi (p-2)/2\right] }}\ .  \label{Eq4.92} 
\end{equation} 
In Fig. 7 the critical temperature of $(S={\frac{1}{2}})$-models, for $d=1$ 
(Fig. 7(a)) and $d=2$ (Fig. 7(b)), is plotted as a function of $p$ and 
compared with the corresponding results recently obtained, for the 
long-range quantum Heisenberg model by means of the EMM, within the 
Tyablikov decoupling, for both cases $d=1$\ and $d=2$ \cite{Nakano95}, the 
modified spin-wave theory \cite{Nakano94a} and MC simulations \cite 
{Vassiliev01}, for $d=2$. Our results appear to be consistent with those 
obtained for the quantum counterpart if we keep in mind that the apparent 
discrepancy is related to the known feature that a quantum spin-$S$ model 
can be reasonably approximated by a classical one only in the large-$S$ 
limit \cite{Stanley71}.

\begin{figure}[h] 
\centerline{ 
\includegraphics*[width=7.5cm]{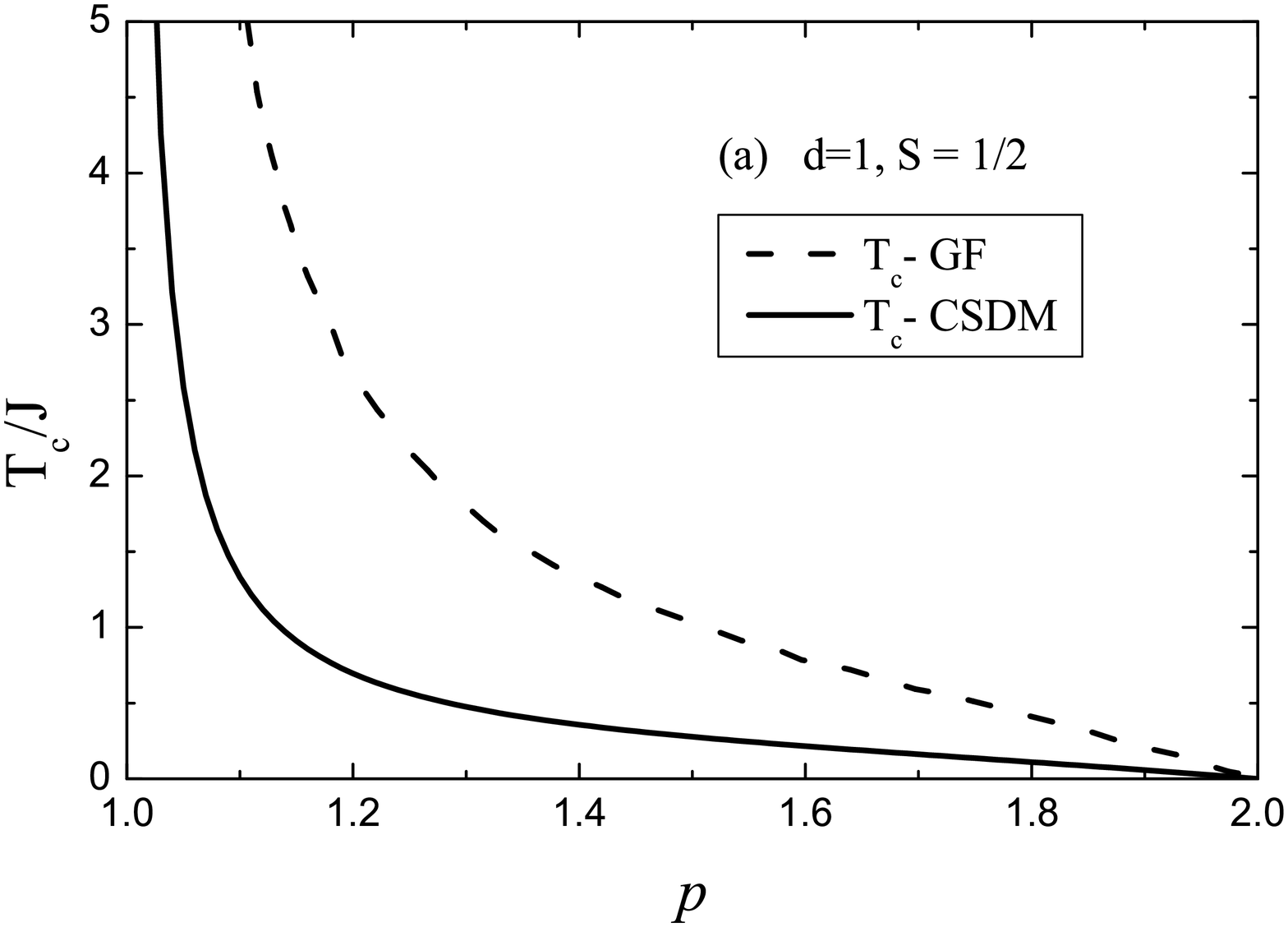} 
\includegraphics*[width=7.5cm]{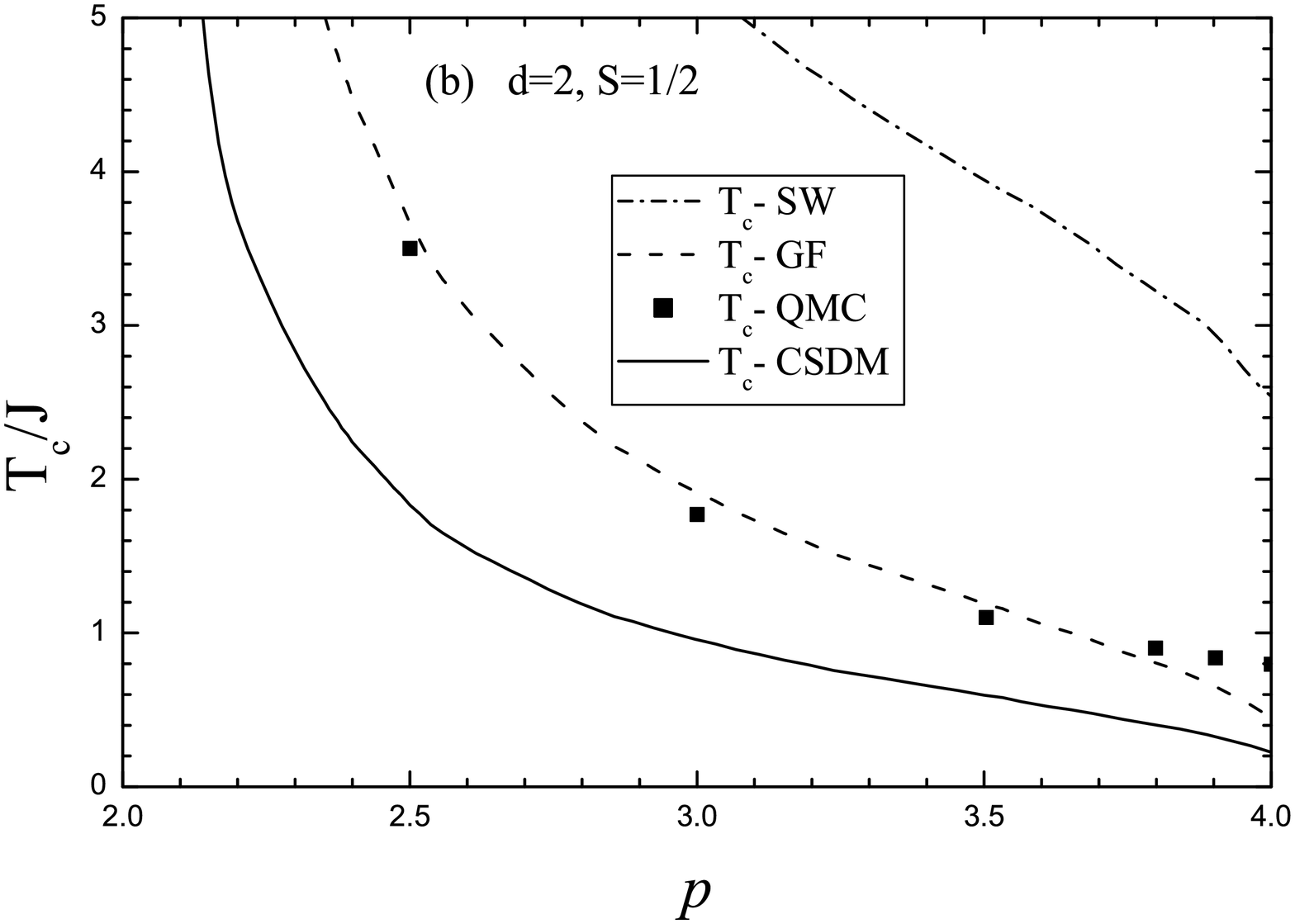}} 
\caption{Critical temperature as a function of the decaying parameter $p$ 
for a one-dimensional (a) and two-dimensional (b) spin-$1/2$ Heisenberg 
ferromagnet. The dashed lines refer to the Green function (GF) method and to 
the modified spin-wave (SW) theory results, respectively. The dots represent 
MC predictions.} 
\end{figure} 

From Eq. (\ref{Eq4.83}), we can easily obtain the behavior of the 
magnetization per spin $m(T)=\sigma ({\overline{T}})S$, with ${\overline{T}}%
=T/JS^{2}$. For all values of $p$ and $d$ for which a phase transition 
occurs, we get
\begin{equation} 
m(T)\sim \left( {\frac{T-T_{c}}{T_{c}}}\right) ^{1/2}\ ,  \label{Eq4.93} 
\end{equation} 
implying $\beta =1/2$ for the magnetization critical exponent.  

The calculations are more delicate to obtain the near criticality behavior
of the paramagnetic susceptibility defined as $\chi =J{\overline{\chi }}%
=J(\sigma /\overline{h})$ in the limit $\overline{h}\rightarrow 0$ with $%
\sigma \rightarrow 0$. In this case, the set of Eqs. (\ref{Eq4.56})-(\ref 
{Eq4.57}), with the appropriate expression for $R({\vec{k}})$ in Eq. (\ref 
{Eq4.58}), reduces to
\begin{eqnarray} 
1 &=&3{\overline{\chi }}{\overline{T}}\int_{1BZ}{\frac{d^{d}k}{(2\pi )^{d}}}{%
\frac{1}{1+{\overline{\chi }}P({\vec{k}})}}\ ,  \label{Eq4.94} \\ 
P({\vec{k}}) &=&\Omega ^{(p)}({\vec{k}})+{\overline{\chi }}{\overline{T}}%
\int_{1BZ}{\frac{d^{d}k^{\prime }}{(2\pi )^{d}}}{\frac{\Omega ^{(p)}({\vec{k}%
}-{\vec{k}}^{\prime })-\Omega ^{(p)}({\vec{k}}^{\prime })}{1+{\overline{\chi 
}}P({\vec{k}}^{\prime })}}\ ,  \label{Eq4.95} 
\end{eqnarray} 
with $P(\vec{k})=\Omega ^{(p)}({\vec{k}})R({\vec{k}})$. Of course, these 
equations can only be solved numerically. However, an estimate of $\chi (T)$ 
as $T\rightarrow T_{c}^{+}$, can be obtained assuming, as usual, the low-$k$ 
behaviors (\ref{Eq4.64}) for $\Omega ^{(p)}(\vec{k})$. Then, a solution of 
Eqs. (\ref{Eq4.94}) and (\ref{Eq4.95}), with ${\overline{\chi }}({\overline{T%
}})\rightarrow \infty $ as ${\overline{T}}\rightarrow {\overline{T}}_{c}^{+}$%
, is found to be
\begin{equation} 
\chi =J{\overline{\chi }}\sim \left( {\frac{T-T_{c}}{T_{c}}}\right) 
^{-\gamma (p,d)}\ ,  \label{Eq4.96} 
\end{equation} 
where the $(p,d)$-dependent susceptibility critical exponent $\gamma (p,d)$ 
is given in Table I for the different regions of the $(p-d)$-plane where LRO 
exists, together with other critical exponents to be obtained below.

\begin{landscape} 
\thispagestyle{empty} 
\begin{table}[h] 
\centering 
\begin{tabular}{c} 
\textbf{Table I} 
\\ 
\small 
\begin{tabular}{|c|c|c|c||ccc|} 
\hline 
  \begin{tabular}{c} \\ Critical \\ exponent \end{tabular}   & 
  $\frac{3}{2}d<p< \left\{ 
     \begin{array}{lll} 
         2d  & , & d\leq 2 \\ 
         d+2 & , & d>2 
     \end{array} \right. $                                & 
  $p=\frac{3}{2},\quad d\leq 4$                           & 
  $d<p< \left\{  
	 \begin{array}{lll} 
         \frac{3}{2}d & , & d<4 \\ 
         d+2          & , & d>4 
     \end{array}  \right. $                               & 
  \begin{tabular}{c}             \\ $2<d<4$  \end{tabular}& 
  \begin{tabular}{c} $p\geq d+2$ \\ $d=4$    \end{tabular}& 
  \begin{tabular}{c}             \\ $d>4$    \end{tabular} 
\\ 
\hline 
$\beta (p,d)$  & $1/2$                & $1/2$      & $1/2$  & $1/2$             & $1/2$      & $1/2$  \\ 
$\gamma (p,d)$ & $\frac{p-d}{2d-p}$   & $1_{\ln }$ & $1$    & $\frac{2}{d-2}$   & $1_{\ln }$ & $1$    \\ 
$\delta (p,d)$ & $\frac{p}{2d-p}$     & $3_{\ln }$ & $3$    & $\frac{d+2}{d-2}$ & $3_{\ln }$ & $3$    \\ 
$\alpha (p,d)$ & $\frac{3d-2p}{2d-p}$ & $0_{\ln }$ & $0$    & $\frac{d-4}{d-2}$ & $0_{\ln }$ & $0$    \\ 
\hline 
\end{tabular} 
\end{tabular} 
\caption{Main ($p,d$)-dependent critical exponents in different domains of 
the ($p-d$)-plane where LRO exists. The double vertical lines separate the 
regions where LRI's are active and SRI regime occurs. The symbol $x_{ln}$ 
denotes the main ($T-T_{c}$)- or $h$- dependence with a logarithmic 
correction (for instance, $\protect\chi (T)\sim (T-T_{c})^{-1}\ln \left[ 
1/(T-T_{c})\right] $, $m(h)\sim h^{1/3}\left| \ln h\right| ^{1/3}$, etc.)} 
\end{table} 

\end{landscape}

We now determine the behavior of the magnetization per spin $m$ 
along the critical isotherm as $h\rightarrow 0$. It is easy to see that, 
from the basic ME's (\ref{Eq4.56})-(\ref{Eq4.58}), the equation of the 
critical isotherm for small $\overline{h}$ and $\sigma $ assumes the form%
\begin{equation} 
1-{\frac{2}{3}}\sigma ^{2}-3{\overline{T}}_{c}Q_{c}\left( {\sigma /\overline{%
h}}\right) =0\ ,  \label{Eq4.97} 
\end{equation} 
where
\begin{equation} 
Q_{c}\left( {\sigma /\overline{h}}\right) =\left( {\sigma /\overline{h}}%
\right) \int_{1BZ}{\frac{d^{d}k}{(2\pi )^{d}}}{\frac{1}{1+{\frac{\sigma }{%
\overline{h}}}P_{c}({\vec{k}})}}\ ,  \label{Eq4.98} 
\end{equation} 
and $P_{c}({\vec{k}})$ is obtained from Eq. (\ref{Eq4.95}) with $\overline{%
\chi }$ replaced by $\sigma /\overline{h}$ and ${\overline{T}}={\overline{T}}%
_{c}$. Eqs. (\ref{Eq4.97}) and (\ref{Eq4.98}) can be solved numerically but 
a suitable estimate of $\sigma $ as ${\overline{h}}\rightarrow 0$ can be 
easily obtained working again in terms of the dominant contribution of $%
\Omega ^{(p)}({\vec{k}})$ as ${\vec{k}}\rightarrow 0$. Then, we get%
\begin{equation} 
m=\sigma S\sim h^{1/\delta (p,d)},  \label{Eq4.99} 
\end{equation} 
where the expressions of the critical exponent $\delta (p,d)$ in the 
different domains of the $(p-d)$-plane are given in Table I.  

Finally, we wish to calculate the critical exponent $\alpha (p,d)$ of the 
specific heat in zero magnetic field as $T\rightarrow T_{c}^{+}$ for values 
of $p$ and $d$ such that a transition to the FM phase takes place. It is 
worth noting that, for this purpose, one cannot use the expression (\ref 
{Eq4.68}) for the reduced internal energy per spin. Indeed, it has a 
physical meaning only in the thermodynamic regimes where the decoupling $%
\langle S_{{\vec{k}}}^{z}S_{-{\vec{k}}}^{z}\rangle \simeq \langle S_{{\vec{k}%
}}^{z}\rangle \langle S_{-{\vec{k}}}^{z}\rangle $ can be used. A 
reliable calculation can be rather performed using the general expression (\ref 
{Eq4.14}) with the decoupling procedure (\ref{Eq4.40}) appropriate near the 
critical point, where $\sigma \rightarrow 0$, which takes into account the 
effects of the longitudinal spin correlations. From this expression, one can 
easily show that the reduced internal energy per spin near criticality 
assumes the form
\begin{equation} 
{\overline{u}}\simeq -{\overline{h}}\sigma -{\frac{1}{2}}\gamma (0)\sigma 
^{2}-{\frac{1}{2}}{\overline{T}}\sigma (1+\sigma ^{2})\int_{1BZ}{\frac{d^{d}k%
}{(2\pi )^{d}}}{\frac{\gamma ({\vec{k}})}{{\overline{\omega }}_{{\vec{k}}}}}%
\ ,  \label{Eq4.100} 
\end{equation} 
where now ${\overline{\omega }}_{{\vec{k}}}$ is given by Eq. (\ref{Eq4.56}) 
with the appropriate $R({\vec{k}})$ in Eq. (\ref{Eq4.58}). With Eq. (\ref 
{Eq4.100}) we have all the ingredients to calculate the zero-field specific 
heat near criticality. Indeed, working at ${\overline{h}}\rightarrow 0$ and 
$\sigma \rightarrow 0$ with $\sigma /{\overline{h}}={\overline{\chi }}({%
\overline{T}})$ as ${\overline{T}}\rightarrow {\overline{T}}_{c}^{+}$ and 
taking into account Eq. (\ref{Eq4.94}) with $P({\vec{k}})$ evaluated as ${%
\vec{k}}\rightarrow 0$, Eq. (\ref{Eq4.100}) can be conveniently expressed, 
in terms of ${\overline{T}}$, $\overline{\chi }$ and $\overline{h}$, as%
\begin{equation} 
{\overline{u}}\simeq {\frac{1}{2}}{\overline{T}}-{\overline{h}}^{2}{%
\overline{\chi }}-{\frac{1}{2}}\gamma (0){\overline{h}}^{2}{\overline{\chi }}%
^{2}-{\frac{1}{6}}(1+\gamma (0){\overline{\chi }}){\overline{\chi }}^{-1}\ .  
\label{Eq4.101} 
\end{equation} 
Then, for the reduced zero-field specific heat ${\overline{C}}_{\overline{h}%
=0}({\overline{T}})=C_{h=0}(T)/S$, we have
\begin{equation} 
C_{\overline{h}=0}({\overline{T}})=\left( {\frac{\partial \overline{u}}{%
\partial {\overline{T}}}}\right) _{\overline{h}=0}\simeq {\frac{1}{2}}+{%
\frac{1}{6}}\overline{\chi }^{-2}{\frac{\partial \overline{\chi }}{\partial {%
\overline{T}}}}\ ,\qquad ({\overline{T}}\rightarrow {\overline{T}}_{c}^{+}).  
\label{Eq4.102} 
\end{equation} 
From this expression one can immediately determine the specific heat 
critical exponent $\alpha (p,d)$. Indeed, with $\overline{\chi }({\overline{T%
}})\sim A({\overline{T}}-{\overline{T}}_{c})^{-\gamma }$ or $\overline{\chi }%
({\overline{T}})\sim A({\overline{T}}-{\overline{T}}_{c})^{-1}\ln \left[ {%
\frac{1}{({\overline{T}}-{\overline{T}}_{c})}}\right] $ (see Table I), Eq. (%
\ref{Eq4.102}) yields
\begin{equation} 
{\overline{C}}_{\overline{h}=0}({\overline{T}})\simeq \left\{ 
\begin{array}{lcl} 
\displaystyle{{\frac{1}{2}}-{\frac{1}{6}}\gamma (p,d)A^{2}({\overline{T}}-{\overline{T}}%
_{c})^{-\alpha (p,d)}} & , & \alpha (p,d)=1-\gamma (p,d)<0, \\
\\ 
\displaystyle{{\frac{1}{2}}-{\frac{1}{6}}A^{2}({\overline{T}}-{\overline{T}}_{c})^{0}\ln 
^{-1}\left[ {\frac{1}{{\overline{T}}-{\overline{T}}_{c}}}\right]} & , & 
\alpha =O_{\ln }\ .  
\end{array} 
\right.  \label{Eq4.103} 
\end{equation} 
From these relations and the exponent $\gamma (p,d)$ found before, one can 
immediately obtain the explicit expression for $\alpha (p,d)$. The different 
values of $\alpha (p,d)$ varying $p$ and $d$ are also given in Table I.  
Other critical exponents can be obtained using the known scaling laws.  

The data collected in Table I show clearly some features which signal the 
degree of efficiency of the approximations made to lowest order in the CSDM.  
First, the critical exponents coincide with those for the classical 
spherical model. Besides, within the region denoted with LRO (SRI) in Fig.  
1, the critical exponents are exactly the same of those for the short-range 
Heisenberg model. This suggests that improving systematically the one-pole 
approximation for $\Lambda _{\vec{k}}(\omega )$ within the spirit of the SDM 
may produce better results beyond the mean-field approximation.  

As a conclusion we note that for $d=1$ with $1<p<2$, when the LRI's are 
active, the explicit expressions of the coefficients in the power laws (\ref 
{Eq4.93}), (\ref{Eq4.96}), (\ref{Eq4.99}) and (\ref{Eq4.103}), as functions 
of $p$, can be found in Ref. \cite{Cavallo02}.  

\paragraph{\textit{B. Low-temperature paramagnetic susceptibility in absence 
of LRO}} 

For obtaining the low-temperature zero-field susceptibility in absence of 
LRO, namely for $p\geq 2d$ and $d\leq 2$ in Fig. 1, one can use again Eqs. (%
\ref{Eq4.94})-(\ref{Eq4.95}) but bearing in mind that now $T_{c}=0$. Then, 
with the same procedure used before for determining the near critical 
behavior of the paramagnetic susceptibility as $T\rightarrow T_{c}^{+}$ 
when LRO occurs, we easily find, for reduced susceptibility as $T\rightarrow 
0 $, 
\begin{equation} 
\overline{\chi }({\overline{T}})\sim \left\{ 
\begin{array}{lcl} 
\exp [A_{d}\Lambda ^{d}/3{\overline{T}}] & , & p=2d \\ 
{\overline{T}}^{-{\frac{p-d}{p-2d}}} & , & 2d<p<2+d \\ 
{\overline{T}}^{-{\frac{2}{2-d}}}\left( \ln \left[ {\frac{1}{{\overline{T}}}}%
\right] \right) ^{\frac{d}{2-d}} & , & p=2+d \\ 
{\overline{T}}^{-{\frac{2}{2-d}}} & , & p>2+d\ .  
\end{array} 
\right.  \label{Eq4.104} 
\end{equation} 
As usual, for $d=1$, all the coefficients can be explicitly calculated and 
Eq. (\ref{Eq4.104}) becomes 
\begin{equation} 
\overline{\chi }({\overline{T}})\simeq \left\{ 
\begin{array}{lcl} 
{\frac{3}{4\pi ^{2}}}\exp \left[ 2\pi ^{2}/9{\overline{T}}\right] & , & p=2 
\\ 
\left[ \left( {\frac{4}{3}}\right) ^{\frac{1}{(p-1)}}{\frac{1}{3A(p)}}\right] 
^{{\frac{(p-1)}{(p-2)}}}{\overline{T}}^{-{\frac{p-1}{p-2}}} & , & 2<p<3 \\ 
\frac{16}{27}{\overline{T}}^{-{2}}\ln \left[ {\frac{1}{{\overline{T}}}}%
\right] & , & p=3 \\ 
\frac{8}{27}\zeta (p-2){\overline{T}}^{-{2}} & , & p>3\ , 
\end{array} 
\right.  \label{Eq4.105} 
\end{equation} 
where
\begin{equation} 
A(p)={\frac{2\left[ 2({\frac{3}{2}})^{\frac{2p-3}{p-1}}-1\right] }{9(p-1)^{3}%
}}\left[ \sin \left( {\frac{\pi }{p-1}}\right) \right] ^{-1}\left( {\frac{1}{%
2}}\eta (p)\right) ^{-\frac{1}{p-1}}\ .  \label{Eq4.106} 
\end{equation} 
The previous results for $d=1$ appear consistent with the ones obtained in Refs.  
\cite{Nakano94a,Nakano95} for the corresponding quantum Heisenberg chain. It 
is worth noting an important aspect of the result in Eq. ({\ref{Eq4.104}}) 
for $p=2d$ and hence in Eq. (\ref{Eq4.105}) for $p=2$. This finding for 
reduced susceptibility, although consistent with the one obtained by means 
of the modified spin-wave theory \cite{Nakano94a} and the EMM for the 
two-time GF's \cite{Nakano95} for quantum model, does not reproduce the well 
known exact Haldane result for $d=1$ \cite{Haldane88} which contains a 
factor proportional to ${\overline{T}}^{-1/2}$. So, we are induced to argue 
that one must expect a $d$-dependent power factor in ${\overline{T}}$ in Eq.  
({\ref{Eq4.104}}) which corrects the pure low temperature exponential 
divergence reproducing ${\overline{T}}^{-1/2}$ for $d=1$. However, the 
lowest-order approximation in the CSDM, used in our calculations, has been 
able to capture the exponential divergence as ${\overline{T}}\rightarrow 0$ 
and, hopefully, one can improve systematically this result by means of 
higher order approximations. 

\subsection{Moment Equations within the two $\protect\delta $-poles 
Approximation for the \\Transverse Spectral Density} 

The next higher order approximation in the CSDM is to use the three exact 
ME's ({\ref{Eq4.33}})-({\ref{Eq4.35}}) and to choose a two $\delta $%
-functions structure for the $\Lambda _{\vec{k}}(\omega )$ involving three 
unknown parameters (see Subsec. 3.2). A reliable possibility, suggested by 
quantum studies \cite 
{Kalashnikov69,Kalashnikov73,Campana79,Caramico80,Nolting89,Nolting91,Hermann97}%
, is
\begin{equation} 
\Lambda _{\vec{k}}(\omega )=2\pi \left\{ {\lambda _{\vec{k}}^{(1)}\delta 
(\omega -\omega _{\vec{k}})+\lambda _{\vec{k}}^{(2)}\delta (\omega +\omega _{%
\vec{k}})}\right\} ,  \label{Eq4.106bis} 
\end{equation} 
so that, Eqs. ({\ref{Eq4.33}})-({\ref{Eq4.34}}) yield
\begin{equation} 
\begin{array}{l} 
\lambda _{\vec{k}}^{(1)}+\lambda _{\vec{k}}^{(2)}=M_{0}(\vec{k})=2Nm, \\ 
\omega _{\vec{k}}(\lambda _{\vec{k}}^{(1)}-\lambda _{\vec{k}}^{(2)})=M_{1}(%
\vec{k}), \\ 
\omega _{\vec{k}}^{2}(\lambda _{\vec{k}}^{(1)}+\lambda _{\vec{k}%
}^{(2)})=M_{2}(\vec{k}), 
\end{array} 
\label{Eq4.107} 
\end{equation} 
where $M_{\nu }({\vec{k}})$ $(\nu =1,2)$ denote the moments in the 
right-hand side of the Eqs. ({\ref{Eq4.34}}) and ({\ref{Eq4.35}}), 
respectively. These moments involve higher order correlation functions of 
the spin variables and hence, to close the set of ME's ({\ref{Eq4.107}}), it 
is necessary to use appropriate decoupling procedures. Below we suggest some 
of them which can be conveniently used for higher order calculations.  

\subsubsection{Full Neglect of Longitudinal Spin Correlations} 

Using in Eqs. ({\ref{Eq4.34}})-({\ref{Eq4.35}}) the Tyablikov-like 
decoupling procedure
\begin{equation} 
\left\langle {S_{\vec{k}}^{+}S_{\vec{q}}^{-}S_{\vec{p}}^{z}}\right\rangle 
\simeq \left\langle {S_{\vec{p}}^{z}}\right\rangle \left\langle {S_{\vec{k}%
}^{+}S_{\vec{q}}^{-}}\right\rangle   \label{Eq4.108} 
\end{equation} 
and the trivial ones $\left\langle {S_{\vec{k}}^{z}S_{\vec{q}}^{z}}%
\right\rangle \simeq \left\langle {S_{\vec{k}}^{z}}\right\rangle 
\left\langle {S_{\vec{q}}^{z}}\right\rangle $ and $\left\langle {S_{\vec{k}%
}^{z}S_{\vec{q}}^{z}S_{\vec{p}}^{z}}\right\rangle \simeq \left\langle {S_{%
\vec{k}}^{z}}\right\rangle \left\langle {S_{\vec{q}}^{z}}\right\rangle 
\left\langle {S_{\vec{p}}^{z}}\right\rangle $, in Eq. ({\ref{Eq4.107}}) we 
have 
\begin{eqnarray}  \label{Eq4.110} 
M_{1}(\vec{k}) &=&2Nm\omega _{\vec{k}}^{(Ty)}+{\frac{{T}}{N}}\sum\limits_{%
\vec{k}^{\prime }}{\left[ {J(\vec{k}^{\prime })-J(\vec{k}-\vec{k}^{\prime })}%
\right] {\frac{{\lambda _{\vec{k}^{\prime }}^{(1)}-\lambda _{\vec{k}^{\prime 
}}^{(2)}}}{{\omega _{\vec{k}^{\prime }}}}}},  \label{Eq4.109} \\ 
M_{2}(\vec{k}) &=&2Nm\left( {\omega _{\vec{k}}^{(Ty)}}\right) ^{2}+2{\omega 
_{\vec{k}}^{(Ty)}\frac{{T}}{N}}\sum\limits_{\vec{k}^{\prime }}{\left[ {J(%
\vec{k}^{\prime })-J(\vec{k}-\vec{k}^{\prime })}\right] {\frac{{\lambda _{%
\vec{k}^{\prime }}^{(1)}-\lambda _{\vec{k}^{\prime }}^{(2)}}}{{\omega _{\vec{%
k}^{\prime }}}},}}  \notag \\ 
&& 
\end{eqnarray} 
which close the set of ME's ({\ref{Eq4.107}}). In the previous equations $%
\omega _{{\vec{k}}}^{(Ty)}$ is given by Eq. ({\ref{Eq4.25}}) and use is made 
of the relation (\ref{Eq4.35bis}) which expresses exactly the transverse 
spin correlation function $\left\langle {S_{\vec{k}}^{+}S_{-\vec{k}}^{-}}%
\right\rangle $ in terms of $\Lambda _{\vec{k}}(\omega )$. From Eqs. ({\ref 
{Eq4.107}}) and ({\ref{Eq4.109}})-({\ref{Eq4.110}}), we can conveniently 
write a set of two self-consistent integral equations for $M_{1}({\vec{k}})$ 
and $M_{2}({\vec{k}})$ and then determine the parameters $\omega _{\vec{k}} 
$ and $\lambda _{\vec{k}}^{(\nu )}$ $(\nu =1,2)$ in terms of $M_{1}({\vec{k}}%
)$ and $M_{2}({\vec{k}})$. Indeed, by Eqs. ({\ref{Eq4.107}}), it is easy to 
show that
\begin{equation} 
{\frac{{M_{1}(\vec{k})}}{{2Nm}}}=\omega _{\vec{k}}^{(Ty)}+{\frac{{T}}{N}}%
\sum\limits_{\vec{k}^{\prime }}{\left[ {J(\vec{k}^{\prime })-J(\vec{k}-\vec{k%
}^{\prime })}\right] {\frac{{M_{1}(\vec{k}^{\prime })}}{{M_{2}(\vec{k}%
^{\prime })}}}},  \label{Eq4.111} 
\end{equation} 
and
\begin{equation} 
{\frac{{M_{2}(\vec{k})}}{{2Nm}}}=\left( {\omega _{\vec{k}}^{(Ty)}}\right) 
^{2}+{2\omega _{\vec{k}}^{(Ty)}\frac{{T}}{N}}\sum\limits_{\vec{k}^{\prime }}{%
\left[ {J(\vec{k}^{\prime })-J(\vec{k}-\vec{k}^{\prime })}\right] {\frac{{%
M_{1}(\vec{k}^{\prime })}}{{M_{2}(\vec{k}^{\prime })}}}},  \label{Eq4.112} 
\end{equation} 
with
\begin{equation} 
\omega _{\vec{k}}^{2}={\frac{{M_{2}(\vec{k})}}{{2Nm}}},  \label{Eq4.113} 
\end{equation} 
and
\begin{equation} 
\lambda _{\vec{k}}^{(\nu )}=Nm+(-1)^{\nu +1}{\frac{{M_{1}(\vec{k})}}{2}}%
\left( {{\frac{{2Nm}}{{M_{2}(\vec{k})}}}}\right) ^{{\frac{1}{2}}},\quad (\nu 
=1,2).  \label{Eq4.114} 
\end{equation} 
Of course, when Eqs. ({\ref{Eq4.111}})-({\ref{Eq4.112}}) are solved and 
hence $M_{1}({\vec{k}})$ and $M_{2}({\vec{k}})$ are known, Eqs. ({\ref 
{Eq4.113}})-({\ref{Eq4.114}}) determine $\Lambda _{\vec{k}}(\omega )$, as 
given by Eq. ({\ref{Eq4.106bis}}), and all the thermodynamic properties 
follow from the general formulas established in Subsec. 4.1. In particular, 
for the dispersion relation $\omega _{\vec{k}}$ of the undamped oscillations 
and the magnetization per spin, we find
\begin{equation} 
\omega _{\vec{k}}=\omega _{\vec{k}}^{(Ty)}\left\{ {1+{\frac{{2}}{{\omega _{%
\vec{k}}^{(Ty)}}}}{\frac{T}{N}}\sum\limits_{\vec{k}^{\prime }}{\left[ {J(%
\vec{k}^{\prime })-J(\vec{k}-\vec{k}^{\prime })}\right] {\frac{{M_{1}(\vec{k}%
^{\prime })}}{{M_{2}(\vec{k}^{\prime })}}}}}\right\} ^{{\frac{1}{2}}} 
\label{Eq4.115} 
\end{equation} 
and
\begin{equation} 
m\simeq S-{\frac{{T}}{N}}\sum\limits_{\vec{k}}{{\frac{{M_{1}(\vec{k})}}{{%
M_{2}(\vec{k})}}.}}  \label{Eq4.116} 
\end{equation} 
We wish to underline that the previous results are physically meaningful 
only for ($T$,$h$)-values in the nearly saturated regimes where, neglecting 
the longitudinal spin correlations, they constitute a reasonable approximation.  

Equations ({\ref{Eq4.111}})-({\ref{Eq4.116}}) show that it is convenient to 
work in terms of the single parameter
\begin{equation} 
x(\vec{k})={\frac{{M_{1}(\vec{k})}}{{M_{2}(\vec{k})}}},  \label{Eq4.117} 
\end{equation} 
satisfying the self-consistent equation
\begin{equation} 
x(\vec{k})={\frac{1}{{\omega _{\vec{k}}^{(Ty)}}}}{\frac{{1+{\frac{{1}}{{%
\omega _{\vec{k}}^{(Ty)}}}}{\frac{T}{N}}\sum\limits_{\vec{k}^{\prime }}{%
\left[ {J(\vec{k}^{\prime })-J(\vec{k}-\vec{k}^{\prime })}\right] x(\vec{k}%
^{\prime })}}}{{1+{\frac{{2}}{{\omega _{\vec{k}}^{(Ty)}}}}{\frac{T}{N}}%
\sum\limits_{\vec{k}^{\prime }}{\left[ {J(\vec{k}^{\prime })-J(\vec{k}-\vec{k%
}^{\prime })}\right] x(\vec{k}^{\prime })}}}}.  \label{Eq4.118} 
\end{equation} 
A full solution of this equation is of course inaccessible and one must 
resort to numerical calculations. However, analytical results can be 
obtained, as usual, by expansion in powers of $T$ with $\frac{1}{N}%
\sum\limits_{\vec{k}^{\prime }}{(...)\rightarrow \int_{1BZ}{d^{d}k^{\prime 
}(...)/}(2\pi )^{d}}$ and appropriate constraints on the values of $h$ 
assuring the convergency of the integrals (see discussion in Subsec. 4.4.1).  
Working, for instance, to first order in $T$, Eqs. ({\ref{Eq4.111}}), ({\ref 
{Eq4.112}}) and ({\ref{Eq4.116}}) yield%
\begin{equation} 
m\simeq S\left\{ {1-{\frac{{T}}{{JS^{2}}}}\mathcal{F}_{1}^{(p)}(\bar{h})}%
\right\} ,  \label{Eq4.119} 
\end{equation} 
which is exactly the same result ({\ref{Eq4.61}}) obtained with the one $%
\delta $-pole ansatz for $\Lambda _{\vec{k}}(\omega )$. Then, from Eq. ({\ref 
{Eq4.118}}), one easily has
\begin{equation} 
x(\vec{k})\simeq \omega _{0\vec{k}}^{-1}\left\{ {1+{\frac{{T}}{{\omega _{0%
\vec{k}}S}}}\left[ {\mathcal{F}_{1}^{(p)}(\bar{h})\Omega ^{(p)}(\vec{k})+%
\mathcal{F}_{2}^{(p)}(\bar{h},\vec{k})}\right] }\right\}  \label{Eq4.120} 
\end{equation} 
where $\mathcal{F}_{\nu }^{(p)}(\bar{h},\vec{k})$ $(\nu =1,2)$ are defined 
by Eqs. ({\ref{Eq4.62}}) and ({\ref{Eq4.63}}) and $\omega _{0{\vec{k}}%
}=\omega _{\vec{k}}^{(Ty)}\left| {_{T=0}}\right. =JS\left( {\bar{h}}+\Omega 
^{(p)}(\vec{k})\right) $. Of course, an estimate of the integrals $\mathcal{F%
}_{\nu }^{(p)}$ can be made using the expansions ({\ref{Eq4.64}}) for $%
\Omega ^{(p)}(\vec{k})$ as $\vec{k}\rightarrow 0$ along the lines outlined 
in Subsec. 4.4.1.  

Now, from Eqs. ({\ref{Eq4.111}}), ({\ref{Eq4.112}}) and ({\ref{Eq4.117}}), a 
straightforward algebra yields
\begin{equation} 
M_{1}(\vec{k})\simeq 2NS\omega _{0\vec{k}}\left\{ {1-{\frac{{T}}{{JS^{2}}}}%
\left[ \mathcal{F}_{1}^{(p)}(\bar{h})+\frac{{JS}}{{\omega _{0\vec{k}}}}%
\left( \mathcal{F}{_{1}^{(p)}(\bar{h})\Omega ^{(p)}(\vec{k})+\mathcal{F}%
_{2}^{(p)}(\bar{h},\vec{k})}\right) \right] }\right\} ,  \label{Eq4.121} 
\end{equation} 
and
\begin{equation} 
M_{2}(\vec{k})\simeq 2NS\omega _{0\vec{k}}^{2}\left\{ {1-{\frac{{T}}{{JS^{2}}%
}}\left[ \mathcal{F}_{1}^{(p)}(\bar{h})+\frac{{JS}}{{\omega _{0\vec{k}}}}%
\left( {\mathcal{F}_{1}^{(p)}(\bar{h})\Omega ^{(p)}(\vec{k})+\mathcal{F}%
_{2}^{(p)}(\bar{h},\vec{k})}\right) \right] }\right\} .  \label{Eq4.122} 
\end{equation} 
Then, Eqs. ({\ref{Eq4.113}}) and ({\ref{Eq4.114}}) provide%
\begin{equation} 
\omega _{\vec{k}}\simeq \omega _{0\vec{k}}\left\{ 1-\frac{{T}}{{\omega _{0%
\vec{k}}S}}\left[ \mathcal{F}{_{1}^{(p)}(\bar{h})\Omega ^{(p)}(\vec{k})+}%
\mathcal{F}{_{2}^{(p)}(\bar{h},\vec{k})}\right] \right\} ,  \label{Eq4.123} 
\end{equation} 
and
\begin{eqnarray} 
\lambda _{\vec{k}}^{(\nu )} &\simeq &NS\left[ {1+(-1)^{\nu +1}}\right] 
\left\{ {1-{\frac{{T}}{{JS}}}\mathcal{F}_{1}^{(p)}(\bar{h})}\right\} 
+O(T^{2})  \notag \\ 
&=&\left\{ 
\begin{array}{lll} 
2Nm+O(T^{2}) & , & \nu =1 \\ 
O(T^{2}) & , & \nu =2.  
\end{array} 
\right.  \label{Eq4.124} 
\end{eqnarray} 
The previous expressions for $\omega _{\vec{k}}$ and $\lambda _{\vec{k}%
}^{(\nu )}$ $(\nu =1,2)$ are the solution of the ME's ({\ref{Eq4.111}}) in 
the nearly saturated regime to the first order in $T$ by fully neglecting the 
longitudinal spin correlations.  

It is worth noting that, by inspection of Eqs. ({\ref{Eq4.119}}), ({\ref 
{Eq4.123}}) and ({\ref{Eq4.124}}), the magnetization and the dispersion 
relation, here obtained to the first order in $T$, coincide with the 
corresponding ones produced by the one-pole ansatz for $\Lambda _{\vec{k}%
}(\omega )$. Of course, possible differences will occur beginning from 
second order in $T$. In any case, all the physical predictions of Subsec.  
4.4.1 about the occurrence of LRO (see Fig. 1) remain qualitatively 
unchanged. All this is a clear signal of the internal consistency of the 
CSDM.  

\subsubsection{Decoupling Procedures for Taking into Account the 
Longitudinal Spin \\Correlations in the Nearly Saturated Regime} 

The previous approximation involving two $\delta $-functions for $\Lambda _{%
\vec{k}}(\omega )$ can be systematically improved using in the ME's ({\ref 
{Eq4.107}}) a closure decoupling procedure which, although based on the near 
saturation relation $S_{j}^{z}\simeq S-S_{j}^{+}S_{j}^{-}/2S$, allows us to 
take properly into account the effect of the correlation function $%
\left\langle {S_{\vec{k}}^{z}S_{-\vec{k}}^{z}}\right\rangle $. To perform 
this program, together with the Tyablikov-like decoupling ({\ref{Eq4.108}}), 
for the higher order longitudinal correlation function in $M_{2}({\vec{k}})$, 
we will use now the Hartree-Fock-like decoupling procedure%
\begin{eqnarray} 
\left\langle {S_{\vec{k}}^{z}S_{\vec{q}}^{z}S_{\vec{p}}^{z}}\right\rangle 
&\simeq &\left\langle {S_{\vec{k}}^{z}S_{\vec{q}}^{z}}\right\rangle 
\left\langle {S_{\vec{p}}^{z}}\right\rangle +\left\langle {S_{\vec{k}}^{z}S_{%
\vec{p}}^{z}}\right\rangle \left\langle {S_{\vec{q}}^{z}}\right\rangle + 
\notag \\ 
&&+\left\langle {S_{\vec{p}}^{z}S_{\vec{q}}^{z}}\right\rangle \left\langle {%
S_{\vec{k}}^{z}}\right\rangle -2\left\langle {S_{\vec{k}}^{z}}\right\rangle 
\left\langle {S_{\vec{q}}^{z}}\right\rangle \left\langle {S_{\vec{p}}^{z}}%
\right\rangle .  \label{Eq4.125} 
\end{eqnarray} 
With this prescription, the moment $M_{2}({\vec{k}})$ in the right-hand side 
of Eq. ({\ref{Eq4.35}}) can be approximated as
\begin{eqnarray} 
M_{2}({\vec{k}}) &\simeq &2h^{2}Nm-4Nm^{3}\left[ {J(0)-J(\vec{k})}\right] 
^{2}+  \notag \\ 
&&+{\frac{2}{N}}\sum\limits_{\vec{k}^{\prime }}{\left[ {J(\vec{k}^{\prime 
})-J(\vec{k}-\vec{k}^{\prime })}\right] \omega _{\vec{k}^{\prime 
}}^{(Ty)}\left\langle {S_{\vec{k}^{\prime }}^{+}S_{-\vec{k}^{\prime }}^{-}}%
\right\rangle }+  \notag \\ 
&&+{\frac{2}{N}}\sum\limits_{\vec{k}^{\prime }}{\left[ {J(\vec{k}^{\prime 
})-J(\vec{k}-\vec{k}^{\prime })}\right] \left\{ {2\omega _{\vec{k}^{\prime 
}}^{(Ty)}+m\left[ {J(0)-J(\vec{k})}\right] }\right\} \left\langle {S_{\vec{k}%
^{\prime }}^{z}S_{-\vec{k}^{\prime }}^{z}}\right\rangle .}  \notag \\ 
&&  \label{Eq4.126} 
\end{eqnarray} 
At this stage, to close the system of ME's ({\ref{Eq4.34}})-({\ref{Eq4.35}}%
), it is necessary to express the sums involving $\left\langle {S_{\vec{k}%
^{\prime }}^{z}S_{-\vec{k}^{\prime }}^{z}}\right\rangle $ in $M_{1}({\vec{k}}%
)$ and $M_{2}({\vec{k}})$ in terms of $\Lambda _{\vec{k}}(\omega )$.  
Essentially, one must find a convenient approximate method to write sums of 
the type
\begin{equation} 
S_{\vec{k}}={\frac{1}{N}}\sum\limits_{\vec{q}}{\Omega _{\vec{k}}(}\vec{q}%
)\left\langle {S_{\vec{q}}^{z}S_{-\vec{q}}^{z}}\right\rangle , 
\label{Eq4.127} 
\end{equation} 
in terms of the original $\Lambda _{\vec{k}}(\omega )$. This is not a simple 
problem and here, within the spirit of the CSDM, we suggest a possible 
procedure which is suitable for exploring nearly saturated regimes.  
Assuming, indeed, $S_{j}^{z}\simeq S-S_{j}^{+}S_{j}^{-}/2S$ and working in 
the Fourier space, with a simple algebra, the sum ({\ref{Eq4.127}}) can be 
written as
\begin{align} 
S_{\vec{k}}& \simeq N\Omega _{\vec{k}}(0)S^{2}-{\frac{{\Omega _{\vec{k}}(0)}%
}{N}}\sum\limits_{\vec{q}}{\left\langle {S_{\vec{q}}^{+}S_{-\vec{q}}^{-}}%
\right\rangle }+  \notag \\ 
& +{\frac{1}{{4S^{2}N^{3}}}}\sum\limits_{\vec{q}_{1},...,\vec{q}_{4}}{\Omega 
_{\vec{k}}(-\vec{q}_{3}-\vec{q}_{4})}\delta _{\vec{q}_{1}+\vec{q}_{2};-\vec{q%
}_{3}-\vec{q}_{4}}\left\langle {S_{\vec{q}_{1}}^{+}S_{\vec{q}_{2}}^{-}S_{%
\vec{q}_{3}}^{+}S_{\vec{q}_{4}}^{-}}\right\rangle ,  \label{Eq4.128} 
\end{align} 
which involves an higher order transverse correlation function. Now, with $%
A=S_{{\vec{q}}_{1}}^{+}S_{q_{2}}^{-}S_{q_{3}}^{+}$ and $B=S_{{\vec{q}}%
_{4}}^{-}$, we introduce the higher order transverse SD
\begin{eqnarray} 
\Lambda _{\vec{q}_{1},...,\vec{q}_{4}}(\omega ) &=&i\left\langle {\left\{ {%
S_{\vec{q}_{1}}^{+}(\tau )S_{\vec{q}_{2}}^{-}(\tau )S_{\vec{q}_{3}}^{+}(\tau 
),S_{\vec{q}_{4}}^{-}}\right\} }\right\rangle _{\omega }  \notag \\ 
&=&i\int_{-\infty }^{+\infty }{d\tau e^{i\omega \tau }}\left\langle {\left\{ 
{S_{\vec{q}_{1}}^{+}(\tau )S_{\vec{q}_{2}}^{-}(\tau )S_{\vec{q}%
_{3}}^{+}(\tau ),S_{\vec{q}_{4}}^{-}}\right\} }\right\rangle .  
\label{Eq4.129} 
\end{eqnarray} 
Then, in view of the general formalism developed in Sec. 2, we have exactly%
%
%

%

%

%

%

%

%

%

%


\begin{equation} 
\left\langle {S_{\vec{q}_{1}}^{+}S_{\vec{q}_{2}}^{-}S_{\vec{q}_{3}}^{+}S_{%
\vec{q}_{4}}^{-}}\right\rangle =T\int_{-\infty }^{+\infty }{{\frac{{d\omega }%
}{{2\pi }}}}{\frac{{\Lambda _{\vec{q}_{1},...,\vec{q}_{4}}(\omega )}}{\omega 
}}.  
\end{equation} 
Hence, in Eq. ({\ref{Eq4.128}}), one can write, again exactly,%
\begin{equation} 
\sum\limits_{\vec{q}_{1},\vec{q}_{2}}{\delta _{\vec{q}_{1}+\vec{q}_{2};-\vec{%
q}_{3}-\vec{q}_{4}}}\left\langle {S_{\vec{q}_{1}}^{+}S_{\vec{q}_{2}}^{-}S_{%
\vec{q}_{3}}^{+}S_{\vec{q}_{4}}^{-}}\right\rangle =T\int_{-\infty }^{+\infty 
}{{\frac{{d\omega }}{{2\pi }}}}{\frac{{\Lambda _{\vec{q}_{3}\vec{q}%
_{4}}(\omega )}}{\omega }},  \label{Eq4.131} 
\end{equation} 
where
\begin{equation} 
\Lambda _{\vec{q}_{3},\vec{q}_{4}}(\omega )=\sum\limits_{\vec{q}_{1},\vec{q}%
_{2}}{\delta _{\vec{q}_{1}+\vec{q}_{2};-\vec{q}_{3}-\vec{q}_{4}}}\Lambda _{%
\vec{q}_{1},...,\vec{q}_{4}}(\omega ).  \label{Eq4.132} 
\end{equation} 
As a next step, we perform in Eq. ({\ref{Eq4.129}}) the natural and 
convenient decoupling procedure:
\begin{eqnarray} 
\Lambda _{\vec{q}_{1},...,\vec{q}_{4}}(\omega )\simeq \left\langle {S_{\vec{q%
}_{1}}^{+}S_{-\vec{q}_{1}}^{-}}\right\rangle \delta _{\vec{q}_{1},-\vec{q}%
_{2}}\left[ {i\left\langle {\left\{ {S_{\vec{q}_{3}}^{+}(\tau ),S_{\vec{q}%
_{4}}^{-}}\right\} }\right\rangle _{\omega }}\right] + &&  \notag \\ 
+\left\langle {S_{\vec{q}_{3}}^{+}S_{-\vec{q}_{3}}^{-}}\right\rangle \delta 
_{\vec{q}_{3},-\vec{q}_{2}}\left[ {i\left\langle {\left\{ {S_{\vec{q}%
_{1}}^{+}(\tau ),S_{\vec{q}_{4}}^{-}}\right\} }\right\rangle _{\omega }}%
\right] &.&  \label{Eq4.133} 
\end{eqnarray} 
Then, the spectral function ({\ref{Eq4.132}}) becomes
\begin{equation} 
\Lambda _{\vec{q}_{3},\vec{q}_{4}}(\omega )\simeq \left[ {\delta _{\vec{q}%
_{3},-\vec{q}_{4}}\sum\limits_{\vec{q}_{1}}{\left\langle {S_{\vec{q}%
_{1}}^{+}S_{-\vec{q}_{1}}^{-}}\right\rangle }+\left\langle {S_{\vec{q}%
_{3}}^{+}S_{-\vec{q}_{3}}^{-}}\right\rangle }\right] \Lambda _{-\vec{q}%
_{4}}(\omega ),  \label{Eq4.134} 
\end{equation} 
and, through Eq. ({\ref{Eq4.131}}), we have that the sum in Eq. ({\ref 
{Eq4.128}}) can be written as
\begin{eqnarray} 
{\frac{1}{{4S^{2}N^{3}}}}\sum\limits_{\vec{q}_{1},...,\vec{q}_{4}}{\Omega _{%
\vec{k}}(-\vec{q}_{3}-\vec{q}_{4})}\delta _{\vec{q}_{1}+\vec{q}_{2};-\vec{q}%
_{3}-\vec{q}_{4}}\left\langle {S_{\vec{q}_{1}}^{+}S_{\vec{q}_{2}}^{-}S_{\vec{%
q}_{3}}^{+}S_{\vec{q}_{4}}^{-}}\right\rangle &\simeq &  \notag \\ 
\simeq {\frac{1}{{4S^{2}N^{3}}}}\sum\limits_{\vec{q},\vec{p}}{\left[ {\Omega 
_{\vec{k}}(0)+\Omega _{\vec{k}}(\vec{q}-\vec{p})}\right] }\left\langle {S_{%
\vec{q}}^{+}S_{-\vec{q}}^{-}}\right\rangle \left\langle {S_{\overrightarrow{p%
}}^{+}S_{-\overrightarrow{p}}^{-}}\right\rangle . &&  \label{Eq4.135} 
\end{eqnarray} 
With this result and Eq. ({\ref{Eq4.128}}), it is easy to show that the 
general sum ({\ref{Eq4.127}}), involving longitudinal correlation functions, 
can be expressed in terms of the transverse ones as
\begin{equation} 
S_{\vec{k}}\simeq Nm^{2}\Omega _{\vec{k}}(0)+{\frac{1}{{4S^{2}N^{3}}}}%
\sum\limits_{\vec{q},\vec{p}}{\left[ {\Omega _{\vec{k}}(\vec{q}-\vec{p})}%
\right] }\left\langle {S_{\vec{q}}^{+}S_{-\vec{q}}^{-}}\right\rangle 
\left\langle {S_{\vec{p}}^{+}S_{-\vec{p}}^{-}}\right\rangle  \label{Eq4.136} 
\end{equation} 
where, in the Fourier representation,
\begin{equation} 
m\simeq S-{\frac{1}{{N^{2}}}}\sum\limits_{\vec{q}}{{\frac{{\left\langle {S_{%
\vec{q}}^{+}S_{-\vec{q}}^{-}}\right\rangle }}{{2S}}}}.  \label{Eq4.137} 
\end{equation} 
Equation ({\ref{Eq4.136}}) is the main result for our next purposes. In 
terms of this expression and the exact relation (\ref{Eq4.35bis}) for ${%
\left\langle {S_{\overrightarrow{k}}^{+}S_{-\overrightarrow{k}}^{-}}%
\right\rangle }$ in terms of $\Lambda _{\overrightarrow{k}}(\omega )$, the 
moments $M_{1}({\vec{k}})$, with $\Omega _{\vec{k}}({\vec{q}})=J({\vec{q}}%
)-J({\vec{k}}-{\vec{q}})$, and $M_{2}({\vec{k})}$, with $\Omega _{\vec{k}}(%
\vec{q})=\left[ {J(\vec{q})-J(\vec{k}-\vec{q})}\right] \left\{ {2\omega _{%
\vec{k}}^{(T)}+m\left[ {J(\vec{q})-J(\vec{k}-\vec{q})}\right] }\right\} $, 
can be easily expressed as 

\begin{eqnarray} 
M_{1}(\vec{k}) &=&2Nm\omega _{\vec{k}}^{(Ty)}{\frac{{T}}{N}}\sum\limits_{%
\vec{k}^{\prime }}{\left[ {J(\vec{k}^{\prime })-J(\vec{k}-\vec{k}^{\prime })}%
\right] }\int_{-\infty }^{+\infty }{{\frac{{d\omega }}{{2\pi }}}}{\frac{{%
\Lambda _{\vec{k}^{\prime }}(\omega )}}{\omega }}+  \notag  \label{Eq4.138} 
\\ 
&&+{\frac{{T^{2}}}{{2S^{2}N^{3}}}}\sum\limits_{k_{1},k_{2}}{\left[ {J(\vec{k}%
_{1}-\vec{k}_{2})-J(\vec{k}-(\vec{k}_{1}-\vec{k}_{2}))}\right] }%
\prod\limits_{\nu =1}^{2}\int_{-\infty }^{+\infty }{{\frac{{d\omega }}{{2\pi 
}}}}{\frac{{\Lambda _{\vec{k}_{\nu }}(\omega )}}{\omega }.}  \notag \\ 
&& 
\end{eqnarray} 
and
\begin{eqnarray} 
M_{2}(\vec{k}) &=&2Nm\left( {\omega _{\vec{k}}^{(Ty)}}\right) ^{2}+  \notag 
\label{Eq4.139} \\ 
&&+{\frac{{2K_{B}T\omega _{\vec{k}}^{(Ty)}}}{N}}\sum\limits_{\vec{k}^{\prime 
}}{\left[ {J(\vec{k}^{\prime })-J(\vec{k}-\vec{k}^{\prime })}\right] }%
\int_{-\infty }^{+\infty }{{\frac{{d\omega }}{{2\pi }}}}{\frac{{\Lambda _{%
\vec{k}^{\prime }}(\omega )}}{\omega }}+  \notag \\ 
&&+{\frac{{T^{2}}}{{2S^{2}N^{3}}}}\sum\limits_{k_{1},k_{2}}{\left[ {J(\vec{k}%
_{1}-\vec{k}_{2})-J(\vec{k}-(\vec{k}_{1}-\vec{k}_{2}))}\right] }\times  
\notag \\ 
&&\times \left\{ {2\omega _{\vec{k}}^{(Ty)}+m\left[ {J(\vec{k}^{\prime })-J(%
\vec{k}-\vec{k}^{\prime })}\right] }\right\} \prod\limits_{\nu 
=1}^{2}\int_{-\infty }^{+\infty }{{\frac{{d\omega }}{{2\pi }}}}{\frac{{%
\Lambda _{\vec{k}_{\nu }}(\omega )}}{\omega }}.  \notag \\ 
&& 
\end{eqnarray} 
It is worth pointing that the previous ME's have been derived without 
assuming a particular functional structure for $\Lambda _{\vec{k}}(\omega )$%
. So, if we consider, for instance, the first two ME's ({\ref{Eq4.33}}) and (%
{\ref{Eq4.34}}) with $M_{1}({\vec{k}})$ given by Eq. ({\ref{Eq4.138}}) and 
assume for $\Lambda _{\vec{k}}(\omega )$ the one $\delta $-pole ansatz ({\ref 
{Eq4.36}}), it is possible to reformulate the analysis of Subsec. 4.4.1 but 
taking now into account the effects of longitudinal spin correlations.  
However, the aim of the last part of this subsection, as a suggestion for 
future more elaborate investigations, is only to show how the basic 
equations of previous subsection are modified using the two $\delta $-poles 
ansatz (4.109) and the expressions ({\ref{Eq4.138}}) and ({\ref{Eq4.139}}) 
for $M_{1}({\vec{k}})$ and $M_{2}({\vec{k}})$, respectively.  

First, we note that Eqs. ({\ref{Eq4.113}}) and ({\ref{Eq4.114}}), formally 
obtained by Eqs. ({\ref{Eq4.107}}), remain unchanged but now, for $M_{1}({%
\vec{k}})$ and $M_{2}({\vec{k}})$, we must consider the system of 
self-consistent equations (to be compared with Eqs. ({\ref{Eq4.111}}) and ({%
\ref{Eq4.112}}))
\begin{eqnarray}  \label{Eq4.140} 
{\frac{{M_{1}(\vec{k})}}{{2Nm}}} &=& \omega _{\vec{k}}^{(Ty)}+{\frac{{T}}{N}}%
\sum\limits_{\vec{k}^{\prime }}{\left[ {J(\vec{k}^{\prime })-J(\vec{k}-\vec{k%
}^{\prime })}\right] }{\frac{{M_{1}(\vec{k}^{\prime })}}{{M_{2}(\vec{k}%
^{\prime })}}}+  \notag \\ 
&& +{\frac{m}{{S^{2}}}}{\frac{{T^{2}}}{{N^{2}}}}\sum\limits_{\vec{k}_{1},%
\vec{k}_{2}}{\left[ {J(\vec{k}_{1}-\vec{k}_{2})-J(\vec{k}-(\vec{k}_{1}-\vec{k%
}_{2}))}\right] }{\frac{{M_{1}(\vec{k}_{1})}}{{M_{2}(\vec{k}_{1})}}\frac{{%
M_{1}(\vec{k}_{2})}}{{M_{2}(\vec{k}_{2})}}},  \notag \\ 
\end{eqnarray} 
and
\begin{eqnarray}  \label{Eq4.141} 
{\frac{{M_{2}(\vec{k})}}{{2Nm}}} &=& \left( {\omega _{\vec{k}}^{(Ty)}}%
\right) ^{2}+{\frac{{2T}}{N}}\sum\limits_{\vec{k}^{\prime }}{\left[ {J(\vec{k%
}^{\prime })-J(\vec{k}-\vec{k}^{\prime })}\right] }{\frac{{M_{1}(\vec{k}%
^{\prime })}}{{M_{2}(\vec{k}^{\prime })}}}+  \notag \\ 
&& +{\frac{m}{{S^{2}}}}{\frac{{T^{2}}}{{N^{2}}}}\sum\limits_{\vec{k}_{1}\vec{%
k}_{2}}{\left[ {J(\vec{k}_{1}-\vec{k}_{2})-J(\vec{k}-(\vec{k}_{1}-\vec{k}%
_{2}))}\right] \times }  \notag \\ 
&& \times \left[ {2\omega _{\vec{k}}^{(Ty)}+mJ(\vec{k}_{1}-\vec{k}_{2})-J(%
\vec{k}-(\vec{k}_{1}-\vec{k}_{2}))}\right] {\frac{{M_{1}(\vec{k}_{1})}}{{%
M_{2}(\vec{k}_{1})}}\frac{{M_{1}(\vec{k}_{2})}}{{M_{2}(\vec{k}_{2})}},} 
\notag \\ 
\end{eqnarray} 
with $m$ given again by Eq. ({\ref{Eq4.116}}) in terms of $M_{1}({\vec{k}})$ 
and $M_{2}({\vec{k}})$. As we see, the appropriate variable for the 
self-consistent problem ({\ref{Eq4.140}})-({\ref{Eq4.141}}) is again $x({%
\vec{k}})=M_{1}({\vec{k}})/M_{2}({\vec{k}})$ for which it is immediate to 
obtain the appropriate equation. Then, when $x({\vec{k}})$ and hence $M_{1}({%
\vec{k}})$ and $M_{2}({\vec{k}})$ will be known (analytically in some 
regimes, and numerically), the parameters which enter $\Lambda _{\vec{k}%
}(\omega )$ and $m$ can be obtained using the Eqs. ({\ref{Eq4.113}}), ({\ref 
{Eq4.114}}) and ({\ref{Eq4.107}}). In particular, for the dispersion 
relation we should have
\begin{eqnarray}  \label{Eq4.142} 
\omega _{\vec{k}} &=& \omega _{\vec{k}}^{(Ty)}\left\{ {1+{\frac{{2}}{{\omega 
_{\vec{k}}^{(Ty)}}}}{\frac{{T}}{N}}\sum\limits_{\vec{k}^{\prime }}{\left[ {J(%
\vec{k}^{\prime })-J(\vec{k}-\vec{k}^{\prime })}\right] }x(\vec{k}^{\prime 
})+}\right.  \notag \\ 
&& +{\frac{m}{{S^{2}}}}{\frac{1}{{\left( {\omega _{\vec{k}}^{(Ty)}}\right) 
^{2}}}}{\frac{{T^{2}}}{{N^{2}}}}\sum\limits_{\vec{k}_{1}\vec{k}_{2}}{\left[ {%
J(\vec{k}_{1})-J(\vec{k}_{1}-\vec{k}_{2})}\right] \times }  \notag \\ 
&& \times \left. \cdot {\left[ {2\omega _{\vec{k}}^{(Ty)}+m\left[ {J(\vec{k}%
_{1}-\vec{k}_{2})-J(\vec{k}-(\vec{k}_{1}-\vec{k}_{2}))}\right] }\right] x(%
\vec{k}_{1})x(\vec{k}_{2})}\right\} ^{{\frac{1}{2}}},  \notag \\ 
\end{eqnarray} 
or
\begin{eqnarray}  \label{Eq4.143} 
\omega _{\vec{k}} &=& \omega _{\vec{k}}^{(d)}\left\{ {1+{\frac{m}{{S^{2}}}}{%
\frac{1}{{\left( {\omega _{\vec{k}}^{(d)}}\right) ^{2}}}}\frac{{T^{2}}}{{%
N^{2}}}}\sum\limits_{\vec{k}_{1},\vec{k}_{2}}{\left[ {J(\vec{k}_{1}-\vec{k}%
_{2})-J\left( {\vec{k}-(\vec{k}_{1}-\vec{k}_{2})}\right) }\right] }\right. {%
\times }  \notag \\ 
&& \times \left. {\left[ {2\omega _{\vec{k}}^{(Ty)}+m\left[ {J(\vec{k}_{1}-%
\vec{k}_{2})-J\left( {\vec{k}-(\vec{k}_{1}-\vec{k}_{2})}\right) }\right] }%
\right] x(\vec{k}_{1})x(\vec{k}_{2})}\right\} ^{{\frac{1}{2}}},  \notag \\ 
\end{eqnarray} 
where we have denoted with $\omega _{\vec{k}}^{(d)}$ the dispersion relation 
({\ref{Eq4.115}}) obtained with full decoupling for the longitudinal spin 
correlation functions. All the remaining thermodynamic properties follow 
from the general formulas established before.  

Eqs. ({\ref{Eq4.138}})-({\ref{Eq4.143}}) and, in particular, the Eqs. ({\ref 
{Eq4.142}}) and ({\ref{Eq4.143}}) for the frequency spectrum $\omega _{\vec{k%
}}$ of the undamped elementary spin oscillations, show in a transparent way 
how the CSDM works taking into account systematically higher order effects 
on the relevant macroscopic quantities. This is just the key idea of the SDM.  

\subsubsection{Suggestions for Future Studies} 

The critical properties and the thermodynamic regimes with near-zero 
magnetization can also be explored within the two $\delta $-functions 
representation ({\ref{Eq4.106bis}}) for $\Lambda _{\vec{k}}(\omega )$. This 
can be performed again on the basis of Eqs. ({\ref{Eq4.107}}) with $M_{2}({%
\vec{k}})$ given by the approximation ({\ref{Eq4.126}}) which preserves the 
effect of the longitudinal correlations as the exact $M_{1}({\vec{k}})$ in 
Eq. ({\ref{Eq4.34}}). But now one must find an appropriate and convenient 
procedure to treat the sums in $M_{\nu }({\vec{k}})$ $(\nu =1,2)$ containing 
the correlation function $\left\langle {S_{\vec{k}}^{z}S_{-\vec{k}}^{z}}%
\right\rangle $ for describing physical situations where $m\geq 0$. A 
possibility consists in extending the decoupling procedure ({\ref{Eq4.40}}) 
as \cite{Kalashnikov73,Kamieniarz77}
\begin{equation} 
{\frac{1}{N}}\sum\limits_{\vec{q}}{\Omega _{\vec{k}}(\vec{q})\left\langle {%
S_{\vec{q}}^{z}S_{-\vec{q}}^{z}}\right\rangle }\approx Nm^{2}\Omega _{\vec{k}%
}(0)-{\frac{1}{2}}\left( {1-{\frac{{m^{2}}}{{S^{2}}}}}\right) {\frac{1}{N}}%
\sum\limits_{\vec{q}}{\Omega _{\vec{k}}(\vec{q})\left\langle {S_{\vec{q}%
}^{+}S_{-\vec{q}}^{-}}\right\rangle }.  
\end{equation} 
The next steps will be very similar to those made in Subsec. 4.4.2, 
but now much more complicated. This analysis is beyond the purpose of the 
present article. Our intention here was rather to give the main ingredients 
and suggestions for future possible analytical and numerical developments 
(also for nonmagnetic classical many-body systems), with the spirit of the 
SDM, for systematic higher order approximations satisfying more and more sum 
rules for the appropriate SD. In the next subsection, using some of the 
general results obtained above which are independent of the functional 
structure assumed for $\Lambda _{\vec{k}}(\omega )$, we will show how one 
can conveniently study the low-temperature damping of the spin oscillations 
along the lines suggested in Sec. 3. This will be performed using the 
general ME's ({\ref{Eq4.33}})-({\ref{Eq4.35}}) with $M_{\nu }({\vec{k}})$ $%
(\nu =1,2)$ in the form ({\ref{Eq4.138}}) and ({\ref{Eq4.139}}) for taking 
into account the effects of longitudinal spin correlations.  

\subsection{Damping of the Spin Excitations in the Low-Temperature Regime 
Via\\ the CSDM} 

At this stage, we have all the ingredients to study the damping of the 
elementary spin oscillations by means of the extended CSDM suggested in Sec. 3.  
For our transverse SD $\Lambda _{\vec{k}}(\omega )$, according to the 
general prescription ({\ref{Eq3.9}}), we assume the modified Gaussian 
structure%
\begin{equation} 
\Lambda _{\vec{k}}(\omega )=2\pi \omega \lambda _{\vec{k}}{\frac{{\exp \left[ 
{{\frac{-{(\omega -\omega _{\vec{k}})^{2}}}{{\Gamma _{\vec{k}}}}}}\right] }}{%
{(\pi \Gamma _{\vec{k}})^{1/2}}}},  \label{Eq4.145} 
\end{equation} 
with the basic condition $\Gamma _{\vec{k}}/\omega _{\vec{k}}^{2}\ll 1$.  
This representation contains three unknown parameters $\lambda _{\vec{k}%
},\omega _{\vec{k}},\Gamma _{\vec{k}}$ and hence we need to consider the 
three closed ME's
\begin{equation} 
\begin{array}{l} 
\displaystyle{\int_{-\infty }^{+\infty }{{\frac{{d\omega }}{{2\pi }}}\Lambda _{\vec{k}}}%
(\omega )=2Nm,} \\ 

\displaystyle{\int_{-\infty }^{+\infty }{{\frac{{d\omega }}{{2\pi }}}\omega \Lambda _{\vec{%
k}}}(\omega )=M_{1}(\vec{k}),} \\
 
\displaystyle{\int_{-\infty }^{+\infty }{{\frac{{d\omega }}{{2\pi }}}\omega ^{2}\Lambda _{%
\vec{k}}}(\omega )=M_{2}(\vec{k}),} 
\end{array} 
\label{Eq4.146} 
\end{equation} 
where $M_{1}({\vec{k}})$ and $M_{2}({\vec{k}})$ are assumed to be given by 
Eqs. ({\ref{Eq4.138}}) and ({\ref{Eq4.139}}), respectively, as obtained by 
means of the decoupling procedures ({\ref{Eq4.108}}), ({\ref{Eq4.125}}) and (%
{\ref{Eq4.136}}). They take into account the longitudinal spin correlation 
effects in the nearly saturated regimes in which we are interested here.  

Inserting the expression ({\ref{Eq4.145}}) in Eqs. (\ref{Eq4.146}) and 
performing standard integrals we have, for the parameters $\lambda _{\vec{k}%
},\omega _{\vec{k}},\Gamma _{\vec{k}}$, the equations
\begin{equation} 
\lambda _{\vec{k}}\omega _{\vec{k}}=2Nm,  \label{Eq4.147} 
\end{equation} 
\begin{eqnarray}  \label{Eq4.148} 
\lambda _{\vec{k}}\left[ {\omega _{\vec{k}}^{2}+{\frac{1}{2}}\Gamma _{\vec{k}%
}}\right] &=& 2Nm\omega _{\vec{k}}^{(Ty)}+{\frac{{T}}{N}}\sum\limits_{\vec{k}%
^{\prime }}{\left[ {J(\vec{k}^{\prime })-J(\vec{k}-\vec{k}^{\prime })}\right] 
}\lambda _{\vec{k}^{\prime }}  \notag \\ 
&& +{\frac{{T^{2}}}{{2S^{2}N^{3}}}}\sum\limits_{\vec{k}_{1},\vec{k}_{2}}{%
\left[ {J(\vec{k}_{1}-\vec{k}_{2})-J(\vec{k}-(\vec{k}_{1}-\vec{k}_{2}))}%
\right] }\lambda _{\vec{k}_{1}}\lambda _{\vec{k}_{2}},  \notag \\ 
\end{eqnarray} 
\begin{eqnarray}  \label{Eq4.149} 
\lambda _{\vec{k}}\omega _{\vec{k}}\left[ {\omega _{\vec{k}}^{2}+{\frac{3}{2}%
}\Gamma _{\vec{k}}}\right] &=& 2Nm\left( {\omega _{\vec{k}}^{(Ty)}}\right) 
^{2}+{\frac{{2\omega _{\vec{k}}^{(Ty)}}}{N}}\sum\limits_{\vec{k}^{\prime }}{%
\left[ {J(\vec{k}^{\prime })-J(\vec{k}-\vec{k}^{\prime })}\right] }\lambda _{%
\vec{k}}+  \notag \\ 
&&+{\frac{{T^{2}}}{{2S^{2}N^{3}}}}\sum\limits_{\vec{k}_{1},\vec{k}_{2}}{%
\left[ {J(\vec{k}_{1}-\vec{k}_{2})-J(\vec{k}-(\vec{k}_{1}-\vec{k}_{2}))}%
\right] \times }  \notag \\ 
&& \times \left[ {2\omega _{\vec{k}}^{(T)}+m\left[ {J(\vec{k}_{1}-\vec{k}%
_{2})-J(\vec{k}-(\vec{k}_{1}-\vec{k}_{2}))}\right] }\right] \lambda _{\vec{k}%
_{1}}\lambda _{\vec{k}_{2}}  \notag \\ 
\end{eqnarray} 
Here, $m$ is given by
\begin{equation} 
m\simeq S-{\frac{{T}}{{2SN^{2}}}}\sum\limits_{\vec{k}}{\int_{-\infty 
}^{+\infty }{{\frac{{d\omega }}{{2\pi }}}{\frac{{\Lambda _{\vec{k}}(\omega )}%
}{\omega }}}=S-}{\frac{{T}}{{2SN^{2}}}}\sum\limits_{\vec{k}}{\lambda _{\vec{k%
}}},  \label{Eq4.150} 
\end{equation} 
and hence, with $\lambda _{\vec{k}}=2Nm/\omega _{\vec{k}}$ from Eq. ({\ref 
{Eq4.147}}), Eq. ({\ref{Eq4.39}}) for the one $\delta $-pole ansatz is 
reproduced. However, now $\omega _{\vec{k}}$ has a much more complicate 
structure.  

It is immediate to see that Eqs. ({\ref{Eq4.148}}) and ({\ref{Eq4.149}}) can 
be rewritten as (with $\mathcal{M}_{\nu }(\vec{k})=M_{\nu }(\vec{k})/(2Nm)$)%
%
%
\begin{equation} 
\begin{array}{ll} 
\displaystyle{\omega _{\vec{k}}+\frac{1}{2}{\Gamma _{\vec{k}}/\omega _{\vec{k}}}=\mathcal{M%
}_{1}(\vec{k})} &  \\ 
\\
\displaystyle{\omega _{\vec{k}}^{2}+\frac{3}{2}\Gamma _{\vec{k}}=\mathcal{M}_{2}(\vec{k}),} 
& 
\end{array} 
\label{Eq4.151} 
\end{equation} 
where
\begin{eqnarray} 
\mathcal{M}_{1}(\vec{k}) &=&\omega _{\vec{k}}^{(Ty)}+\frac{{T}}{N}%
\sum\limits_{\vec{k}^{\prime }}\frac{{\left[ {J(\vec{k}^{\prime })-J(\vec{k}-%
\vec{k}^{\prime })}\right] }}{{\omega _{\vec{k}^{\prime }}}}  \notag \\ 
&&+\frac{m}{{S^{2}}}\frac{{T^{2}}}{{N^{2}}}\sum\limits_{\vec{k}_{1},\vec{k}%
_{2}}\frac{{\left[ {J(\vec{k}_{1}-\vec{k}_{2})-J(\vec{k}-(\vec{k}_{1}-\vec{k}%
_{2}))}\right] }}{{\omega _{\vec{k}_{1}}\omega _{\vec{k}_{2}}}}, 
\label{Eq4.152} 
\end{eqnarray} 
and
\begin{eqnarray} 
\mathcal{M}_{2}(\vec{k}) &=&\left( {\omega _{\vec{k}}^{(Ty)}}\right) ^{2}+{%
2\omega _{\vec{k}}^{(Ty)}}\frac{{T}}{N}\sum\limits_{\vec{k}^{\prime }}\frac{{%
\left[ {J(\vec{k}^{\prime })-J(\vec{k}-\vec{k}^{\prime })}\right] }}{{\omega 
_{\vec{k}^{\prime }}}}+  \notag \\ 
&&+\frac{m}{{S^{2}}}\frac{{T^{2}}}{{N^{2}}}\sum\limits_{\vec{k}_{1},\vec{k}%
_{2}}\frac{{\left[ {J(\vec{k}_{1}-\vec{k}_{2})-J(\vec{k}-(\vec{k}_{1}-\vec{k}%
_{2}))}\right] }}{{\omega _{\vec{k}_{1}}\omega _{\vec{k}_{2}}}}\times  
\notag \\ 
&&\times {\left[ {2\omega _{\vec{k}}^{(Ty)}+m\left[ {J(\vec{k}_{1}-\vec{k}%
_{2})-J(\vec{k}-(\vec{k}_{1}-\vec{k}_{2}))}\right] }\right] .} 
\label{Eq4.153} 
\end{eqnarray} 
Now, consistently with the near saturation condition and with $\Gamma _{\vec{%
k}}/\omega _{\vec{k}}^{2}\ll 1$, for which the concept of damped oscillations 
has a physical meaning (see Subsec. 3.2), we show below that a solution of 
the self-consistent problem ({\ref{Eq4.151}})-({\ref{Eq4.153}}) exists in 
the low-temperature regime. By simple manipulations of Eq. ({\ref{Eq4.151}}%
), we find
\begin{equation} 
\frac{1}{2}\Gamma _{\vec{k}}\left( {1-{\frac{{\Gamma _{\vec{k}}}}{{2\omega _{%
\vec{k}}^{2}}}}}\right) =\mathcal{M}_{2}(\vec{k})-\mathcal{M}_{1}^{2}(\vec{k}%
),  \label{Eq4.154} 
\end{equation} 
and, for $\Gamma _{\vec{k}}/\omega _{\vec{k}}^{2}\ll 1$, we have formally%
\begin{equation} 
\Gamma _{\vec{k}}\simeq \mathcal{M}_{2}({\vec{k}})-\mathcal{M}_{1}^{2}({\vec{%
k}}).  \label{Eq4.155} 
\end{equation} 
Thus, from Eqs. ({\ref{Eq4.151}}), it follows
\begin{equation} 
\omega _{{\vec{k}}}^{2}\simeq 3\mathcal{M}_{1}^{2}({\vec{k}})-2\mathcal{M}%
_{2}({\vec{k}}).  \label{Eq4.156} 
\end{equation} 
Of course, since $\Gamma _{\vec{k}}$ is expressed in terms of $\omega _{\vec{%
k}}$ through $\mathcal{M}_{\nu }({\vec{k}})$ $(\nu =1,2)$, our problem is 
reduced to solve the self-consistent Eq. ({\ref{Eq4.156}}) for the 
dispersion relation $\omega _{\vec{k}}$ under physical conditions $\Gamma _{%
\vec{k}}\geq 0$ and $\Gamma _{\vec{k}}/\omega _{\vec{k}}^{2}\ll 1$.  

Focusing on the low-temperature limit and neglecting in Eqs. ({\ref{Eq4.152}}%
)-({\ref{Eq4.153}}) terms $O(T^{n},n\geq 3)$, with $m|_{T=0}=S$ and $\omega 
_{{\vec{k}}}^{2}|_{T=0}=[h+S(J(0)-J({\vec{k}}))]^{2}=\omega _{{0\vec{k}}%
}^{2} $, Eqs. ({\ref{Eq4.155}}) and ({\ref{Eq4.156}}) (with $\omega _{{\vec{k%
}}}\geq 0$) yield (as $N\rightarrow \infty $)
\begin{equation} 
\Gamma _{\vec{k}}\simeq T^{2}\int\limits_{1BZ}{{\frac{{d^{d}k_{1}}}{{(2\pi 
)^{d}}}}}\int\limits_{1BZ}{{\frac{{d^{d}k_{2}}}{{(2\pi )^{d}}}}}{\frac{{\Phi 
(\vec{k};\vec{k}_{1},\vec{k}_{2})}}{{\left[ {h+S(J(0)-J(\vec{k}_{1}))}\right] %
\left[ {h+S(J(0)-J(\vec{k}_{2}))}\right] }},}  \label{Eq4.157} 
\end{equation} 
and
\begin{equation} 
\omega _{\vec{k}}\simeq \left[ {h+S(J(0)-J(\vec{k}))}\right] -{\frac{{T}}{S}}%
\left[ \mathcal{F}{_{1}^{(p)}(\bar{h})\Omega ^{(p)}(\vec{k})+}\mathcal{F}{%
_{2}^{(p)}(\bar{h},\vec{k})}\right] +O(T^{2}),  \label{Eq4.158} 
\end{equation} 
where
\begin{eqnarray} 
\Phi (\vec{k};\vec{k}_{1},\vec{k}_{2})=J^{2}\left\{ \left[ {\Omega ^{(p)}(%
\vec{k}}_{1}{-\vec{k}_{2})-\Omega ^{(p)}(\vec{k}-(\vec{k}_{1}-\vec{k}_{2}))}%
\right] ^{2}-\right.  \notag \\ 
+\left. {\left[ {\Omega ^{(p)}(\vec{k}_{1})-\Omega ^{(p)}(\vec{k}-\vec{k}%
_{1})}\right] \left[ {\Omega ^{(p)}(\vec{k}_{2})-\Omega ^{(p)}(\vec{k}-\vec{k%
}_{2})}\right] }\right\} {.}  \label{Eq4.159} 
\end{eqnarray} 
Of course, the solution ({\ref{Eq4.157}})-({\ref{Eq4.158}}) satisfies the 
conditions assumed at beginning. It is worth noting that, to the leading order 
in $T$, the frequency spectrum of the spin excitations coincides with the one 
obtained by means of the one $\delta $-function ansatz for $\Lambda _{\vec{k}%
}(\omega )$ (see Eq. (\ref{Eq4.60})). A similar situation occurs for $m$ for 
which the expression ({\ref{Eq4.119}}) was found. Now, using the extension 
of the CSDM including the modified Gaussian functional structures for $\Lambda _{%
\vec{k}}(\omega )$, we have been able to determine the low-temperature 
damping of the classical oscillations.  

The integrals involved in the previous solution are very complicated for 
general $d$ and $p$. An estimate could be obtained, as usual, using for $J({%
\vec{k}})$ or $\Omega ^{(p)}(\vec{k})$ the dominant contributions as $|{\vec{%
k}}|\rightarrow 0$ (see Eq. ({\ref{Eq4.64}})). It is instructive, however, 
to present analytical results for case $d=1$ and SRI's ($p=\infty $) for 
which the integrals can be exactly performed. On finds indeed, with $%
J(0)-J(k)=2J(1-\cos k)$ $(0\leq k\leq 2\pi $)
\begin{eqnarray} 
\Gamma _{k} &\simeq &16J^{2}T^{2}(1-\cos k)^{2}\left[ {\int_{0}^{\pi }{{%
\frac{{dk}}{\pi }}{\frac{1}{{h+2SJ(1-\cos k)}}}}}\right] ^{2}  \notag \\ 
&\simeq &{\frac{16J^{2}T^{2}}{h(h+4SJ)}}(1-\cos k)^{2},  \label{Eq4.160} 
\end{eqnarray} 
and
\begin{equation} 
\omega _{k}\simeq h+4JS\sin ^{2}\left( {{\frac{k}{2}}}\right) \left\{ {1-{%
\frac{{T}}{{4JS^{2}}}}\left[ {1-\left( {{\frac{h}{{h+4SJ}}}}\right) }\right] 
}\right\} .  \label{Eq4.161} 
\end{equation} 
Notice that, due to the approximation used for $S_{i}^{z}$ under nearly 
saturation conditions, the previous results break down at any given low 
temperature where the external magnetic field becomes sufficiently small (see 
discussion in Subsec. 4.4.1). In particular, for $d=1$ and $p=\infty $, they 
have a physical meaning only if $T\ll h^{1/2}$ for very low values of $h$ 
(see Eq. ({\ref{Eq4.160}})).  

It is interesting to compare the results ({\ref{Eq4.160}}) and ({\ref 
{Eq4.161}}), here obtained by means of the extended CSDM, with those 
obtained by different methods applied directly to the Heisenberg 
ferromagnetic chain with short-range interactions. First, the Gaussian peak 
width $\Gamma _{k}^{1/2}$ exhibits the correct low-temperature behavior ($%
\propto T$) found by other authors \cite{Balucani82,Reither80}. Furthermore, 
our result for $\omega _{k}$ is consistent with the corresponding exact one 
obtained in Ref. \cite{Reither80}. Of course, further comparisons for the cases $%
d\geq 1$ and $p\neq \infty $ could be performed by means of numerical 
calculation taking also into account that we find $\Gamma _{\overrightarrow{k%
}}^{1/2}\propto T$ for any $d$ and $p$ with $h\neq 0$.  

\section{Conclusions} 

In this article we began with a review of the general formalism of the 
two-time GF's techniques in classical statistical physics in a form quite 
parallel to the well known version in quantum many-body theory. Particular 
emphasis has been given to the spectral properties and to the concept of SD 
which plays a relevant role for our purposes. The EMM for calculation of 
classical two-time GF's is also presented but we have mainly focused on the 
general ingredients of the less known classical version of the SDM. This 
method, introduced by Kalashnikov and Fradkin \cite 
{Kalashnikov69,Kalashnikov73} many years ago in quantum statistical physics, 
has been extensively tested by several authors for a wide variety of quantum 
many-body systems \cite 
{Kalashnikov69,Kalashnikov73,Campana79,Caramico80,Nolting89,Nolting91,Hermann97}%
. We have explained in detail, also with a pedagogical aim, how the CSDM 
works in classical many-body physics for obtaining systematic 
nonperturbative approximations to explore the macroscopic properties of a 
wide variety of systems, also including phase transitions and critical 
phenomena, without the explicit calculation of the (generalized) partition 
function. Besides, general insights and key ideas, for an implementation of 
the method to investigate the dispersion relation and the damping of the 
classical excitations on the same footing, are also given. The possibility 
to formulate a many-body theory in the framework of the classical 
statistical mechanics offers the remarkable advantage to draw insights 
directly from the wide baggage of powerful mathematical techniques and 
approximation methods developed in many years of research activity in 
quantum many-body theory which have sensibly contributed to the great 
progress in condensed matter physics and, more generally, in quantum 
statistical physics. It is also worth noting that the classical version of 
both the EMM and the SDM, with the use of commuting dynamical variables, 
allows to avoid the intrinsic difficulties related to the noncommutability 
of the operators which are the basic elements of the quantum framework. This 
constitutes a very favorable opportunity to have physically relevant 
information about the macroscopic properties of microscopic models when the 
quantum fluctuations do not play an important role. Of course, the use of 
classical dynamical variables and the machinery of a reliable classical 
many-body framework is also particularly advantageous for numerical 
simulations.  

Next, the effectiveness of the CSDM has been explicitly checked with the 
study of the thermodynamic properties of a highly nontrivial classical $d$%
-dimensional spin-$S$ Heisenberg FM model with LRI's decaying as $r^{-p}$ ($%
p>d$) in the presence of a magnetic field. However, the method is general 
and may be conveniently used for studying other classical many-body systems 
especially when a reliable nonperturbative method is required. Focusing on 
the selected spin model and working in some detail to lowest order in the 
CSDM, a rich physical scenario has been obtained. The main magnetic 
quantities in the nearly saturated regime have been calculated analytically 
as functions of $d$ and $p$. The FM LRO at finite temperature has been 
shown, in a transparent way, to take place in a wide region of the $(p-d)$%
-plane consistently with exact and numerical predictions. The thermodynamic 
regimes with zero and nearly zero magnetization have been also investigated by 
means of a reliable decoupling procedure, as suggested by quantum studies 
\cite{Kalashnikov73,Kamieniarz77}, which takes into account the effect of 
longitudinal spin correlations. This allowed us to estimate the critical 
temperature, the critical properties in terms of $d$ and $p$ beyond the mean 
field approximation and the low-temperature behavior of the paramagnetic 
susceptibility when LRO is absent.  

All this supports our conviction that the CSDM, as in the quantum case, 
constitutes a very effective method for describing the macroscopic 
properties of classical many-body systems. It is indeed surprising that, 
already to the lowest order of approximation, the method is able to capture, 
in a relatively simple way, the relevant physics of an highly untrivial spin 
model, consistently with MC simulations \cite{Romano89,Romano90,Vassiliev01} 
and exact results \cite 
{Kunz76,Frolich78,Roger81,Pfister81,Bruno01,Kosterlitz76,Haldane88}. It is 
also worth pointing that the method provides the possibility to perform, 
systematically and in a transparent way, higher-order approximations with 
the aim to improve the results here presented. Some suggestions along this 
direction, also for exploring the damping of the classical excitations, have 
been given in Secs. 4.5 and 4.6, but new ideas can be profitably drawn from 
the big experience acquired working by means of the quantum version of the 
SDM.  

In conclusion, we believe that the CSDM and the formalism of the two-time 
GF's in classical statistical mechanics constitute a powerful tool to study 
efficaciously the equilibrium and transport properties of classical 
many-body systems, also in nonextensivity conditions \cite{Cavallo01}.  
Further developments and applications are, of course, desirable and we 
warmly hope that the present article will stimulate further and more elaborate 
and meaningful studies.




\end{document}